\pdfoutput=1
\documentclass[12pt,a4paper]{article}

\usepackage{ifthen} 
\newboolean{pdflatex}
\setboolean{pdflatex}{true} 

\newboolean{articletitles}
\setboolean{articletitles}{true} 

\newboolean{uprightparticles}
\setboolean{uprightparticles}{false} 

\def\paperauthors{LHCb collaboration} 
\def\paperasciititle{ADS_GLW_Run12_PAPER} 
\def\papertitle{Measurement of $C\!P$ observables in $B^\pm \to D^{(*)} K^\pm$ and $B^\pm \to D^{(*)} \pi^\pm$ decays using two-body $D$ final states} 
\def\paperkeywords{{High Energy Physics}, {LHCb}} 
\def\papercopyright{\the\year\ CERN for the benefit of the LHCb collaboration} 
\def\paperlicence{CC BY 4.0 licence}
\def\paperlicenceurl{https://creativecommons.org/licenses/by/4.0/}

\usepackage[top=1in, bottom=1.25in, left=1in, right=1in]{geometry}

\usepackage{microtype}
\usepackage{lineno}  
\usepackage{xspace} 
\usepackage{caption} 

\usepackage{graphicx}  
\usepackage{color}
\usepackage{colortbl}
\graphicspath{{./figs/}} 

\usepackage{amsmath} 
\usepackage{amssymb}
\usepackage{amsfonts}
\usepackage{upgreek} 

\newcommand*\patchAmsMathEnvironmentForLineno[1]{%
\expandafter\let\csname old#1\expandafter\endcsname\csname #1\endcsname
\expandafter\let\csname oldend#1\expandafter\endcsname\csname
end#1\endcsname
 \renewenvironment{#1}%
   {\linenomath\csname old#1\endcsname}%
   {\csname oldend#1\endcsname\endlinenomath}%
}
\newcommand*\patchBothAmsMathEnvironmentsForLineno[1]{%
  \patchAmsMathEnvironmentForLineno{#1}%
  \patchAmsMathEnvironmentForLineno{#1*}%
}
\AtBeginDocument{%
\patchBothAmsMathEnvironmentsForLineno{equation}%
\patchBothAmsMathEnvironmentsForLineno{align}%
\patchBothAmsMathEnvironmentsForLineno{flalign}%
\patchBothAmsMathEnvironmentsForLineno{alignat}%
\patchBothAmsMathEnvironmentsForLineno{gather}%
\patchBothAmsMathEnvironmentsForLineno{multline}%
\patchBothAmsMathEnvironmentsForLineno{eqnarray}%
}

\usepackage{hyperxmp}

\usepackage[pdftex,
            pdfauthor={\paperauthors},
            pdftitle={\paperasciititle},
            pdfkeywords={\paperkeywords},
            pdfcopyright={Copyright (C) \papercopyright},
            pdflicenseurl={\paperlicenceurl}]{hyperref}

\usepackage[colorinlistoftodos,textsize=scriptsize]{todonotes}

\usepackage[bottom,flushmargin,hang,multiple]{footmisc}

\usepackage[all]{hypcap} 

\usepackage{xspace} 
\usepackage{upgreek}

\def\lhcb   {\mbox{LHCb}\xspace}

\def\MagUp {\mbox{\em Mag\kern -0.05em Up}\xspace}

\ifthenelse{\boolean{uprightparticles}}%
{
 
 \def\Pgamma      {\ensuremath{\upgamma}\xspace}

 \def\Ppi         {\ensuremath{\uppi}\xspace}

 \def\PDelta      {\ensuremath{\Delta}\xspace}                 
 \def\PXi         {\ensuremath{\Xi}\xspace}                 
 \def\PLambda     {\ensuremath{\Lambda}\xspace}                 
 \def\PSigma      {\ensuremath{\Sigma}\xspace}                 
 \def\POmega      {\ensuremath{\Omega}\xspace}                 
 \def\PUpsilon    {\ensuremath{\Upsilon}\xspace}

 \def\PB      {\ensuremath{\mathrm{B}}\xspace}                 
                  
 \def\PD      {\ensuremath{\mathrm{D}}\xspace}

 \def\PK      {\ensuremath{\mathrm{K}}\xspace}

 \def\Pb      {\ensuremath{\mathrm{b}}\xspace}                 
 \def\Pc      {\ensuremath{\mathrm{c}}\xspace}

 \def\Pi      {\ensuremath{\mathrm{i}}\xspace}

 \def\Ps      {\ensuremath{\mathrm{s}}\xspace}

 \def\thebaroffset{0.0em}
}
{
 
 \def\Pgamma      {\ensuremath{\gamma}\xspace}

 \def\Ppi         {\ensuremath{\pi}\xspace}

 \mathchardef\PDelta="7101
 \mathchardef\PXi="7104
 \mathchardef\PLambda="7103
 \mathchardef\PSigma="7106
 \mathchardef\POmega="710A
 \mathchardef\PUpsilon="7107
                  
 \def\PB      {\ensuremath{B}\xspace}                 
                  
 \def\PD      {\ensuremath{D}\xspace}

 \def\PK      {\ensuremath{K}\xspace}

 \def\Pb      {\ensuremath{b}\xspace}                 
 \def\Pc      {\ensuremath{c}\xspace}

 \def\Pi      {\ensuremath{i}\xspace}

 \def\Ps      {\ensuremath{s}\xspace}

 \def\thebaroffset{0.18em}
}
\newcommand{\offsetoverline}[2][\thebaroffset]{\kern #1\overline{\kern -#1 #2}}%

\makeatletter
\ifcase \@ptsize \relax
  \newcommand{\miniscule}{\@setfontsize\miniscule{4}{5}}
\or
  \newcommand{\miniscule}{\@setfontsize\miniscule{5}{6}}
\or
  \newcommand{\miniscule}{\@setfontsize\miniscule{5}{6}}
\fi
\makeatother

\DeclareRobustCommand{\optbar}[1]{\shortstack{{\miniscule (\rule[.5ex]{1.25em}{.18mm})}
  \\ [-.7ex] $#1$}}

\def\g      {{\ensuremath{\Pgamma}}\xspace}

\def\squark    {{\ensuremath{\Ps}}\xspace}

\def\cquark    {{\ensuremath{\Pc}}\xspace}

\def\bquark    {{\ensuremath{\Pb}}\xspace}

\def\pion   {{\ensuremath{\Ppi}}\xspace}
\def\piz    {{\ensuremath{\pion^0}}\xspace}
\def\pip    {{\ensuremath{\pion^+}}\xspace}
\def\pim    {{\ensuremath{\pion^-}}\xspace}
\def\pipm   {{\ensuremath{\pion^\pm}}\xspace}
\def\pimp   {{\ensuremath{\pion^\mp}}\xspace}

\def\kaon    {{\ensuremath{\PK}}\xspace}

\def\KorKbar {\kern \thebaroffset\optbar{\kern -\thebaroffset \PK}{}\xspace}

\def\Kp      {{\ensuremath{\kaon^+}}\xspace}
\def\Km      {{\ensuremath{\kaon^-}}\xspace}
\def\Kpm     {{\ensuremath{\kaon^\pm}}\xspace}
\def\Kmp     {{\ensuremath{\kaon^\mp}}\xspace}
\def\KS      {{\ensuremath{\kaon^0_{\mathrm{S}}}}\xspace}

\def\Dbar    {{\ensuremath{\offsetoverline{\PD}}}\xspace}
\def\D       {{\ensuremath{\PD}}\xspace}
\def\Db      {{\ensuremath{\Dbar}}\xspace}
\def\DorDbar {\kern \thebaroffset\optbar{\kern -\thebaroffset \PD}\xspace}
\def\Dz      {{\ensuremath{\D^0}}\xspace}
\def\Dzb     {{\ensuremath{\Dbar{}^0}}\xspace}
\def\Dp      {{\ensuremath{\D^+}}\xspace}
\def\Dm      {{\ensuremath{\D^-}}\xspace}

\def\DpDm    {\ensuremath{\Dp {\kern -0.16em \Dm}}\xspace}
\def\Dstar   {{\ensuremath{\D^*}}\xspace}

\def\Dstarz  {{\ensuremath{\D^{*0}}}\xspace}

\def\Dstarp  {{\ensuremath{\D^{*+}}}\xspace}

\def\B       {{\ensuremath{\PB}}\xspace}
\def\Bbar    {{\ensuremath{\offsetoverline{\PB}}}\xspace}

\def\BorBbar {\kern \thebaroffset\optbar{\kern -\thebaroffset \PB}\xspace}
\def\Bz      {{\ensuremath{\B^0}}\xspace}
\def\Bzb     {{\ensuremath{\Bbar{}^0}}\xspace}
\def\Bd      {{\ensuremath{\B^0}}\xspace}

\def\BdorBdbar {\kern \thebaroffset\optbar{\kern -\thebaroffset \Bd}\xspace}
\def\Bu      {{\ensuremath{\B^+}}\xspace}
\def\Bub     {{\ensuremath{\B^-}}\xspace}
\def\Bp      {{\ensuremath{\Bu}}\xspace}
\def\Bm      {{\ensuremath{\Bub}}\xspace}
\def\Bpm     {{\ensuremath{\B^\pm}}\xspace}
\def\Bmp     {{\ensuremath{\B^\mp}}\xspace}
\def\Bs      {{\ensuremath{\B^0_\squark}}\xspace}

\def\BsorBsbar {\kern \thebaroffset\optbar{\kern -\thebaroffset \Bs}\xspace}

\def\Y#1S{\ensuremath{\PUpsilon{(#1S)}}\xspace}

\def\Lz          {{\ensuremath{\PLambda}}\xspace}

\def\LorLbar     {\kern \thebaroffset\optbar{\kern -\thebaroffset \PLambda}\xspace}

\def\Lc          {{\ensuremath{\Lz^+_\cquark}}\xspace}

\def\Lb           {{\ensuremath{\Lz^0_\bquark}}\xspace}

\def\to                 {\ensuremath{\rightarrow}\xspace}

\def\CP                {{\ensuremath{C\!P}}\xspace}

\def\AT#1     {\ensuremath{A_{\mathrm{T}}^{#1}}\xspace}           

\def\C#1      {\ensuremath{\mathcal{C}_{#1}}\xspace}                       
\def\Cp#1     {\ensuremath{\mathcal{C}_{#1}^{'}}\xspace}                    
\def\Ceff#1   {\ensuremath{\mathcal{C}_{#1}^{\mathrm{(eff)}}}\xspace}        
\def\Cpeff#1  {\ensuremath{\mathcal{C}_{#1}^{'\mathrm{(eff)}}}\xspace}       
\def\Ope#1    {\ensuremath{\mathcal{O}_{#1}}\xspace}                       
\def\Opep#1   {\ensuremath{\mathcal{O}_{#1}^{'}}\xspace}                    

\newcommand{\nospaceunit}[1]{\ensuremath{\text{#1}}}       
\newcommand{\aunit}[1]{\ensuremath{\text{\,#1}}}       

\newcommand{\tev}{\aunit{Te\kern -0.1em V}\xspace}
\newcommand{\gev}{\aunit{Ge\kern -0.1em V}\xspace}
\newcommand{\mev}{\aunit{Me\kern -0.1em V}\xspace}
\newcommand{\kev}{\aunit{ke\kern -0.1em V}\xspace}
\newcommand{\ev}{\aunit{e\kern -0.1em V}\xspace}
 
\newcommand{\mevc}{\ensuremath{\aunit{Me\kern -0.1em V\!/}c}\xspace}
\newcommand{\gevc}{\ensuremath{\aunit{Ge\kern -0.1em V\!/}c}\xspace}
\newcommand{\mevcc}{\ensuremath{\aunit{Me\kern -0.1em V\!/}c^2}\xspace}
\newcommand{\gevcc}{\ensuremath{\aunit{Ge\kern -0.1em V\!/}c^2}\xspace}

\def\mum  {\ensuremath{\,\upmu\nospaceunit{m}}\xspace}

\def\fb   {\ensuremath{\aunit{fb}}\xspace}
\def\invfb   {\ensuremath{\fb^{-1}}\xspace}

\newcommand{\chisq}{\ensuremath{\chi^2}\xspace}

\newcommand{\chisqip}{\ensuremath{\chi^2_{\text{IP}}}\xspace}

\def\gsim{{~\raise.15em\hbox{$>$}\kern-.85em
          \lower.35em\hbox{$\sim$}~}\xspace}
\def\lsim{{~\raise.15em\hbox{$<$}\kern-.85em
          \lower.35em\hbox{$\sim$}~}\xspace}

\def\pt         {\ensuremath{p_{\mathrm{T}}}\xspace}

\def\ptot       {\ensuremath{p}\xspace}

\def\evtgen     {\mbox{\textsc{EvtGen}}\xspace}

\def\geant      {\mbox{\textsc{Geant4}}\xspace}

\def\photos     {\mbox{\textsc{Photos}}\xspace}

\def\pythia     {\mbox{\textsc{Pythia}}\xspace}

\def\tell1  {TELL1\xspace}
\def\ukl1   {UKL1\xspace}

\usepackage{cite} 
\usepackage{mciteplus}

\usepackage{longtable} 
\usepackage{placeins}
\allowdisplaybreaks
\usepackage{rotating}
\usepackage{float}
\usepackage{booktabs}

\begin{document}

\renewcommand{\thefootnote}{\fnsymbol{footnote}}
\setcounter{footnote}{1}


\begin{titlepage}
\pagenumbering{roman}

\vspace*{-1.5cm}
\centerline{\large EUROPEAN ORGANIZATION FOR NUCLEAR RESEARCH (CERN)}
\vspace*{1.5cm}
\noindent
\begin{tabular*}{\linewidth}{lc@{\extracolsep{\fill}}r@{\extracolsep{0pt}}}
\ifthenelse{\boolean{pdflatex}}
{\vspace*{-1.5cm}\mbox{\!\!\!\includegraphics[width=.14\textwidth]{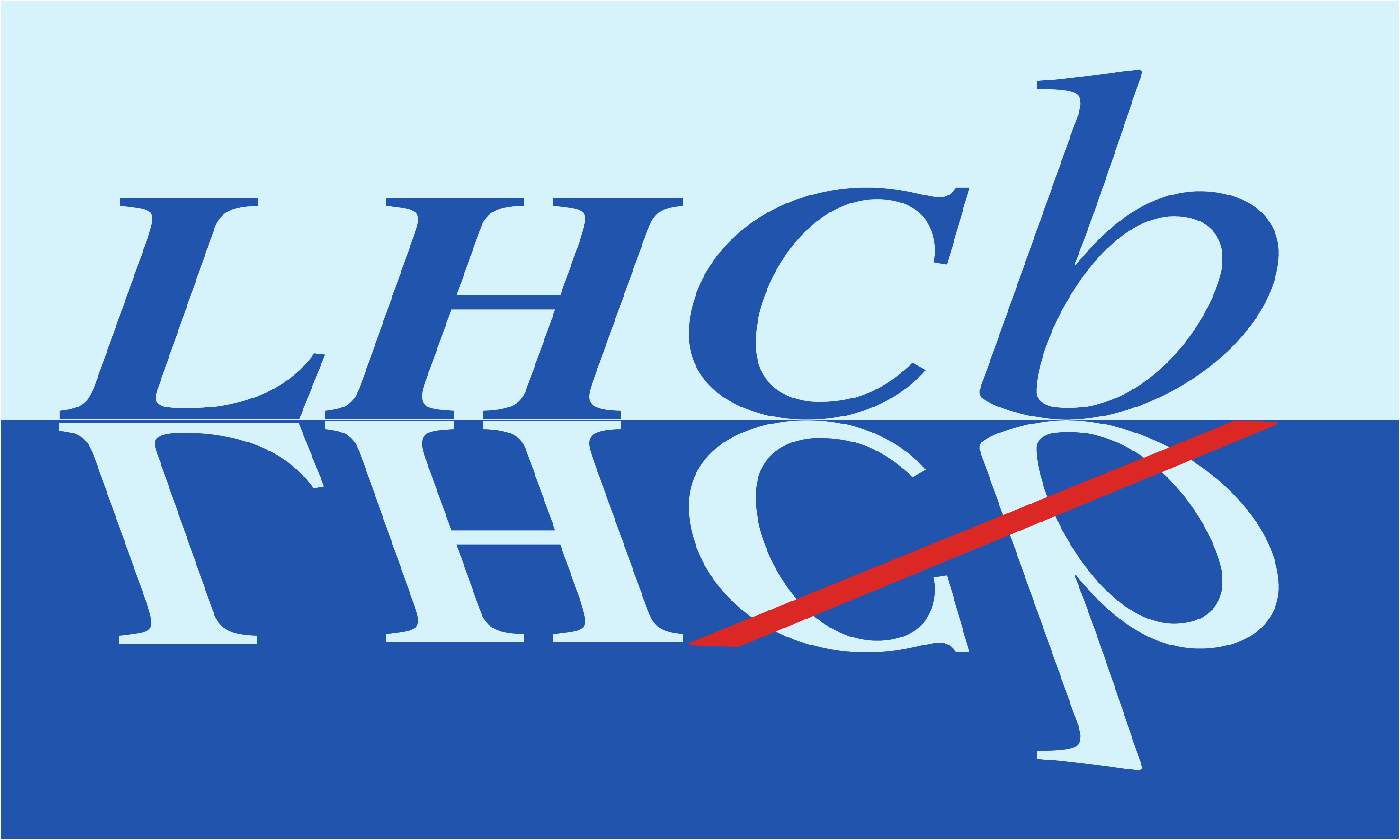}} & &}%
{\vspace*{-1.2cm}\mbox{\!\!\!\includegraphics[width=.12\textwidth]{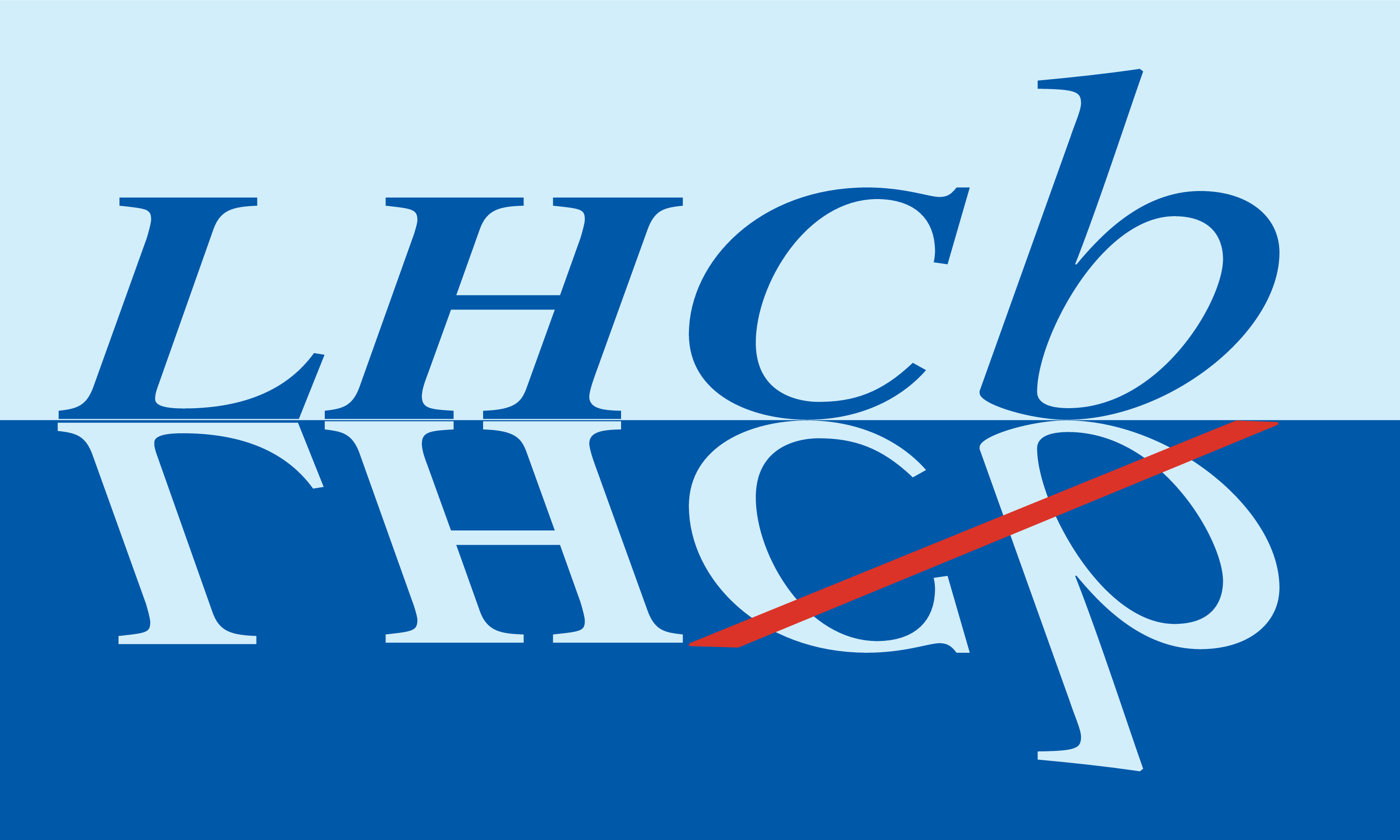}} & &}%
\\
 & & CERN-EP-2020-225 \\  
 & & LHCb-PAPER-2020-036 \\  
 & & April 15, 2021 \\ 
 & & \\
\end{tabular*}

\vspace*{2.0cm}

{\normalfont\bfseries\boldmath\huge
\begin{center}
  \papertitle 
\end{center}
}

\vspace*{2.0cm}

\begin{center}
\paperauthors\footnote{Authors are listed at the end of this paper.}
\end{center}

\vspace{\fill}

\begin{abstract}
  \noindent
  Measurements of $C\!P$ observables in $B^\pm \rightarrow D^{(*)} K^\pm$ and $B^\pm \rightarrow D^{(*)} \pi^\pm$ decays are presented, where $D^{(*)}$ indicates a neutral $\D$ or $\Dstar$ meson that is an admixture of meson and anti-meson states. Decays of the $\Dstar$ meson to the $\D\piz$ and $\D\gamma$ final states are partially reconstructed without inclusion of the neutral pion or photon. Decays of the $\D$ meson are reconstructed in the $\Kpm \pimp$, $\Kp\Km$, and $\pip\pim$ final states. The analysis uses a sample of charged $B$ mesons produced in proton-proton collisions and collected with the LHCb experiment, corresponding to integrated luminosities of 2.0, 1.0, and 5.7\invfb taken at centre-of-mass energies of 7, 8, and 13\tev, respectively. The measurements of partially reconstructed $\Bpm \to \Dstar \Kpm$ and $\Bpm \to \Dstar \pipm$ with $D \to \Kmp \pipm$ decays are the first of their kind, and a first observation of the $\Bpm \to (D \pi^0)_{D^*} \pipm$ decay is made with a significance of 6.1 standard deviations. All $C\!P$ observables are measured with world-best precision, and in combination with other LHCb results will provide strong constraints on the CKM angle $\gamma$.
\end{abstract}

\vspace*{2.0cm}

\begin{center}
Published in JHEP 04 (2021) 081.
\end{center}

\vspace{\fill}

{\footnotesize 
\centerline{\copyright~\papercopyright. \href{\paperlicenceurl}{\paperlicence}.}}
\vspace*{2mm}

\end{titlepage}


\newpage
\setcounter{page}{2}
\mbox{~}
%
%
%
%


\renewcommand{\thefootnote}{\arabic{footnote}}
\setcounter{footnote}{0}

\cleardoublepage


\pagestyle{plain} 
\setcounter{page}{1}
\pagenumbering{arabic}


\section{Introduction}
\label{sec:Introduction}
 
Overconstraining the Unitarity Triangle (UT) derived from the Cabibbo-Kobayashi-Maskawa (CKM) quark-mixing matrix is central to testing the Standard Model description of charge-parity (\CP) violation~\cite{Cabibbo:1963yz,*Kobayashi:1973fv}. The least well-known angle of the UT is \mbox{$\gamma \equiv \text{arg}(-V_{ud}V_{ub}^{*}/V_{cd}V_{cb}^{*})$}, which has been determined with a precision of about $5^\circ$ from a combination of measurements~\cite{LHCb-CONF-2018-002,HFLAV18} and recently with a standalone precision of $5.2^\circ$ by LHCb using $\Bm \to D h^-$ ($h^- \in \{\pim,\Km\}$) with $D \to \KS h^+ h^-$ decays~\cite{LHCb-PAPER-2020-019}.\footnote{ The inclusion of charge-conjugate processes is implied throughout except in discussions of asymmetries.} The angles $\alpha$ and $\beta$ are measured with $4.5^\circ$ and $< 1^\circ$ precision, respectively~\cite{Charles:2004jd,Bona:2006ah}. Among the UT angles, \g is unique in that it does not depend on any top-quark coupling, and can thus be measured in $B$-hadron decays that are dominated by tree-level contributions. In such decays, the interpretation of physical observables (rates and \CP asymmetries) in terms of the underlying UT parameters is subject to negligible theoretical uncertainties~\cite{Brod2014}. Any disagreement between measurements of \g and the value inferred from global CKM fits performed without any \g information would thus invalidate the Standard Model description of \CP violation.

The most powerful method for determining $\gamma$ in decays dominated by tree-level contributions is through the measurement of relative partial widths in $\Bm \to D\Km$ decays, where $D$ represents an admixture of the \Dz and \Dzb states. The amplitude for the $\Bm \to \Dz \Km$ decay, which at the quark level proceeds via a $b \to c\bar{u}s$ transition, is proportional to the CKM matrix element $V_{cb}$. The corresponding amplitude for the $\Bm \to \Dzb \Km$ decay, which proceeds via a $b\to u\bar{c}s$ transition, is proportional to $V_{ub}$. By studying hadronic \D decays accessible to both \Dz and \Dzb mesons, phase information can be determined from the interference between these two amplitudes. 
The degree of the resulting \CP violation depends on the size of $r_B^{DK}$, the ratio of the magnitudes of the $\Bm \to \Dzb \Km$ and $\Bm \to \Dz \Km$ amplitudes. 
The relatively large value of $r_B^{DK} \approx 0.10$~\cite{HFLAV18} in $\Bm \to D\Km$ decays allows the determination of the relative phase of the two interfering amplitudes. This relative phase has both \CP-violating (\g) and \CP-conserving ($\delta_B^{DK}$) contributions; a measurement of the decay rates for both \Bp and \Bm mesons gives sensitivity to \g. Similar interference effects also occur in $\Bm \to D\pim$ decays, albeit with lower sensitivity to the phases due to additional Cabibbo-suppression which decreases the amplitude ratio relative to $\Bm \to D\Km$ decays by around a factor of 30.

The $\Bm \to \Dstar \Km$ decay, in which the vector $\Dstar$ meson\footnote{ $\Dstar$ represents an admixture of the $D^{*}(2007)^{0}$ and $\Db^{*}(2007)^{0}$ states.} decays to either the $\D \piz$ or $\D \gamma$ final state, also exhibits \CP-violating effects when hadronic \D decays accessible to both \Dz and \Dzb mesons are studied. In this decay, the exact strong-phase difference of $\pi$ between $\Dstar \to \D \piz$ and $\Dstar \to \D \gamma$ decays can be exploited to measure \CP observables for states with opposite \CP eigenvalues~\cite{PhysRevD.70.091503}. The amount of \CP violation observed in $\Bm \to \Dstar \Km$ depends on the size of $r_B^{\Dstar \!K}$, and measurement of the phase for both \Bp and \Bm allows \g and $\delta_B^{\Dstar \!K}$ to be determined. 

The study of $\Bm \to D^{(*)}\Km$ decays for measurements of $\g$ was first suggested for \CP eigenstates of the \D decay, for example the \CP-even $\D \to\Kp\Km$ and $\D \to\pip\pim$ decays, labelled herein as GLW modes~\cite{Gronau:1990ra,Gronau1991172}. Higher sensitivity to $\g$ can be achieved using non-\CP eigenstates such as $D \to \Kp \pim$, where the $\Dz \to \Kp \pim$ and $\Dzb \to \Kp \pim$ decays are related by the amplitude magnitude ratio $r_D^{K\pi}$ and the strong-phase \mbox{difference $\delta_D^{K \pi}$}. The similar magnitude of $r_D^{K\pi}$ and $r_B^{DK}$ leads to significant interference between the two possible suppressed decay paths (favoured $B$ decay followed by suppressed $D$ decay, and suppressed $B$ decay followed by favoured $D$ decay), resulting in large \CP asymmetries. These decays are herein referred to as ADS modes~\cite{Atwood:1996ci}. In this work, the GLW $D \to \Kp \Km$ and $D \to \pip \pim$ modes are considered, as well as the ADS $D \to \Kp \pim$ mode; the favoured $\D \to \Km\pip$ decay is used for normalisation purposes and to define shape parameters in the fit to data. The $\Bm \to D^{(*)} \Km$ and $\Bm \to D^{(*)} \pim$ GLW modes have previously been studied by the LHCb collaboration~\cite{LHCb-PAPER-2017-021}, as have the $\Bm \to D \Km$ and $\Bm \to D \pim$ ADS modes~\cite{LHCb-PAPER-2016-003}. This paper reports updated and improved results for these modes, and a first measurement of the $\Bm \to D^* \Km$ and $\Bm \to D^* \pim$ ADS modes at LHCb. A sample of charged $B$ mesons produced in proton-proton ($pp$) collisions and collected with the LHCb experiment is used, corresponding to integrated luminosities of 2.0, 1.0, and 5.7\invfb taken at centre-of-mass energies of \mbox{$\sqrt{s}$ = 7, 8, and 13\tev}, respectively. The small $\Dstar - \D$ mass difference and the conservation of angular momentum in \mbox{$\Dstar \to \D\piz$} and \mbox{$\Dstar \to \D\gamma$} decays results in distinctive signatures for the $\Bm \to \Dstar h^-$ signal in the $\D h^-$ invariant mass, enabling yields to be obtained with a partial reconstruction technique. Since the reconstruction efficiency for low momentum neutral pions and photons is relatively low in LHCb~\cite{LHCb-DP-2014-002}, the partial reconstruction method provides significantly larger yields compared to full reconstruction. However, the statistical sensitivity per signal decay is reduced since several signal and background components in the same region of $Dh^-$ invariant mass must be distinguished.

A total of 28 measurements of \CP observables are reported, nine of which correspond to the fully reconstructed $\Bm \to D h^-$ decays while the remaining 19 relate to the partially reconstructed $\Bm \to D^* h^-$ decays. 
A summary of all measured \CP observables is provided in Tables~\ref{tab:Observables_FullReco} and~\ref{tab:Observables_PartReco}. 
The \CP observables for the decay $\Bmp\to X$ with $D\to f$ are defined in terms of partial rates, which are related to the underlying parameters $\gamma,\,r_B^X,\,\delta_B^X,\,r_D^f$, and $\delta_D^f$. Including $D$-mixing effects~\cite{Rama:2013voa}, the partial rates for $f \in \{\Kpm\pimp, \Kp\Km, \pip\pim\}$ are
\begin{eqnarray}
\Gamma(\B^\mp\to \left[[f]_D h^\mp\right]_{\!X}) &\propto& (r_D^f)^2+(r_B^X)^2+2r_D^f r_B^X\cos(\delta_B^X+\delta_D^f\mp\gamma) \label{masterADS}\\
&& -\alpha y(1+(r_B^X)^2)r_D^f\cos\delta_D^f -\alpha y(1+(r_D^f)^2)r_B^X\cos(\delta_B^X\mp\gamma)  \nonumber  \\
&&+ \alpha x(1-(r_B^X)^2)r_D^f\sin\delta_D^f-\alpha x(1-(r_D^f)^2)r_B^X\sin(\delta_B^X\mp\gamma)\,, \nonumber  
\end{eqnarray}
where $x$ and $y$ are the charm mixing parameters, and $\alpha$ is an analysis-specific coefficient that quantifies the decay-time acceptance of the candidate $D$ mesons. It is noted that $r_D^f=1$ and $\delta_D^f=0$ for the GLW modes, so the \CP observables are unaffected by charm mixing. The favoured mode partial widths are similarly defined,
\begin{eqnarray}
\Gamma(\B^\mp\to\left[[\Kmp\pipm]_D h^\mp\right]_{\!X}) &\propto& 1+(r_D^{K\pi})^2(r_B^X)^2+2r_D^{K\pi} r_B^X\cos(\delta_B^X-\delta_D^{K\pi}\mp\gamma) \label{masterFAV}\\
&& -\alpha y(1+(r_B^X)^2)r_D^{K\pi}\cos\delta_D^{K\pi} -\alpha y(1+(r_D^{K\pi})^2)r_B^X\cos(\delta_B^X\mp\gamma)  \nonumber  \\
&&- \alpha x(1-(r_B^X)^2)r_D^{K\pi}\sin\delta_D^{K\pi}+\alpha x(1-(r_D^{K\pi})^2)r_B^X\sin(\delta_B^X\mp\gamma)\,, \nonumber  
\end{eqnarray}
although the mixing effects are negligible. The GLW modes $D \to \Kp \Km$ and $D \to \pip \pim$ are described using common \CP observables in the analysis, accounting for small differences due to the charm \CP asymmetry difference $\Delta A_{\CP}$~\cite{HFLAV18}. In addition to the \CP observables, the branching fractions $\mathcal{B}(\Bm \to \Dstarz \pim)$ and $\mathcal{B}(\Dstarz \to \Dz \piz)$ are measured. 

\begingroup
\renewcommand*{\arraystretch}{2.1}
\begin{table}[!h]
\centering
\caption{The nine \CP observables measured using $\Bm \to D h^-$ decays, defined in terms of $B$ meson
decay widths. Where indicated, $h^+h^-$ represents an average of the $D \to \Kp \Km$ and
$D \to \pip\pim$ modes. The $R$ observables represent partial width ratios and double
ratios. The \mbox{$A$ observables} represent \CP asymmetries.}
\small
\begin{tabular}{l l}
\toprule
Observable & Definition \\ \midrule 

$A_K^{CP}$ & $\frac{\Gamma(\Bm \to [h^+h^-]_{D} \Km) \phantom{0}-\phantom{0} \Gamma(\Bp \to [h^+h^-]_{D} \Kp )}{\Gamma(\Bm \to [h^+h^-]_{D} \Km) \phantom{0}+\phantom{0} \Gamma(\Bp \to [h^+h^-]_{D} \Kp )}$ \\

$A_\pi^{CP}$ & $\frac{\Gamma(\Bm \to [h^+h^-]_{D} \pim) \phantom{0}-\phantom{0} \Gamma(\Bp \to [h^+h^-]_{D} \pip )}{\Gamma(\Bm \to [h^+h^-]_{D} \pim) \phantom{0}+\phantom{0} \Gamma(\Bp \to [h^+h^-]_{D} \pip )}$ \\

$A_K^{K\pi}$ & $\frac{\Gamma(\Bm \to [\Km \pip ]_{D} \Km) \phantom{0}-\phantom{0} \Gamma(\Bp \to [\Kp \pim]_{D} \Kp )}{\Gamma(\Bm \to [\Km \pip ]_{D} \Km) \phantom{0}+\phantom{0} \Gamma(\Bp \to [\Kp \pim]_{D} \Kp )}$ \\

$R^{CP}$ & $\frac{\Gamma(\Bm \to [h^+h^-]_{D} \Km) \phantom{0}+\phantom{0} \Gamma(\Bp \to [h^+h^-]_{D} \Kp )}{\Gamma(\Bm \to [h^+h^-]_{D} \pim) \phantom{0}+\phantom{0} \Gamma(\Bp \to [h^+h^-]_{D} \pip )} \times \frac{1}{R_{K/\pi}^{K\pi}}$ \\

$R_{K/\pi}^{K\pi}$ & $\frac{\Gamma(\Bm \to [\Km \pip ]_{D} \Km) \phantom{0}+\phantom{0} \Gamma(\Bp \to [\Kp \pim]_{D} \Kp )}{\Gamma(\Bm \to [\Km \pip ]_{D} \pim) \phantom{0}+\phantom{0} \Gamma(\Bp \to [\Kp \pim]_{D} \pip )}$ \\

$R_{K^-}^{\pi K}$ & $\frac{\Gamma(\Bm \to [\Kp \pim]_{D} \Km)}{\Gamma(\Bm \to [\Km \pip]_{D} \Km )}$ \\

$R_{\pi^-}^{\pi K}$ & $\frac{\Gamma(\Bm \to [\Kp \pim]_{D} \pim)}{\Gamma(\Bm \to [\Km \pip]_{D} \pim )}$ \\

$R_{K^+}^{\pi K}$ & $\frac{\Gamma(\Bp \to [\Km \pip]_{D} \Kp)}{\Gamma(\Bp \to [\Kp \pim]_{D} \Kp )}$ \\

$R_{\pi^+}^{\pi K}$ & $\frac{\Gamma(\Bp \to [\Km \pip]_{D} \pip)}{\Gamma(\Bp \to [\Kp \pim]_{D} \pip )}$ \\

\bottomrule

\end{tabular}
\label{tab:Observables_FullReco}
\end{table}
\endgroup

\begingroup
\renewcommand*{\arraystretch}{2.1}
\begin{table}[!htbp]
\centering
\caption{The 19 \CP observables measured using $\Bm \to D^* h^-$ decays, defined in terms of $B$ meson decay widths. Where indicated, $h^+h^-$ represents an average of the $D \to \Kp \Km$ and
$D \to \pip\pim$ modes. The $R$ observables represent partial width ratios and double
ratios. The $A$ observables represent \CP asymmetries. The $R_{K/\pi}^{K\pi,\g/\piz}$ observable
is an average over the $D^* \to D \piz$ and $D^* \to D\g$ modes.}
\small
\begin{tabular}{l l}
\toprule
Observable & Definition \\ \midrule

$A_K^{CP,\gamma}$ & $\frac{\Gamma(\Bm \to ([h^+h^-]_{D} \g)_\Dstar \Km) \phantom{0}-\phantom{0} \Gamma(\Bp \to ([h^+h^-]_{D} \g)_\Dstar \Kp )}{\Gamma(\Bm \to ([h^+h^-]_{D} \g)_\Dstar \Km) \phantom{0}+\phantom{0} \Gamma(\Bp \to ([h^+h^-]_{D} \g)_\Dstar \Kp)}$ \\

$A_K^{CP,\piz}$ & $\frac{\Gamma(\Bm \to ([h^+h^-]_{D} \piz)_\Dstar \Km) \phantom{0}-\phantom{0} \Gamma(\Bp \to ([h^+h^-]_{D} \piz)_\Dstar \Kp )}{\Gamma(\Bm \to ([h^+h^-]_{D} \piz)_\Dstar \Km) \phantom{0}+\phantom{0} \Gamma(\Bp \to ([h^+h^-]_{D} \piz)_\Dstar \Kp)}$ \\

$A_K^{K\pi,\gamma}$ & $\frac{\Gamma(\Bm \to ([\Km \pip ]_{D} \g)_\Dstar \Km) \phantom{0}-\phantom{0} \Gamma(\Bp \to ([\Kp \pim]_{D} \g)_\Dstar \Kp )}{\Gamma(\Bm \to ([\Km \pip ]_{D} \g)_\Dstar \Km) \phantom{0}+\phantom{0} \Gamma(\Bp \to ([\Kp \pim]_{D} \g)_\Dstar \Kp)}$ \\

$A_K^{K\pi,\piz}$ & $\frac{\Gamma(\Bm \to ([\Km \pip ]_{D} \piz)_\Dstar \Km) \phantom{0}-\phantom{0} \Gamma(\Bp \to ([\Kp \pim]_{D} \piz)_\Dstar \Kp )}{\Gamma(\Bm \to ([\Km \pip ]_{D} \piz)_\Dstar \Km) \phantom{0}+\phantom{0} \Gamma(\Bp \to ([\Kp \pim]_{D} \piz)_\Dstar \Kp)}$ \\

$R^{CP,\gamma}$ & $\frac{\Gamma(\Bm \to ([h^+h^-]_{D} \g)_\Dstar \Km) \phantom{0}+\phantom{0} \Gamma(\Bp \to ([h^+h^-]_{D} \g)_\Dstar \Kp )}{\Gamma(\Bm \to ([h^+h^-]_{D} \g)_\Dstar \pim) \phantom{0}+\phantom{0} \Gamma(\Bp \to ([h^+h^-]_{D} \g)_\Dstar \pip)} \times \frac{1}{R_{K/\pi}^{K\pi,\gamma/\pi^0}}$ \\

$R^{CP,\piz}$ & $\frac{\Gamma(\Bm \to ([h^+h^-]_{D} \piz)_\Dstar \Km) \phantom{0}+\phantom{0} \Gamma(\Bp \to ([h^+h^-]_{D} \piz)_\Dstar \Kp )}{\Gamma(\Bm \to ([h^+h^-]_{D} \piz)_\Dstar \pim) \phantom{0}+\phantom{0} \Gamma(\Bp \to ([h^+h^-]_{D} \piz)_\Dstar \pip)} \times \frac{1}{R_{K/\pi}^{K\pi,\gamma/\pi^0}}$ \\ 

$R_{K/\pi}^{K\pi,\gamma/\pi^0}$ &  $\frac{\Gamma(\Bm \to ([\Km \pip ]_{D} \g/\piz)_\Dstar \Km) \phantom{0}+\phantom{0} \Gamma(\Bp \to ([\Kp \pim]_{D} \g/\piz)_\Dstar \Kp )}{\Gamma(\Bm \to ([\Km \pip ]_{D} \g/\piz)_\Dstar \pim) \phantom{0}+\phantom{0} \Gamma(\Bp \to ([\Kp \pim]_{D} \g/\piz)_\Dstar \pip)}$ \\ 

$R_{K^-}^{\pi K,\g}$ & $\frac{\Gamma(\Bm \to ([\Kp \pim]_{D} \g)_\Dstar \Km)}{\Gamma(\Bm \to ([\Km \pip]_{D} \g)_\Dstar \Km )}$ \\

$R_{K^-}^{\pi K,\piz}$ & $\frac{\Gamma(\Bm \to ([\Kp \pim]_{D} \piz)_\Dstar \Km)}{\Gamma(\Bm \to ([\Km \pip]_{D} \piz)_\Dstar \Km )}$ \\

$R_{K^+}^{\pi K,\g}$ & $\frac{\Gamma(\Bp \to ([\Km \pip]_{D} \g)_\Dstar \Kp)}{\Gamma(\Bp \to ([\Kp \pim]_{D} \g)_\Dstar \Kp )}$ \\

$R_{K^+}^{\pi K,\piz}$ & $\frac{\Gamma(\Bp \to ([\Km \pip]_{D} \piz)_\Dstar \Kp)}{\Gamma(\Bp \to ([\Kp \pim]_{D} \piz)_\Dstar \Kp )}$ \\

$A_\pi^{CP,\gamma}$ & $\frac{\Gamma(\Bm \to ([h^+h^-]_{D} \g)_\Dstar \pim) \phantom{0}-\phantom{0} \Gamma(\Bp \to ([h^+h^-]_{D} \g)_\Dstar \pip )}{\Gamma(\Bm \to ([h^+h^-]_{D} \g)_\Dstar \pim) \phantom{0}+\phantom{0} \Gamma(\Bp \to ([h^+h^-]_{D} \g)_\Dstar \pip)}$ \\

$A_\pi^{CP,\piz}$ & $\frac{\Gamma(\Bm \to ([h^+h^-]_{D} \piz)_\Dstar \pim) \phantom{0}-\phantom{0} \Gamma(\Bp \to ([h^+h^-]_{D} \piz)_\Dstar \pip )}{\Gamma(\Bm \to ([h^+h^-]_{D} \piz)_\Dstar \pim) \phantom{0}+\phantom{0} \Gamma(\Bp \to ([h^+h^-]_{D} \piz)_\Dstar \pip)}$ \\

$A_\pi^{K\pi,\gamma}$ & $\frac{\Gamma(\Bm \to ([\Km \pip ]_{D} \g)_\Dstar \pim) \phantom{0}-\phantom{0} \Gamma(\Bp \to ([\Kp \pim]_{D} \g)_\Dstar \pip )}{\Gamma(\Bm \to ([\Km \pip ]_{D} \g)_\Dstar \pim) \phantom{0}+\phantom{0} \Gamma(\Bp \to ([\Kp \pim]_{D} \g)_\Dstar \pip)}$\\

$A_\pi^{K\pi,\piz}$ & $\frac{\Gamma(\Bm \to ([\Km \pip ]_{D} \piz)_\Dstar \pim) \phantom{0}-\phantom{0} \Gamma(\Bp \to ([\Kp \pim]_{D} \piz)_\Dstar \pip )}{\Gamma(\Bm \to ([\Km \pip ]_{D} \piz)_\Dstar \pim) \phantom{0}+\phantom{0} \Gamma(\Bp \to ([\Kp \pim]_{D} \piz)_\Dstar \pip)}$ \\

$R_{\pi^-}^{\pi K,\g}$ & $\frac{\Gamma(\Bm \to ([\Kp \pim]_{D} \g)_\Dstar \pim)}{\Gamma(\Bm \to ([\Km \pip]_{D} \g)_\Dstar \pim )}$ \\

$R_{\pi^-}^{\pi K,\piz}$ & $\frac{\Gamma(\Bm \to ([\Kp \pim]_{D} \piz)_\Dstar \pim)}{\Gamma(\Bm \to ([\Km \pip]_{D} \piz)_\Dstar \pim )}$ \\

$R_{\pi^+}^{\pi K,\g}$ & $\frac{\Gamma(\Bp \to ([\Km \pip]_{D} \g)_\Dstar \pip)}{\Gamma(\Bp \to ([\Kp \pim]_{D} \g)_\Dstar \pip )}$ \\

$R_{\pi^+}^{\pi K,\piz}$ & $\frac{\Gamma(\Bp \to ([\Km \pip]_{D} \piz)_\Dstar \pip)}{\Gamma(\Bp \to ([\Kp \pim]_{D} \piz)_\Dstar \pip )}$ \\
\bottomrule
\end{tabular}
\label{tab:Observables_PartReco}
\end{table}
\endgroup

\section{LHCb detector and simulation}
\label{sec:detector}

The \lhcb detector~\cite{LHCb-DP-2008-001,LHCb-DP-2014-002} is a single-arm forward
spectrometer covering the \mbox{pseudorapidity} range $2<\eta <5$,
designed for the study of particles containing \bquark or \cquark
quarks. The detector includes a high-precision tracking system
consisting of a silicon-strip vertex detector surrounding the $pp$
interaction region, a large-area silicon-strip detector located
upstream of a dipole magnet with a bending power of about
$4{\mathrm{\,Tm}}$, and three stations of silicon-strip detectors and straw
drift tubes placed downstream of the magnet.
The tracking system provides a measurement of the momentum, \ptot, of charged particles with
a relative uncertainty that varies from 0.5\% at low momentum to 1.0\% at 200\gevc. The minimum distance of a track to a primary $pp$ collision vertex (PV), the impact parameter (IP), 
is measured with a resolution of $(15+29/\pt)\mum$,
where \pt is the component of the momentum transverse to the beam, in\,\gevc.
Different types of charged hadrons are distinguished using information
from two ring-imaging Cherenkov (RICH) detectors. Photons, electrons, and hadrons are identified by a calorimeter system consisting of
scintillating-pad and preshower detectors, an electromagnetic calorimeter,
and a hadronic calorimeter. Muons are identified by a
system composed of alternating layers of iron and multiwire
proportional chambers.
The online event selection is performed by a trigger, 
which consists of a hardware stage, based on information from the calorimeter and muon
systems, followed by a software stage, where a full event
reconstruction is applied. The events considered in the analysis are triggered at the hardware level either when one of the final-state tracks of the signal decay deposits enough energy in the calorimeter system, or when one of the other particles in the event, not reconstructed as part of the signal candidate, fulfils any trigger requirement.
At the software stage, it is required that at least one particle should have high \pt and high \chisqip, where \chisqip is defined as the difference in the PV fit \chisq with and without the inclusion of that particle. A multivariate algorithm~\cite{Gligorov:2012qt} is used to identify displaced vertices consistent with being a two-, three-, or four-track $b$-hadron decay. The PVs are fitted with and without the $B$ candidate tracks, and the PV that gives the smallest \chisqip is associated with the $B$ candidate.

Simulation is required to model the invariant mass distributions of the signal and background contributions and determine their selection efficiencies. In the simulation, $pp$ collisions are generated using \pythia~\cite{Sjostrand:2007gs,*Sjostrand:2006za} with a specific \lhcb configuration~\cite{LHCb-PROC-2010-056}. Decays of unstable particles are described by \evtgen~\cite{Lange:2001uf}, in which final-state radiation is generated using \photos~\cite{Golonka:2005pn}. The interaction of the generated particles with the detector, and its response, are implemented using the \geant toolkit~\cite{Allison:2006ve, *Agostinelli:2002hh} as described in Ref.~\cite{LHCb-PROC-2011-006}. Some subdominant sources of background are generated with a fast simulation~\cite{Cowan:2016tnm} that mimics the geometric acceptance and tracking efficiency of the \lhcb detector as well as the dynamics of the decay via \evtgen.

\section{Event selection}
\label{sec:Selection}

After reconstruction of a \D-meson candidate from two oppositely charged particles, the same event selection is applied to all $\Bm \to D^{(*)} h^-$ channels in both data and simulation. Since the neutral pion or photon from the vector \Dstar decay is not reconstructed, partially reconstructed $\Bm \to \Dstar h^-$ candidates and fully reconstructed $\Bm \to \D h^-$ candidates contain the same reconstructed particles, and thus appear in the same sample. These decays are distinguished using their reconstructed invariant mass $m(Dh^-)$, as described in Sec.~\ref{sec:Fit}.

The reconstructed \D-meson candidate mass is required to be within $\pm 25 \mevcc$ of the known \Dz mass~\cite{PDG2020}; this range corresponds to approximately three times the mass resolution. 
The kaon or pion originating directly from the \Bm decay, subsequently referred to as the companion particle, is required to have \pt in the range 0.5--10\gevc and $p$ in the range 5--100\gevc. 
These requirements ensure that the track is within the kinematic coverage of the RICH detectors, which provide particle identification (PID) information used to create independent samples of $\Bm \to D^{(*)} \pim$ and $\Bm \to D^{(*)} \Km$ decays. Details of the calibration procedure used to determine PID requirement efficiencies are given in Sec.~\ref{sec:Fit}. A kinematic fit is performed to each decay chain, with vertex constraints applied to both the \Bm and \D decay products, and the \D candidate constrained to its known mass~\cite{Hulsbergen:2005pu}. The \Bm meson candidates with invariant masses in the interval \mbox{4900--5900\mevcc} are retained. This range includes the partially reconstructed $\Bm \to (D \g)_\Dstar h^-$ and $\Bm \to (D \piz)_\Dstar h^-$ decays, which fall at $m(Dh^-)$ values below the known \Bm meson mass. 

A boosted decision tree (BDT) classifier, implemented using the gradient boost algorithm~\cite{Roe} in the \textsc{scikit-learn} library~\cite{Scikit-learn-paper}, is employed to achieve further background suppression. 
The BDT classifier is trained using simulated \mbox{$\Bm \to [\Km \pip]_{D}\Km$} decays and a background sample of $\Km\pip\Km$ combinations in data with invariant mass in the range 5900--7200\mevcc. The BDT classifier is also used on all other $D$ decay modes, and provides equivalent performance across samples. The input to the BDT classifier is a set of features that characterise the signal decay. These features can be divided into two categories:
(1) properties of any particle and (2) properties of composite particles (the \D and \Bm candidates). Specifically
\begin{enumerate}
\item{$p$, \pt, and \chisqip;}
\item{decay time, flight distance, decay vertex quality, radial distance between the decay vertex and the PV, and the angle between the particle's momentum vector and the line connecting the production and decay vertices.}
\end{enumerate} 
In addition, a feature that estimates the imbalance of \pt around the \Bm candidate momentum vector is also used. It is defined as
\begin{equation*}
I_{\pt} = \frac{\pt(\Bm) - \Sigma \pt}{\pt(\Bm) + \Sigma \pt}\,,
\end{equation*}
where the sum is taken over charged tracks inconsistent with originating from the PV which lie within a cone around the \Bm candidate, excluding tracks used to make the signal candidate.
The cone is defined by a circle with a radius of 1.5 in the plane of pseudorapidity and azimuthal angle defined in radians. Including the $I_{\pt}$ feature in the BDT classifier training gives preference to \Bm candidates that are isolated from the rest of the event.

Since no PID information is used in the BDT classifier, the efficiencies for \mbox{$\Bm \to D^{(*)}\Km$} and $\Bm \to D^{(*)} \pim$ decays are similar, with insignificant variations arising from small differences in the decay kinematics. 
The requirement applied to the BDT classifier response is optimised by minimising the expected relative uncertainty on $R_K^{\pi K}$ (see Table~\ref{tab:Observables_FullReco} for definition), as measured using the invariant mass fit described in Sec.~\ref{sec:Fit}. The purity of the sample is further improved by requiring that all kaons and pions in the \D decay are positively identified by the RICH~\cite{LHCb-DP-2012-003,LHCb-DP-2018-001}; this selection has an efficiency of about 90\% per final-state particle.

Peaking background contributions from charmless decays that result in the same final state as the signal are suppressed by requiring that the flight distance of the \D candidate from the \Bm decay vertex is larger than two times its uncertainty. Peaking background from $\Bm \to [h_1^- h^+]_D h_2^-$ signal decays, where $h_1^-$ and $h_2^-$ are exchanged, are vetoed by requiring that the $h_2^- h^+$ invariant mass is more than 25\mevcc away from the known \Dz mass. A veto is also applied on candidates consistent with containing a fully reconstructed $D^* \to D \g/\piz$ candidate, in order to statistically decouple this measurement from a possible analysis of fully reconstructed $\Bm \to D^* h^-$ decays; the veto is found to be 98.5\% efficient on data. 

Background from favoured decays misidentified as ADS decays is reduced by application of the $D$ mass window and $D$ decay product PID requirements detailed above. To further reduce this background, an additional veto on the $D$ mass, calculated with both decay products misidentified, is applied to the favoured and ADS samples, where candidates are required to fall further than 15\mevcc away from the known $\Dz$ mass.

\section{Invariant-mass fit}
\label{sec:Fit}

\begin{figure}[!h]
  \begin{center}
   \includegraphics[width=0.85\linewidth]{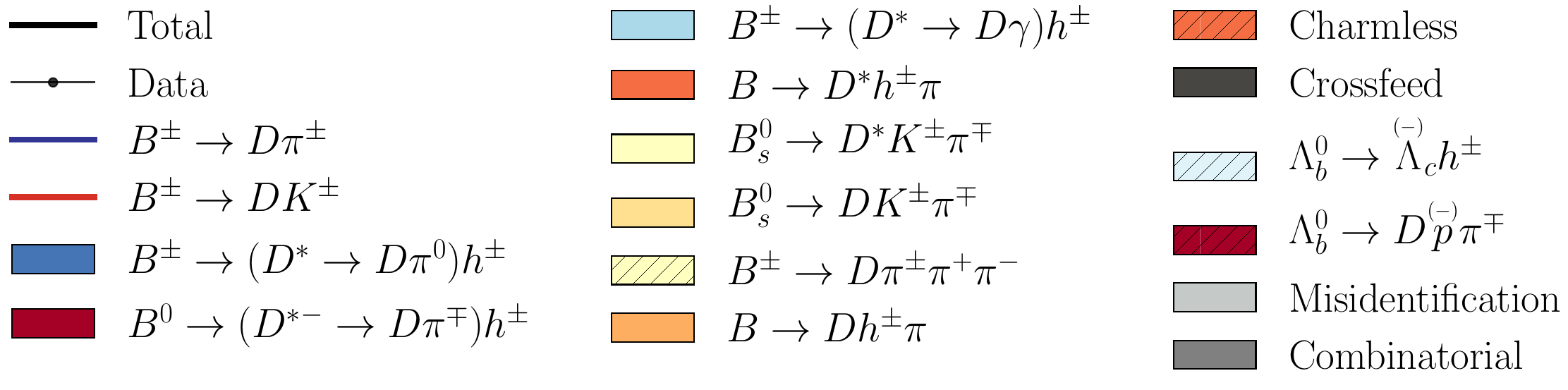} \\
   \end{center}
   \caption{Legend indicating the invariant-mass fit components shown in Figs.~\ref{fig:fit_kpi}$-$\ref{fig:fit_pipi}.
  \label{fig:Fit_Legend}}
\end{figure}

\begin{figure}[!h]
  \begin{center}
  \includegraphics[width=0.49\linewidth]{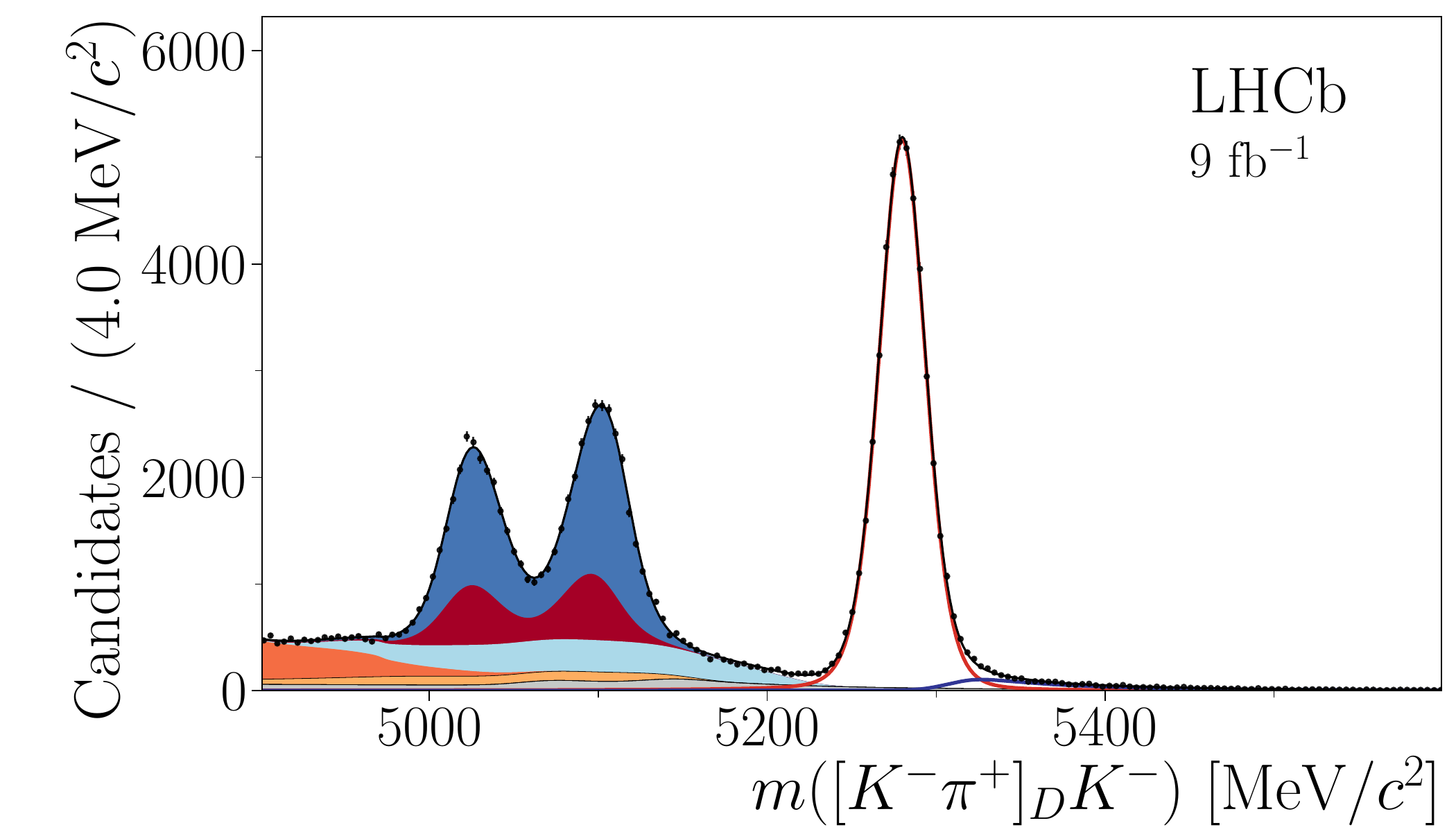}
   \includegraphics[width=0.49\linewidth]{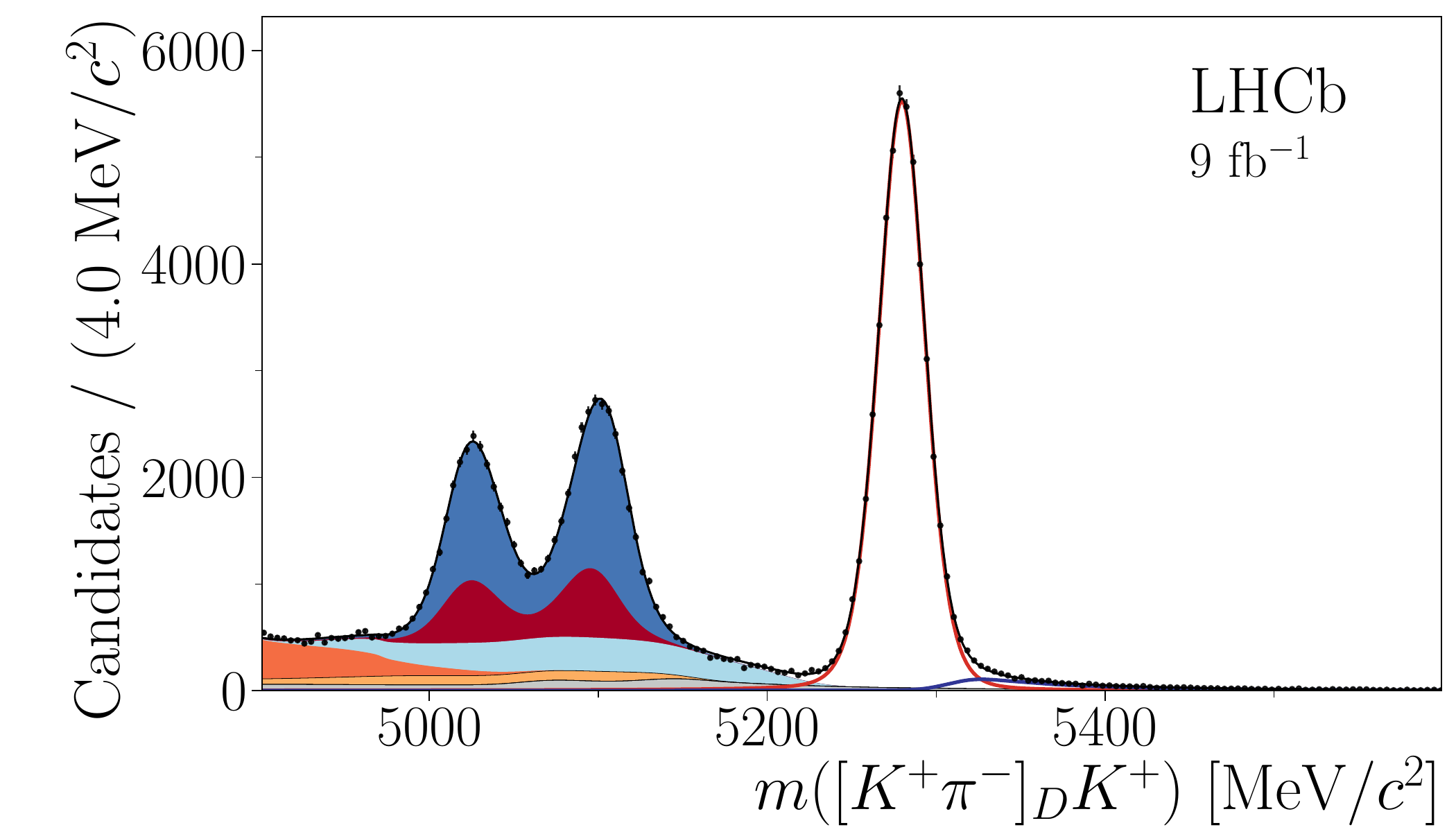}\\ \vspace{0.5cm}
   \includegraphics[width=0.49\linewidth]{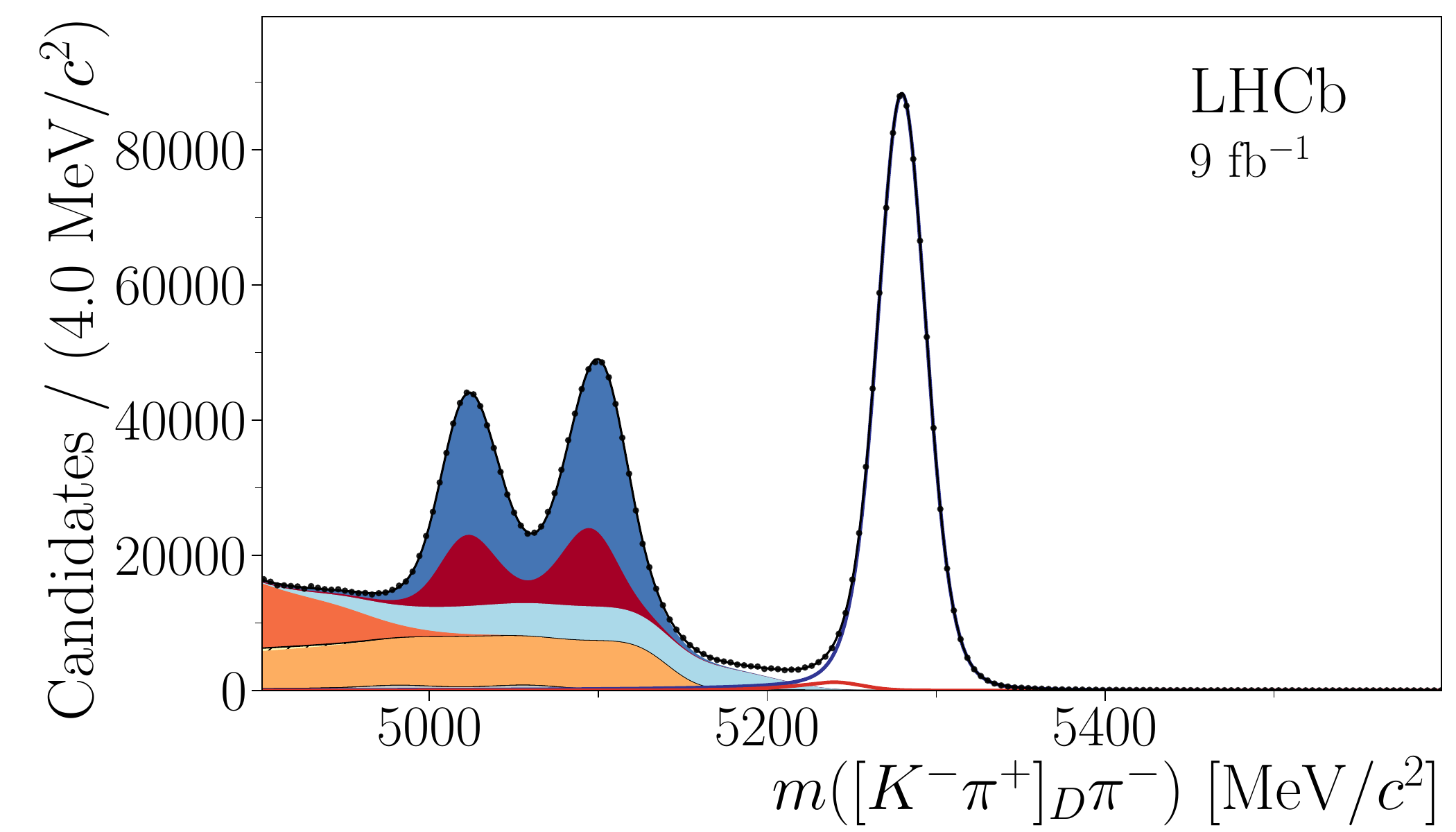}
   \includegraphics[width=0.49\linewidth]{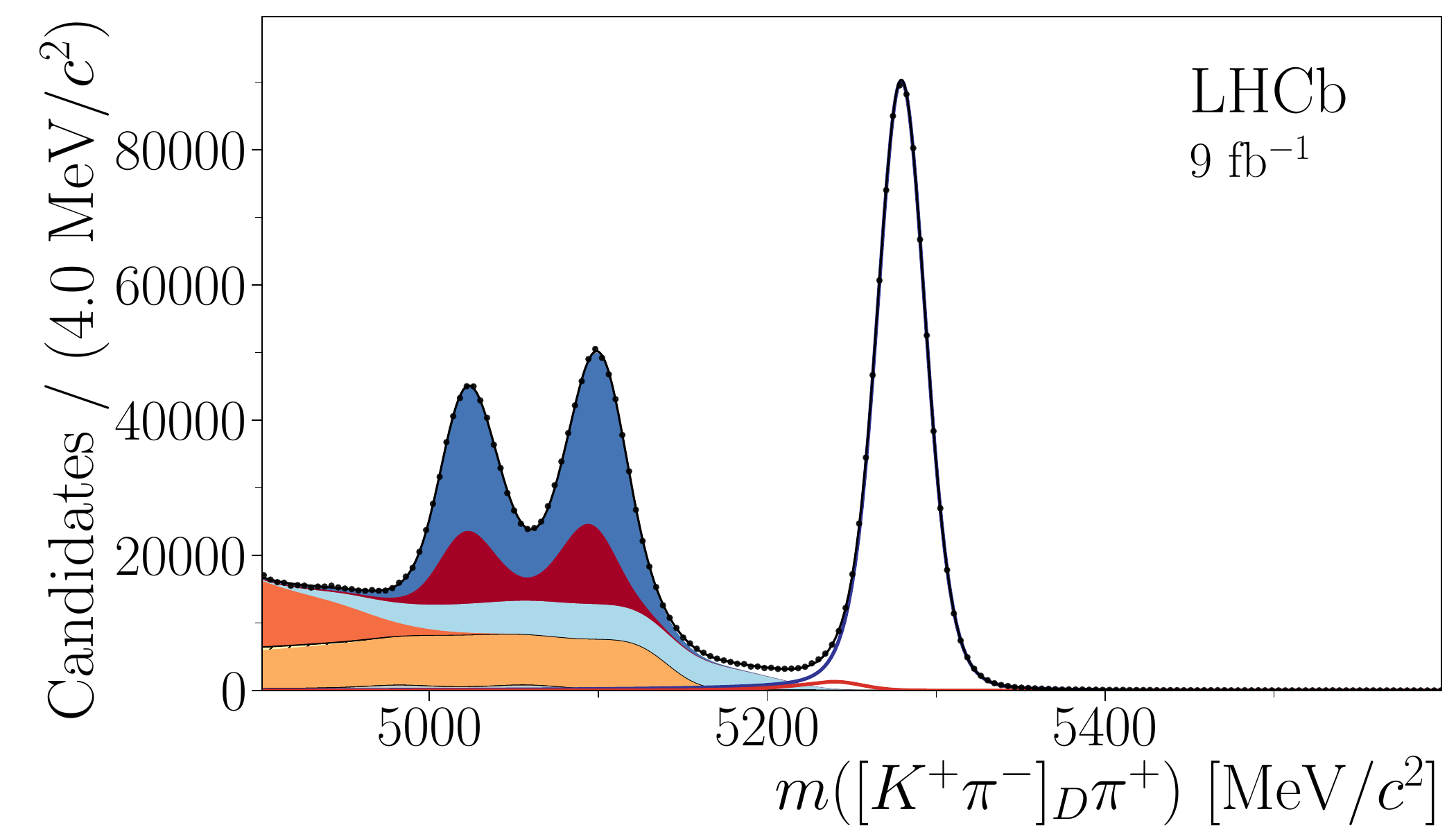}\\
   \end{center}
   \caption{Invariant-mass distribution of selected $\Bpm \to [\Kpm \pimp]_{D}h^{\pm}$ candidates. The result of the fit is shown by the solid navy line, and each component is listed in a legend provided in Fig.~\ref{fig:Fit_Legend}.
\label{fig:fit_kpi}}
\end{figure}

\begin{figure}[!h]
  \begin{center}
   \includegraphics[width=0.49\linewidth]{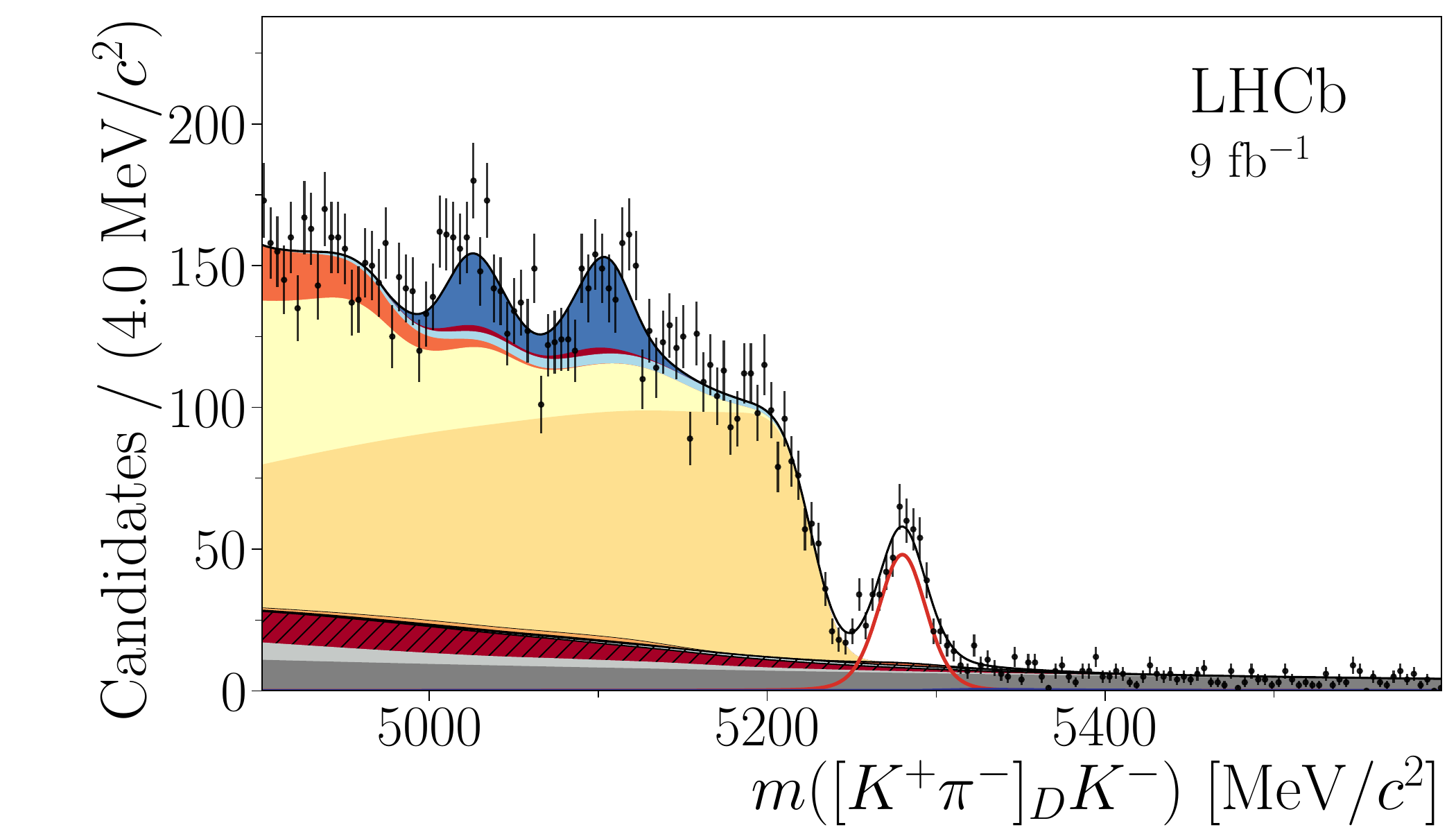}
   \includegraphics[width=0.49\linewidth]{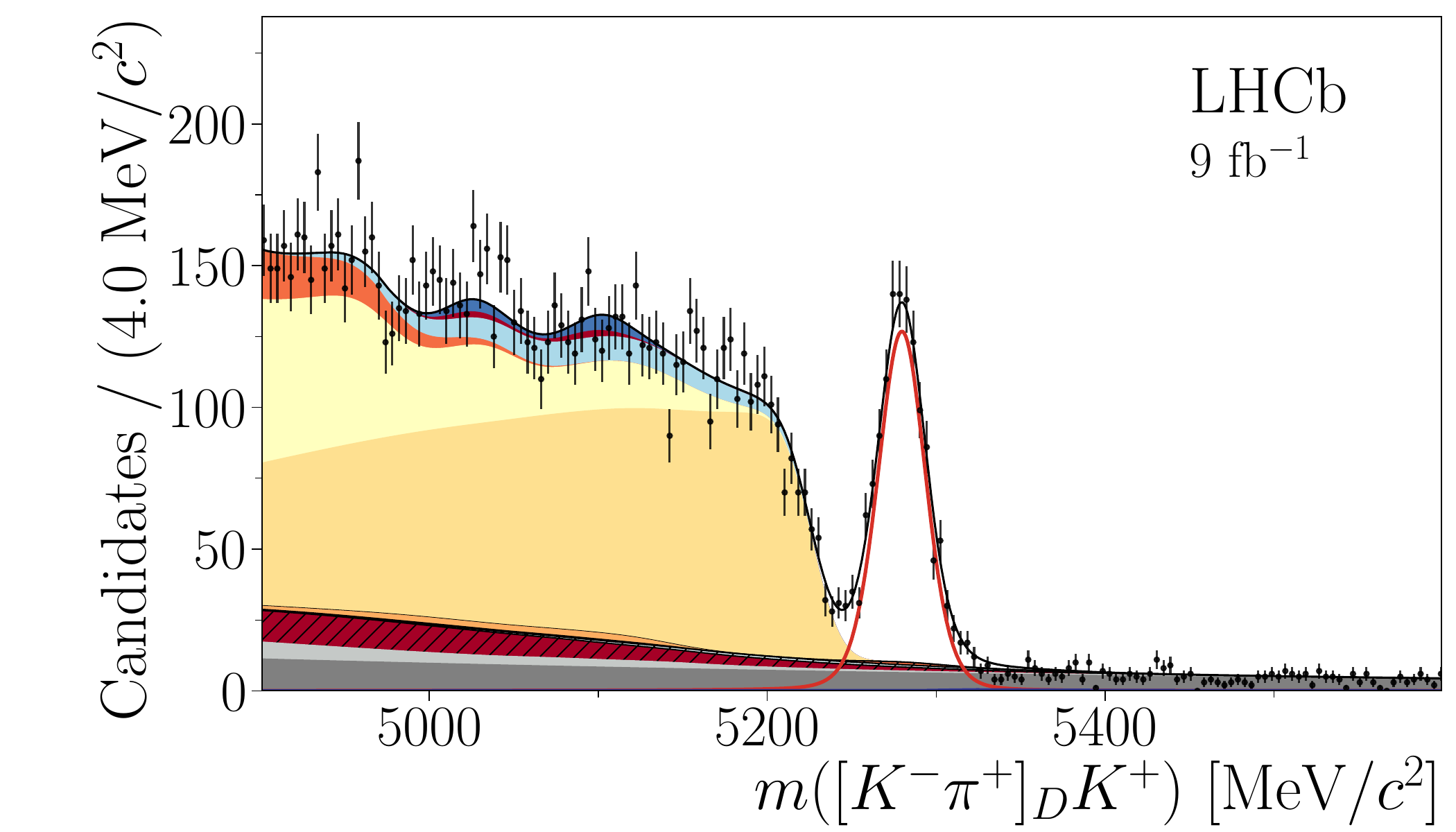}\\ \vspace{0.5cm}
   
   \includegraphics[width=0.49\linewidth]{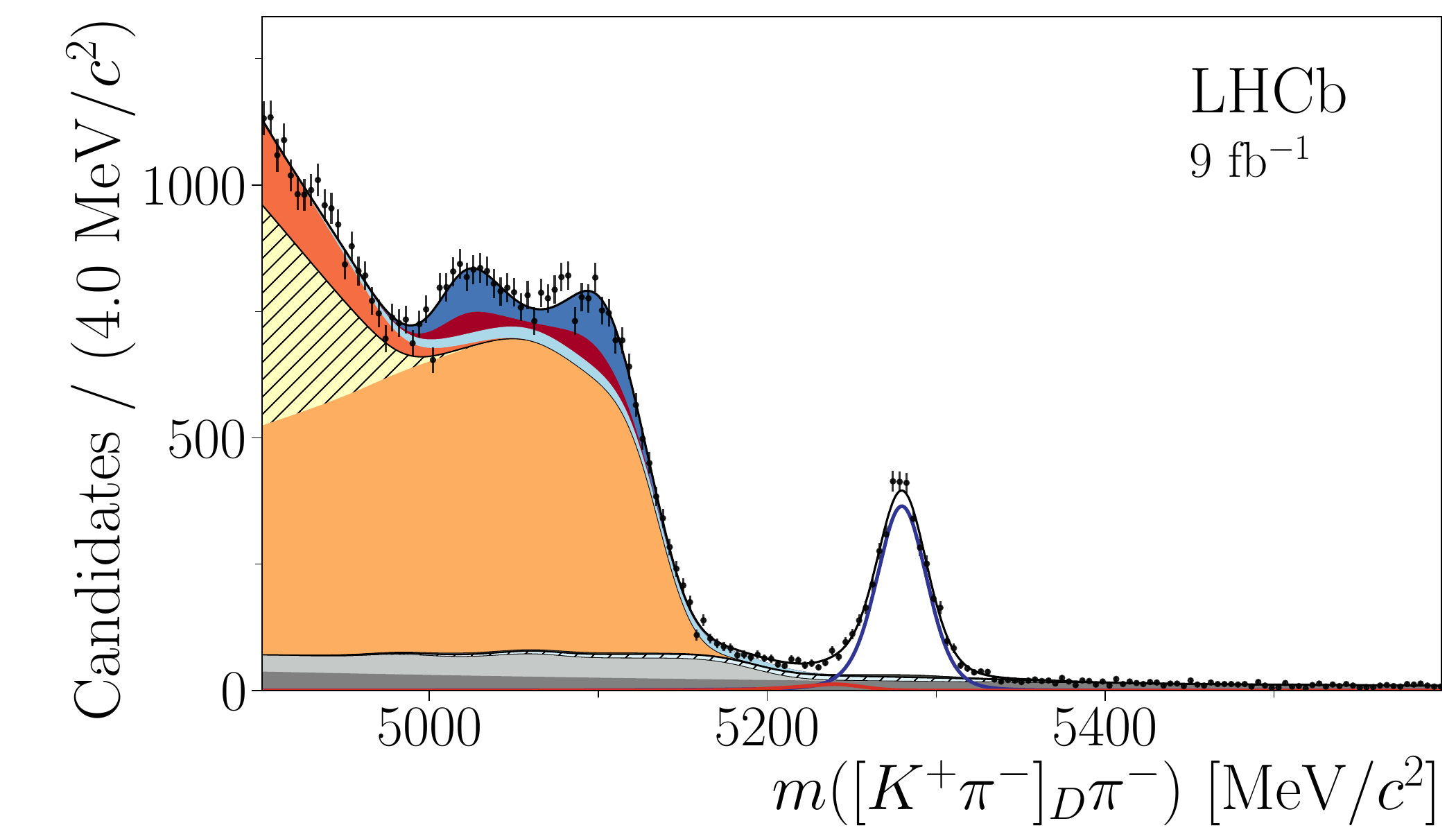}
   \includegraphics[width=0.49\linewidth]{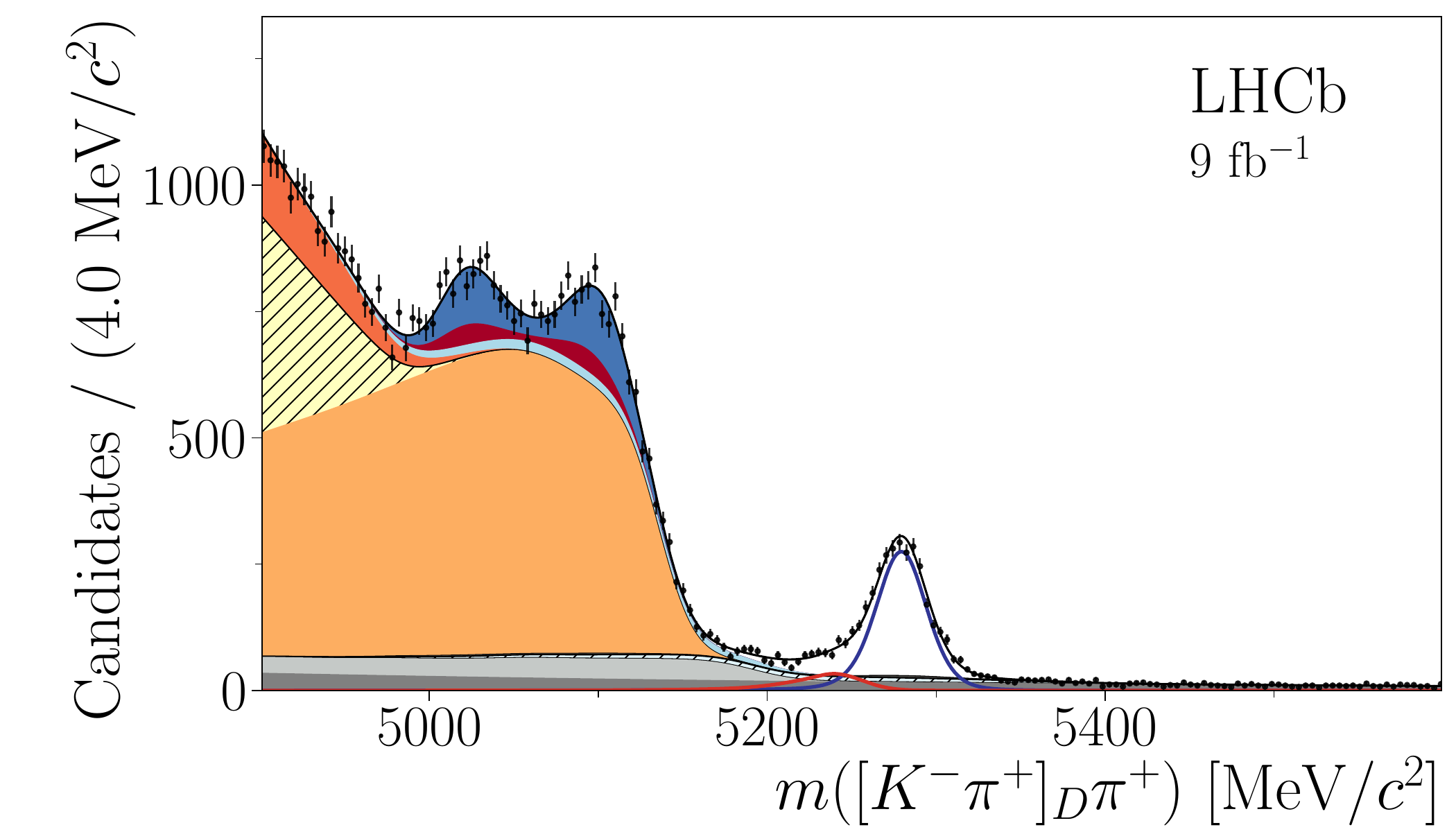}\\
   \end{center}
   \caption{Invariant-mass distribution of $\Bpm \to [\Kmp\pipm]_D h^\pm$ candidates with the fit result overlaid. A legend is provided in Fig.~\ref{fig:Fit_Legend}.
\label{fig:fit_pik}}
\end{figure}

\begin{figure}[!t]
  \begin{center}
   \includegraphics[width=0.49\linewidth]{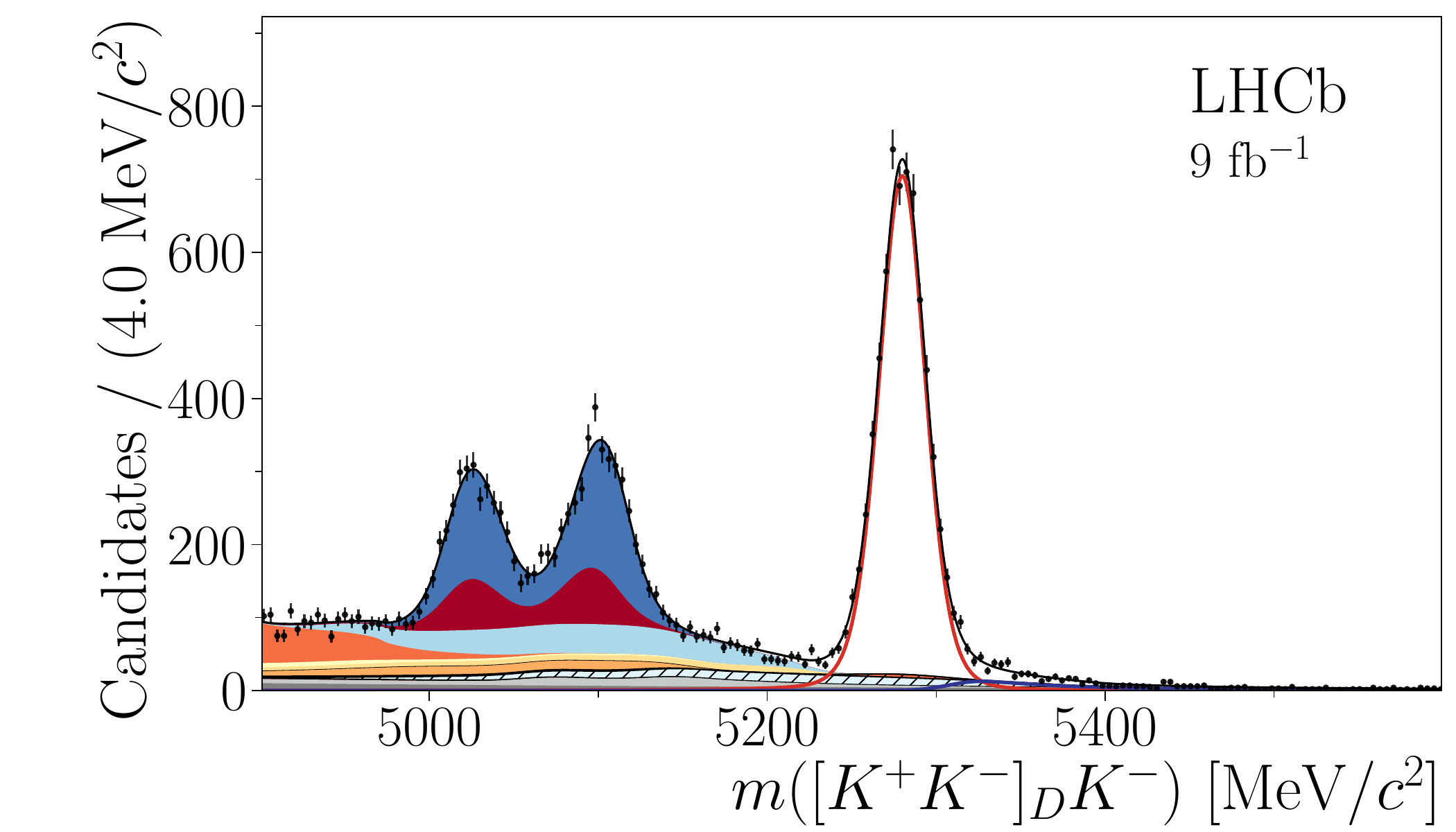}
   \includegraphics[width=0.49\linewidth]{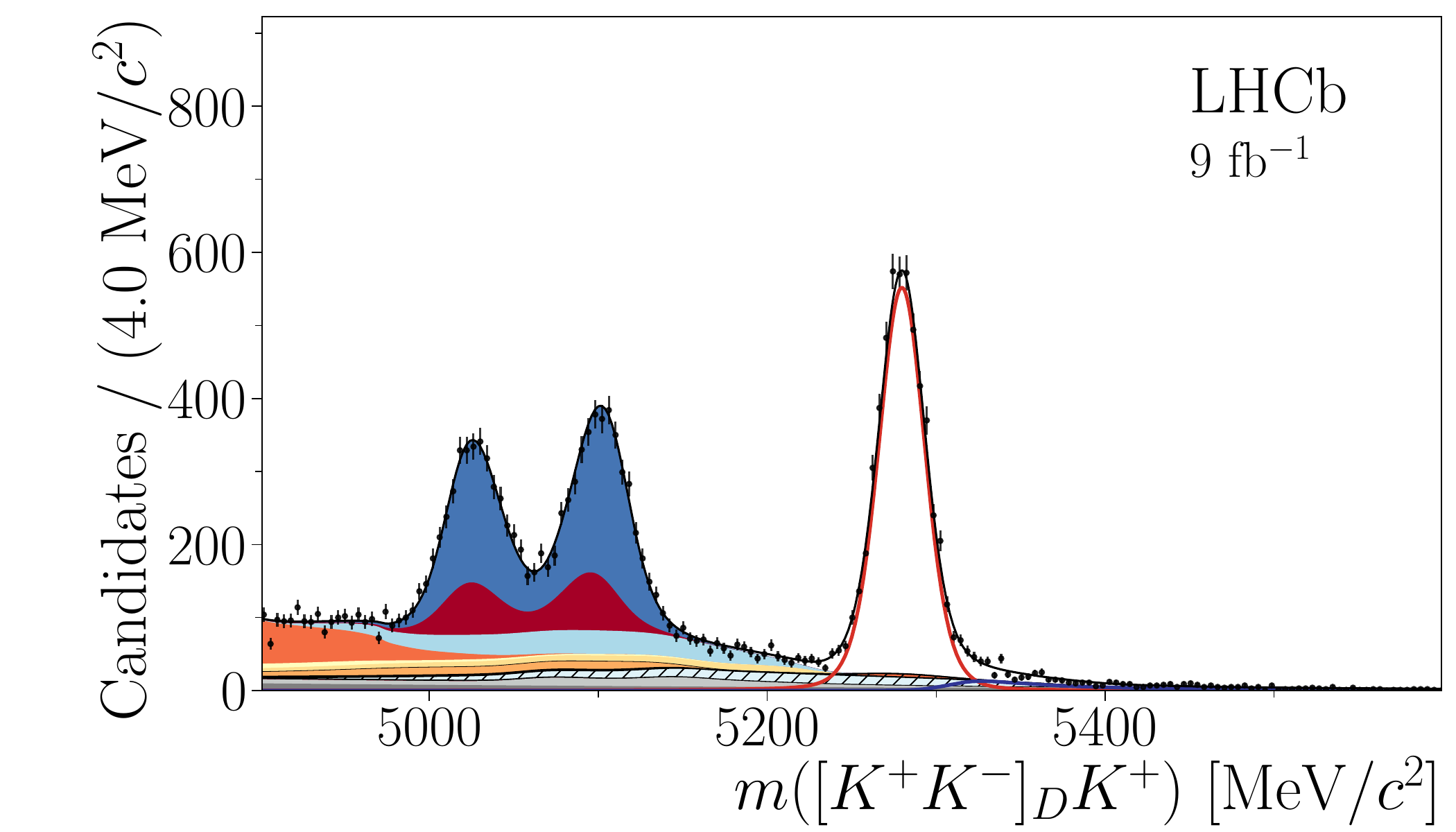}\\ \vspace{0.5cm}
   
   \includegraphics[width=0.49\linewidth]{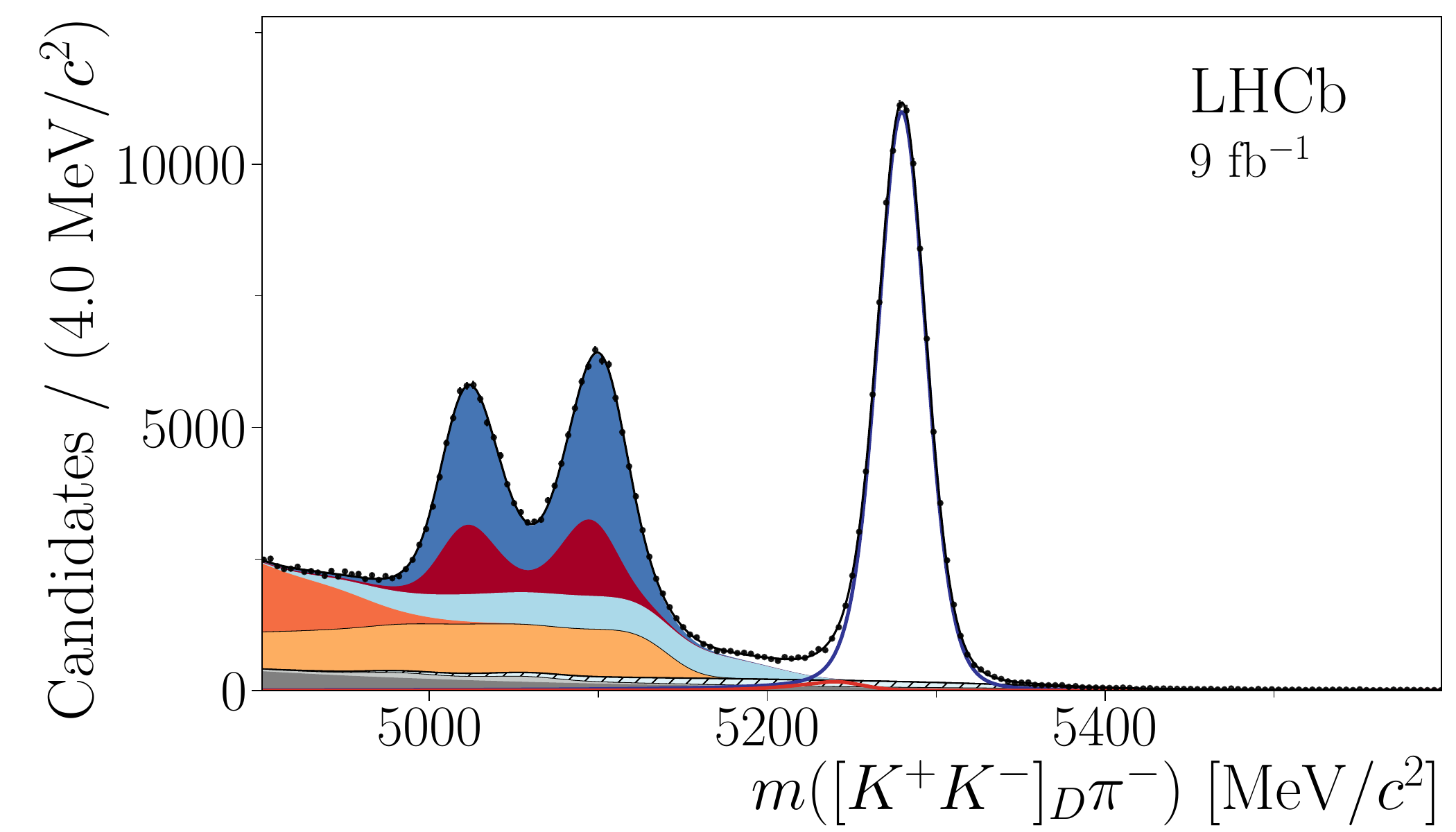}
   \includegraphics[width=0.49\linewidth]{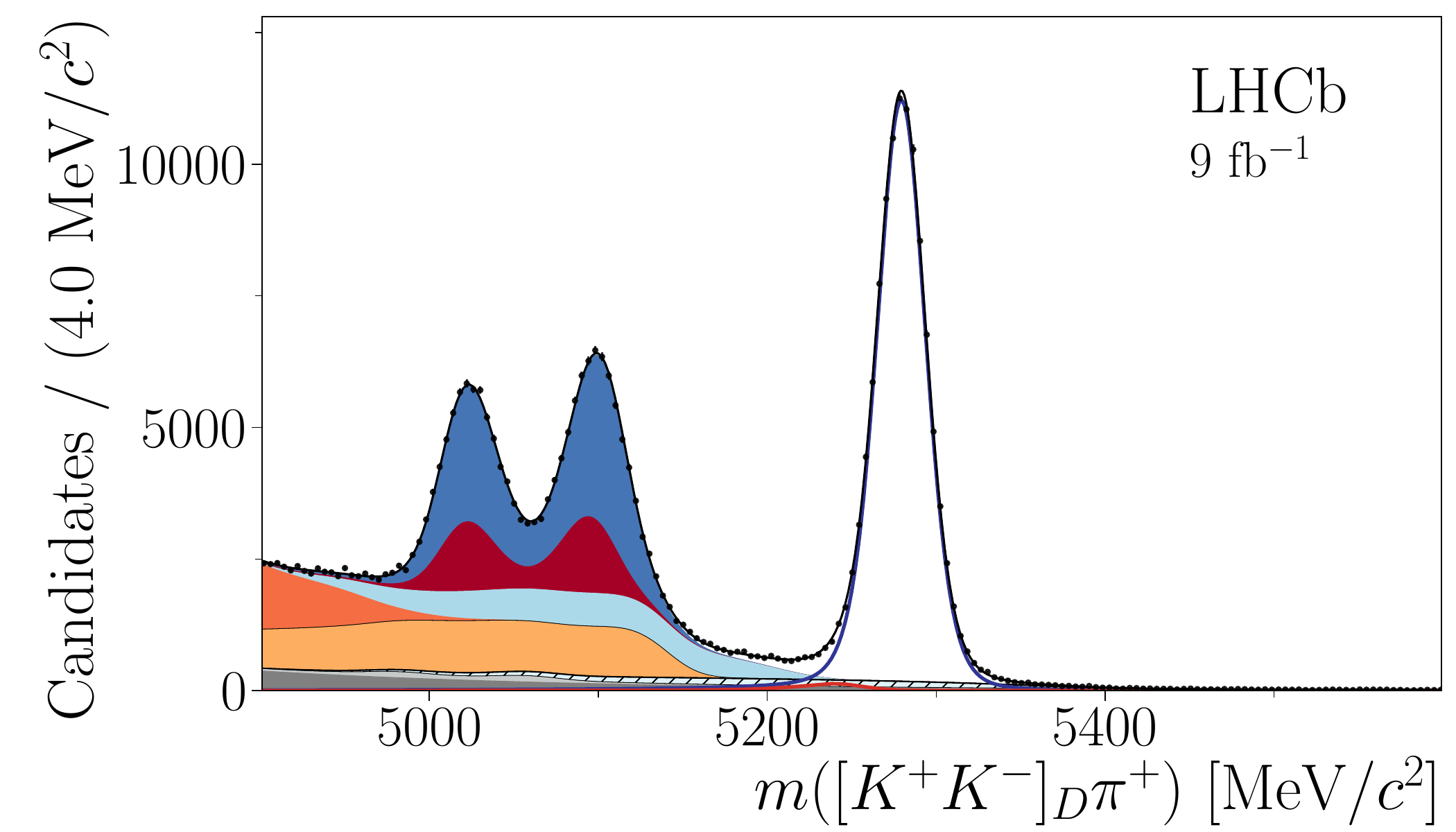}
   \end{center}
   \caption{Invariant-mass distribution of $\Bpm \to [\Kp\Km]_D h^\pm$ candidates with the fit result overlaid. A legend is provided in Fig.~\ref{fig:Fit_Legend}. 
  \label{fig:fit_kk}}
\end{figure}

\begin{figure}[!h]
  \begin{center}
   \includegraphics[width=0.49\linewidth]{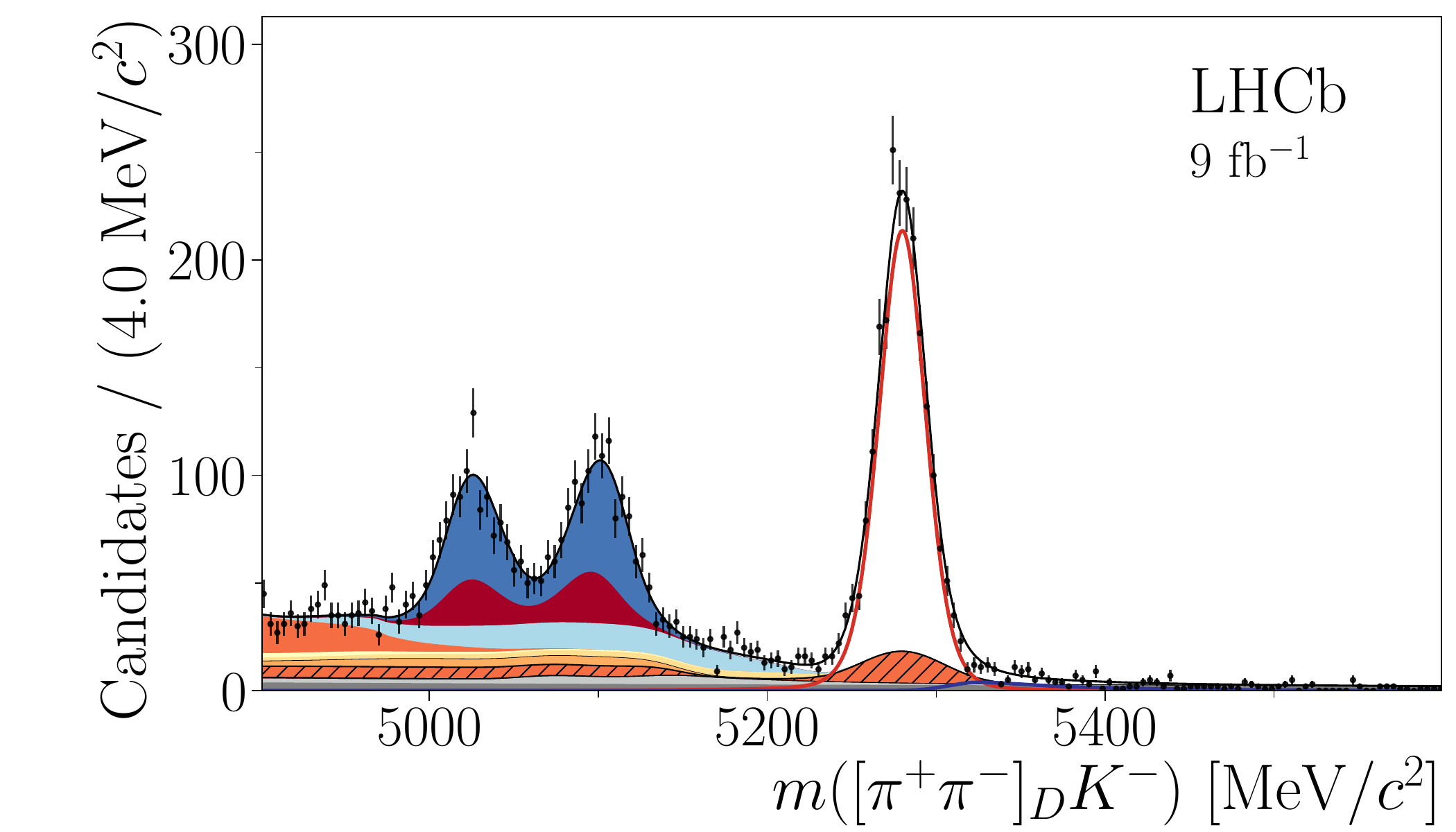}
   \includegraphics[width=0.49\linewidth]{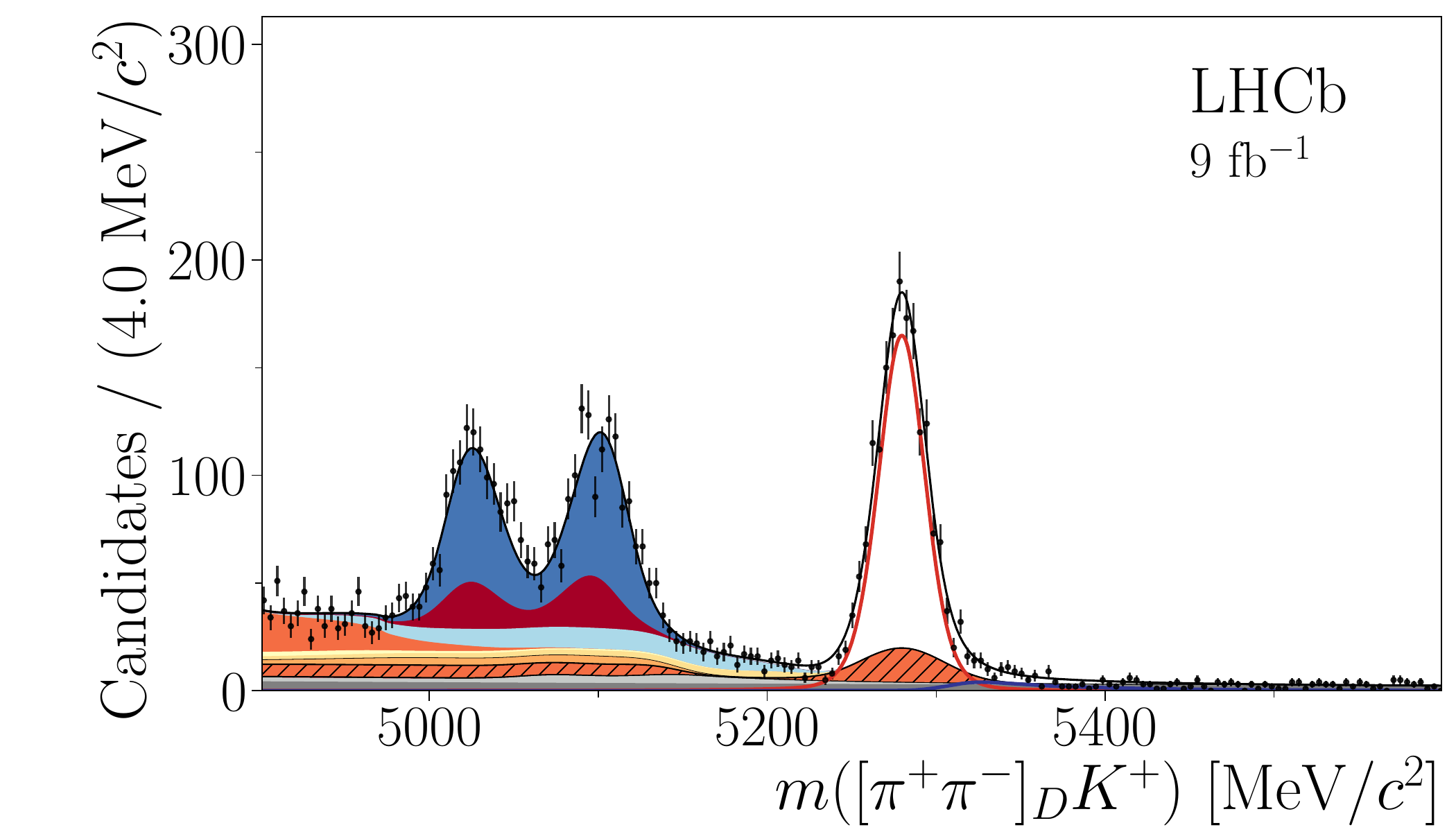}\\ \vspace{0.5cm}
   
   \includegraphics[width=0.49\linewidth]{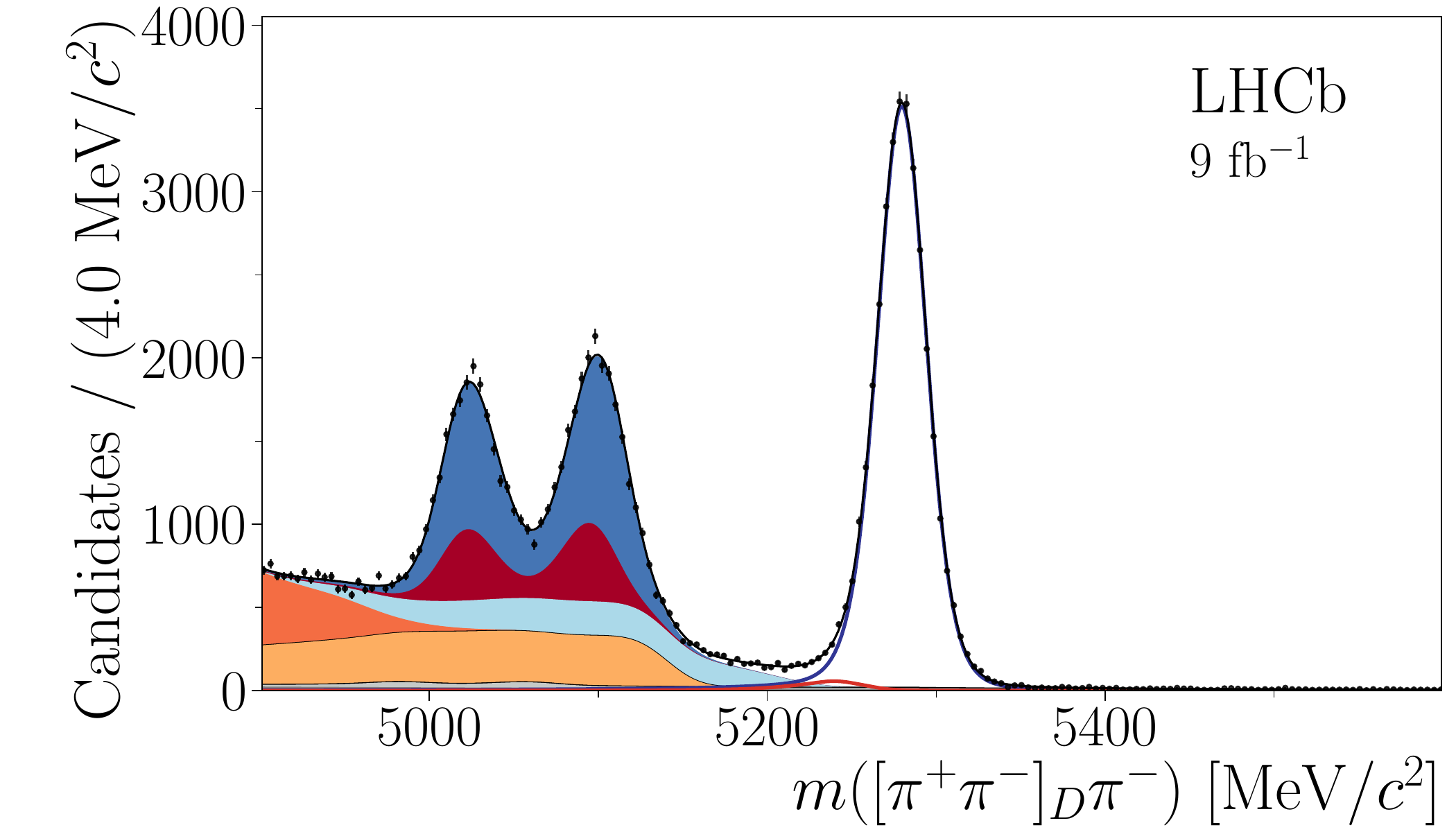}
   \includegraphics[width=0.49\linewidth]{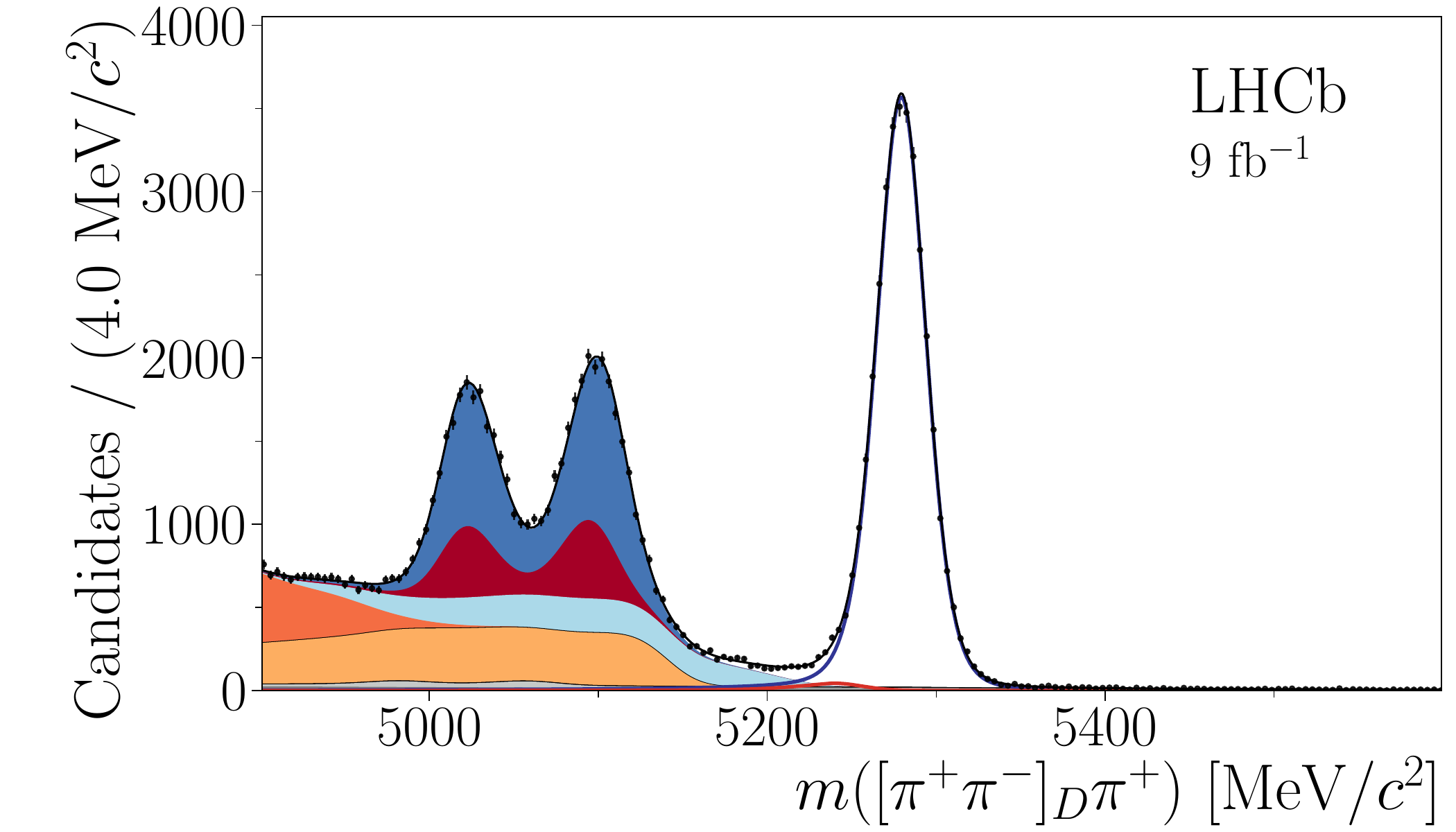}
   \end{center}
   \caption{Invariant-mass distribution of $\Bpm \to [\pip\pim]_D h^\pm$ candidates with the fit result overlaid. A legend is provided in Fig.~\ref{fig:Fit_Legend}. 
  \label{fig:fit_pipi}}
\end{figure}

The values of the 28 \CP observables and two branching fractions are determined using a binned extended maximum likelihood fit to the $m(Dh^-)$ distribution in data. Distinguishing between \Bp and \Bm candidates, companion particle hypotheses, and the four \D decay product final states, yields 16 independent samples which are fitted simultaneously. The invariant-mass spectra and results of the fit are shown in Figs.~\ref{fig:fit_kpi}$-$\ref{fig:fit_pipi}, where the 16 subsamples are displayed separately. Although the fit is performed to data in the 4900--5900\mevcc range, the 4900--5600\mevcc region is displayed to focus on the signal components. A legend listing each fit component is provided in Fig.~\ref{fig:Fit_Legend}. The $\chi^2$ per degree of freedom of the fit is 7652/7875, indicating that the data are well-described.

\FloatBarrier

\subsection{Fit components}

The total probability density function (PDF) is built from six signal functions, one for each of the \mbox{$\Bm \to \D \pim$}, \mbox{$\Bm \to \D \Km$}, \mbox{$\Bm \to (\D \piz)_{\Dstar} \pim$}, \mbox{$\Bm \to (\D \piz)_{\Dstar} \Km$}, \mbox{$\Bm \to (\D \gamma)_{\Dstar} \pim$}, and \mbox{$\Bm \to (\D \gamma)_{\Dstar} \Km$} decays. In addition, there are PDFs that describe the misidentified signal and background components, combinatorial background, background from $B$ decays to charmless final states, background from favoured decays misidentified as ADS decays, and background from other partially reconstructed decays.  
All PDFs are identical for \Bp and \Bm decays. In cases where shape parameters are derived from simulation and fixed in the fit to data, the parameter uncertainties are considered as a source of systematic uncertainty.

\subsection*{\normalsize\boldmath$\Bm \to \D\pim$ decays}

The $\Bm \to D\pim$ signal component is modelled using a sum of a double-sided Hypatia PDF~\cite{Santos:2013ky} and a
Johnson SU PDF~\cite{JohnsonSU}. Both PDFs have a common mean which is shared across all samples. The width of the Johnson SU component varies freely in the favoured and GLW modes, while the ADS mode shares this parameter with the favoured mode. The Hypatia width is related to the Johnson SU width with a single freely varying parameter, which is shared across all $D$ decay modes. The relative fraction of the Hypatia PDF, and the kurtosis ($\gamma$) and skewness ($\delta$) parameters of the Johnson SU PDF, are shared across all $D$ modes and vary freely in the data fit. All other shape parameters are fixed to the values found in fits to simulated samples of $\Bm \to D\pim$ signal decays.

The contribution from $\Bm \to D\pim$ decays misidentified as $\Bm \to D\Km$ is shifted to higher invariant masses in the $D\Km$ samples. These misidentified candidates are modelled with the sum of two Crystal Ball PDFs~\cite{Skwarnicki:1986xj} with a common mean. All shape parameters are fixed to the values found in simulation.

\subsection*{\normalsize\boldmath$\Bm \to \D\Km$ decays}

In the $D\Km$ samples, the $\Bm \to D\Km$ signal is described using the sum of a Hypatia PDF and Johnson SU PDF. The Hypatia and Johnson SU widths are related to the corresponding $\Bm \to D \pim$ signal values with a single freely varying ratio, while the Hypatia fraction and Johnson SU $\gamma$ and $\delta$ parameters are shared with the $\Bm \to D \pim$ signal PDF. All other shape parameters are fixed to the values found in fits to simulated samples of $\Bm \to D\Km$ signal decays.

Misidentified $\Bm \to D\Km$ decays are displaced to lower masses in the $D\pim$ samples. These candidates are modelled with the sum of two Crystal Ball PDFs with a common mean. The mean, widths, and tail parameters are fixed to the values found in simulation.

\subsection*{\normalsize\boldmath$\Bm \to (\D \piz)_{\Dstar} \pim$ decays}

In partially reconstructed decays involving a vector meson, the $\D h^-$ invariant-mass distribution depends upon the spin and mass of the missing particle. 
For $\Bm \to (\D \piz)_{\Dstar} \pim$ decays, the missing neutral pion has spin-parity $0^{-}$. The distribution is described by a parabola exhibiting a minimum, whose range is defined by the kinematic endpoints of the decay. It is convolved with a Gaussian resolution function, yielding
\begin{equation}
f(m) = \int_a^b \! \left(\mu - \frac{a+b}{2}\right)^2\left(\frac{1-\xi}{b-a}\mu + \frac{b\xi - a}{b - a}\right)e^{-\frac{(\mu - m)^2}{2\sigma^2}}\text{d}\mu \\
\label{eq:RooHORNSdini}\,.
\end{equation}
The resulting distribution has a characteristic double-peaked shape, visible in Figs.~\ref{fig:fit_kpi}$-$\ref{fig:fit_pipi} as dark blue filled regions appearing to the left of the fully reconstructed $\Bm \to D h^-$ peaks. The lower and upper endpoints of the parabola are $a$ and $b$, respectively, while the relative height of the lower and upper peaks is determined by the $\xi$ parameter. When $\xi = 1$, both peaks are of equal height, and a deviation of $\xi$ from unity accounts for mass-dependent reconstruction and selection efficiency effects. The values of $a$, $b$, and $\xi$ are taken from simulation, while the Gaussian resolution $\sigma$ is allowed to vary freely in the favoured and GLW samples; the ADS widths are shared with the favoured mode.

Partially reconstructed $\Bm \to (\D \piz)_{\Dstar} \pim$ decays, where the companion pion is misidentified as a kaon, are parameterised with a semi-empirical PDF, formed from the sum of Gaussian and error functions. The parameters of this PDF are fixed to the values found in fits to simulated events.

\subsection*{\normalsize\boldmath$\Bm \to (\D \piz)_{\Dstar} \Km$ decays}

Equation~\ref{eq:RooHORNSdini} is also used to describe partially reconstructed $\Bm \to (\D \piz)_{\Dstar} \Km$ decays, where the width $\sigma$ in each of the $D\Km$ samples is related to the $D\pim$ width by a freely varying ratio which is shared with the $\Bm \to D \Km$ signal PDFs. The kinematic endpoints $a$ and $b$ are determined from a fit to simulated events, and the $\xi$ parameter is shared with the $\Bm \to (\D \piz)_{\Dstar} \pim$ PDF. 

Partially reconstructed $\Bm \to (\D \piz)_{\Dstar} \Km$ decays, where the companion kaon is misidentified as a pion, are parameterised with a semi-empirical PDF, formed from the sum of Gaussian and error functions. The parameters of this PDF are fixed to the values found in fits to simulated events.

\subsection*{\normalsize\boldmath$\Bm \to (\D \gamma)_{\Dstar} \pim$ decays}

Partially reconstructed $\Bm \to (\D \gamma)_{\Dstar} \pim$ decays involve a missing particle of zero mass and spin-parity $1^{-}$. The $\D\pim$ invariant-mass distribution is described by a parabola exhibiting a maximum, convolved with a Gaussian resolution function. The functional form of this component is
\begin{equation}
f(m) = \int_a^b \! -(\mu-a)(\mu-b)\left(\frac{1-\xi}{b-a}\mu + \frac{b\xi - a}{b - a}\right)e^{-\frac{(\mu-m)^2}{2\sigma^2}}\text{d}\mu
\label{eq:RooHILLdini}\,.
\end{equation}
This distribution exhibits a broad single peak, as opposed to the double-peaked \mbox{$\Bm \to (\D \piz)_{\Dstar} \pim$} distribution described by Eq.~\ref{eq:RooHORNSdini}. In Figs.~\ref{fig:fit_kpi}$-$\ref{fig:fit_pipi}, this component is visible as the light blue filled region to the left of the fully reconstructed $\Bm \to D h^-$ peaks. The values of $a$, $b$, $\xi$, and $\sigma$ are fixed using fits to simulated events. The difference between the invariant-mass distributions of $\Bm \to (\D \gamma)_{\Dstar} \pim$ and $\Bm \to (\D \piz)_{\Dstar} \pim$ decays enables their statistical separation in the fit, and hence the determination of \CP observables for each mode. 

Partially reconstructed $\Bm \to (\D \gamma)_{\Dstar} \pim$ decays where the companion pion is misidentified as a kaon are treated in an equivalent manner to misidentified $\Bm \to (\D \piz)_{\Dstar} \pim$ decays, as described above.

\subsection*{\normalsize\boldmath$\Bm \to (\D \gamma)_{\Dstar} \Km$ decays}

Equation~\ref{eq:RooHILLdini} is also used to describe partially reconstructed $\Bm \to (\D \gamma)_{\Dstar} \Km$ decays, where the width $\sigma$ in each of the $D\Km$ samples is related to the $D\pim$ width by a common ratio shared with the $\Bm \to D \Km$ signal PDFs. The kinematic endpoints $a$ and $b$ are derived from a fit to simulated events, and the $\xi$ parameter is shared with the $\Bm \to (\D \gamma)_{\Dstar} \pim$ PDF. Partially reconstructed $\Bm \to (\D \gamma)_{\Dstar} \Km$ decays where the companion kaon is misidentified as a pion are treated in an equivalent manner to misidentified $\Bm \to (\D \piz)_{\Dstar} \Km$ decays.

\subsection*{\normalsize\textbf{Combinatorial background}}

An exponential PDF is used to describe the combinatorial background. Independent and freely varying exponential parameters and yields are used to model this component in each subsample.

\subsection*{\normalsize\textbf{Charmless background}}

Charmless $\Bm \to h_1^- h_2^- h^+$ decays, where $h_{1}^-$, $h_{2}^-$, and $h^+$ each represent a charged kaon or pion, peak at the \Bm mass and cannot be distinguished effectively from the fully reconstructed $\Bm \to \D h^-$ signals.
A Crystal Ball PDF is used to model this component, with shape parameters fixed to the values found in a fit to simulated $\Bm \to \Km \pip \pim$ decays.

The charmless contribution is determined from fits to the \Bm mass spectrum in the \D-mass sidebands, without the kinematic fit of the decay chain.
The charmless background yields are determined independently for \Bp and \Bm candidates and are then fixed in the analysis. Their uncertainties contribute to the systematic uncertainties of the final results. The largest charmless contribution is in the $\Bm \to [\pip \pim]_{D}\Km$ sample, which has a yield corresponding to 10\% of the measured signal yield. 

Partially reconstructed charmless decays of the type $B \to h_{1}^- h_{2}^- h^+ X$, where $X$ is a charged pion, neutral pion, or photon that has not been reconstructed, contribute at low invariant masses. Their contributions are determined relative to the fully reconstructed charmless components using a freely varying ratio which is shared across all $D$ decay modes. A parabola exhibiting a minimum convolved with a Gaussian resolution function is used to model this component, with shape parameter values taken from simulation.

\subsection*{\normalsize\textbf{Partially reconstructed background}}

Several additional partially reconstructed $b$-hadron decays contribute at low invariant-mass values.
The dominant contributions are from $\Bm \to \D h^- \piz$ and $\Bzb \to (\D \pip)_{\Dstarp} \pim$ decays, where a neutral pion or positively charged pion is not reconstructed.\footnote{ When considering partially reconstructed background sources, the production fractions $f_{u}$ and $f_{d}$ are taken to be equal~\cite{PDG2020}.}
The invariant-mass distribution of these sources depends upon the spin and mass of the missing particle, as with the $\Bm \to \Dstar h^-$ signals.
In both cases, the missing particle has spin-parity $0^{-}$, such that the $D h^-$ distribution is described using Eq.~\ref{eq:RooHORNSdini}, with shape parameter values taken from simulation. The Dalitz structure of $\Bm \to \D h^- \piz$ decays is modelled using \textsc{Laura++}~\cite{Laura++} and the amplitude model from Ref.~\cite{LHCb-PAPER-2014-070}.

The yields of the $\Bzb \to (\D \pip)_{\Dstarp} \pim$ and $\Bzb \to (\D \pip)_{\Dstarp} \Km$ contributions, where the $\pip$ meson is not reconstructed, are fixed relative to the corresponding $\Bm \to \D \pim$ yields using branching fractions~\cite{PDG2020} and efficiencies obtained from simulation. The \CP asymmetries of these modes are fixed to zero in all subsamples, as no \CP violation is expected in a time-integrated measurement. 

The yields of the $\Bm \to \D \pim \piz$ decay vary freely in the favoured and GLW subsamples allowing for potential \CP violation, and the total rate in the ADS mode is fixed to \mbox{$(4.0 \pm 1.3) \times 10^{-3}$} relative to the favoured mode yield. This estimate is based on the expectation that $r_B^{D\pi\piz}$ is similar in size to $r_B^{D\pi}$, and thus much smaller than the $D$ decay amplitude ratio $r_D^{K\pi}$~\cite{PDG2020}. No \CP violation is permitted in the data fit, but the fixed values of $r_{B^-}^{D\pi\piz}$ and $r_{B^-}^{D\pi\piz}$ are independently varied to determine the systematic uncertainty to account for potential \CP violation. In the ADS mode, a separate contribution from colour-suppressed \mbox{$\Bz \to [\Kp \pim]_D \pip \pim$} decays is also included, where the $\pip$ meson is not reconstructed. The yield of this component varies freely in the fit, and shape parameters are taken from a fit to simulated samples generated using \textsc{Laura++} and the amplitude model from Ref.~\cite{LHCb-PAPER-2014-070}.  

In the favoured $DK^-$ sample, the yield of the $\Bm \to \D \Km \piz$ component varies freely in both the $\Bm$ and $\Bp$ subsamples, allowing for the presence of a $B^0 \to D \Km \pip$ contribution. In the GLW and ADS modes, average values and uncertainties from Ref.~\cite{HFLAV18} are used to estimate the expected rates and \CP asymmetries, accounting for the presence of $B^0 \to D \Km \pip$. These quantities are fixed in the invariant-mass fit, and are considered as sources of systematic uncertainty. The distribution is modelled using a fit to simulated events generated using \textsc{Laura++} and the amplitude model from Ref.~\cite{LHCb-PAPER-2015-017}.   

 Contributions from partially reconstructed \mbox{$\Bm \to (\D \piz/\gamma)_{\Dstar} h^- \piz$} and \mbox{$\Bzb \to (\D \pip)_{\Dstarp} h^- \piz$} decays occur at the lowest values of invariant mass, where two particles are not reconstructed. These decays are described by the sum of several parabolas convolved with resolution functions according to Eqs.~\ref{eq:RooHORNSdini} and~\ref{eq:RooHILLdini}, with shape parameters fixed to the values found in fits to simulated samples. The yields and \CP asymmetries of these contributions vary freely in each subsample.     

In the $\Bm \to [\Kp \Km]_{D} h^-$ samples, $\Lb \to [p^+ \Km \pip]_{\Lc} h^-$ decays contribute to the background when the pion is missed and the proton is misidentified as the second kaon. 
The PDF describing this component is fixed from simulation, but the yields in the $\Bm \to [\Kp \Km]_{D} \pim$ and $\Bm \to [\Kp \Km]_{D} \Km$ subsamples vary freely. A contribution from $\Lb \to [p^+ \pim \pip]_{\Lc} h^-$ in the ADS $D \pim$ mode is also modelled using the same wide PDF, with a yield fixed relative to the $\Lb \to [p^+ \Km \pip]_{\Lc} \pim$ yield measured in the $\Bm \to [\Kp \Km]_{D} \pim$ sample using $\Lc$ branching fractions~\cite{PDG2020}. 

In the ADS $D\Km$ sample, a contribution from $\Lb \to [\Km \pip]_D p \pim$ decays is modelled, where the proton is misidentified as a kaon and the pion is not reconstructed. A fixed shape is used to describe this contribution, with shape parameters determined using a fit to a sample of $\Lb \to [\Km \pip]_D p \pim$ decays in data taken from Ref.~\cite{LHCb-PAPER-2016-061}. The rate of this mode is fixed in the invariant-mass fit relative to the favoured $\Bm \to D \pim$ yield using branching fractions~\cite{PDG2020}, efficiencies derived from simulation, and the $\Lb$ production fraction relative to $\Bp$ as measured at LHCb~\cite{LHCb-PAPER-2011-018}. The \CP asymmetry is assumed to be zero for this favoured decay in the invariant-mass fit.

In the ADS $D\Km$ sample, and to a lesser extent in the GLW $D\Km$ samples, \mbox{$\Bs \to \D \Km \pip$} decays in which the companion pion is not reconstructed contribute to the background. The PDF describing this component is fixed from fits to simulated samples generated according to the amplitude model from Ref.~\cite{LHCb-PAPER-2014-036}. The yield of this component varies freely in the ADS mode, and the GLW mode yields are fixed relative to that using $D$ branching fractions~\cite{PDG2020}. Contributions from $\Bs \to (D \g/\piz)_\Dstar \Km \pip$ are also modelled using simulated samples generated with a longitudinal polarisation fraction $f_{\rm L} = 0.9 \pm 0.1$; as this quantity has not yet been measured for $\Bs \to (D \g/\piz)_\Dstar \Km \pip$ decays, the value used is based on the $B^- \to D^* K^{*-}$ measurement~\cite{Aubert:2003ae} with an additional systematic uncertainty assigned to account for potential differences. The yield of this component is fixed relative to the freely varying $\Bs \to \D \Km \pip$ yield assuming the same branching fraction with 25\% uncertainty, and adjusting for the relative efficiency as determined using simulation. The \CP asymmetries of the $\Bs \to D^{(*)} \Km \pip$ contributions are assumed to be zero, as no \CP violation is expected for these modes in a time-integrated measurement.

In the ADS $D\pim$ sample, a contribution from favoured $\Bp \to [\Kp \pim]_D \pip \pip \pim$ decays is modelled, where the two $\pip$ produced in the $\Bp$ decay are not reconstructed. The rate of this contribution varies freely, with a fixed shape determined from a fit to simulated $\Bp \to [\Kp \pim]_D (\pip \pip \pim)_{a_1(1270)^+}$ decays. Only the $a_1(1270)^+$ contribution is simulated as this is the dominant $3\pi$ resonance observed in this mode~\cite{LHCb-PAPER-2011-040}. The \CP asymmetry is assumed to be zero for this favoured decay in the invariant-mass fit.

For all partially reconstructed background contributions considered in the fit, components accounting for particle misidentification are also taken into account. They are parameterised with  semi-empirical PDFs formed from the sum of Gaussian and error functions. The parameters of each of these PDFs are fixed to the values found in fits to simulated events.

\subsection*{\normalsize\textbf{Background from favoured decays in the ADS samples}}

The favoured and ADS signal modes have identical final states aside from the relative charge of the companion hadron and the kaon produced in the $D$ decay. It is possible to misidentify both $D$ decay products, such that a favoured decay is reconstructed as an ADS signal candidate. Given the much larger rate of favoured decays relative to the ADS signals, this crossfeed background must be reduced to a manageable level with specific requirements. Using simulated favoured $\Bm \to D \pim$ decays, a combination of the $D$ invariant-mass window, $D$ decay product PID requirements, and the veto on the $D$ mass calculated with both decay products misidentified, is found to accept doubly misidentified decays at the $10^{-4}$ level relative to correctly identified decays. This relative efficiency is used in the fit to fix the quantity of favoured background in the ADS subsamples relative to the corresponding favoured yields. The same relative efficiency is employed for the $D\pim$ and $D\Km$ samples, as well as for the $\Bm \to D h^-$ and $\Bm \to D^* h^-$ signals. The crossfeed components are modelled using fixed shapes derived from favoured simulated samples reconstructed as ADS decays. Due to the double misidentification, these components are found to be wider than their corresponding correctly identified signals.

\subsection{PID efficiencies}

In the $D\Km$ subsamples, the rates of contributions from misidentified $\Bm \to D^{(*)} \pim$ decays are determined by the fit to data via a single freely varying parameter. The $\Bm \to D^{(*)}\Km$ contributions in the $D\pim$ subsample are not well separated from background,
so the expected yield is determined using a PID calibration procedure with approximately 40 million $\Dstarp \to [\Km \pip]_{D} \pip$ decays. 
This decay is identified using kinematic variables only, and thus provides a pure sample of \Kmp and \pipm particles unbiased in the PID variables. 
The PID efficiency is parameterised as a function of particle momentum and pseudorapidity, as well as the charged-particle multiplicity in the event. 
The effective PID efficiency of the signal is determined by weighting the calibration sample such that the distributions of these variables match those of selected $\Bm \to [\Km \pip]_D \pim$ signal decays. It is found that around 70\% of $\Bm \to \D\Km$ decays pass the companion kaon PID requirement and are placed in the $D\Km$ sample, with negligible statistical uncertainty due to the size of the calibration sample; the remaining 30\% fall into the $D\pim$ sample. 
With the same PID requirement, 99.6\% of the \mbox{$\Bm \to \D \pim$} decays are correctly identified, as measured by the fit to data.
These efficiencies are also taken to represent \mbox{$\Bm \to (\D \piz)_{\Dstar} h^-$} and \mbox{$\Bm \to (\D \gamma)_{\Dstar} h^-$} signal decays in the fit, with a correction of 0.98 applied to the kaon efficiency to account for small differences in companion kinematics. The related systematic uncertainty on the kaon efficiency is determined by the size of the signal samples used, and thus increases for the lower yield modes. 

\subsection{Production and detection asymmetries}

In order to measure \CP asymmetries, the detection asymmetries for \Kpm and \pipm mesons must be taken into account. 
A detection asymmetry of $(-0.96 \pm 0.13)$\% is assigned for each kaon in the final state, primarily due to the fact that the nuclear interaction length of \Km mesons is shorter than that of \Kp mesons. 
It is taken from Ref.~\cite{LHCb-PAPER-2016-054}, where the charge asymmetries in $\Dm\to\Kp\pim\pim$ and $\Dm\to\KS\pim$ calibration samples are compared after weighting to match the kinematics of the signal kaons in favoured $\Bm \to D \pim$ decays. An additional correction of $(-0.17 \pm 0.08) \%$ is applied to the kaon detection asymmetry, to account for the asymmetry introduced by the hadronic hardware trigger. The detection asymmetry for pions is smaller, and is taken to be $(-0.17 \pm 0.10)$\%~\cite{LHCb-PAPER-2016-054}.

The \CP asymmetry in the favoured $\Bm \to [\Km \pip]_{D}\pim$ decay is fixed to \mbox{$(+0.09 \pm 0.05)\%$}, calculated using the average value and uncertainty on \g from Ref.~\cite{HFLAV18} and the assumption that $r_B^{D\pi}$ is below 0.02 with uniform probability; no assumption is made about the strong phase $\delta_B^{D\pi}$. 
This enables the effective production asymmetry, defined as $A^{\rm eff}_{\Bpm} = \textstyle{\frac{\sigma^\prime(\Bm)-\sigma^\prime(\Bp)}{\sigma^\prime(\Bm)+\sigma^\prime(\Bp)}}$, where $\sigma^\prime$ is the $B$-meson production cross-section, to be measured and simultaneously subtracted from the charge asymmetry measurements in other modes. 

To correct for left-right asymmetry effects in the LHCb detector, similarly sized data samples are collected in two opposite magnet polarity configurations. The analysis is performed on the total dataset summed over both polarities, where no residual left-right asymmetry effects remain to be corrected. 

\subsection{Yields and selection efficiencies}

The total yield for each mode is a sum of the number of correctly identified and misidentified candidates; their values are given in Table~\ref{tab:Signal_Yields}. To obtain the observable $R_{K/\pi}^{K\pi}$ ($R_{K/\pi}^{K\pi,\piz/\gamma}$) in the fit, which is defined in Table~\ref{tab:Observables_FullReco} (\ref{tab:Observables_PartReco}), the ratio of yields is corrected for the relative efficiency with which $\Bm \to \D\Km$ and $\Bm \to \D\pim$ ($\Bm \to \Dstar \Km$ and \mbox{$\Bm \to \Dstar \pim$}) decays are reconstructed and selected. The relative efficiencies are found to be close to unity, where the $\Bm \to D^{(*)} \pim$ efficiencies are around 2\% larger than $\Bm \to D^{(*)} \Km$. The uncertainties assigned on these efficiency corrections take into account the size of the simulated samples and the imperfect modelling of the relative pion and kaon absorption in the detector material. To determine the branching fraction \mbox{$\mathcal{B}(\Dstarz \to \Dz \piz)$}, the yields of the \mbox{$\Bm \to (\D \piz)_{\Dstar} \pim$} and \mbox{$\Bm \to (\D \gamma)_{\Dstar} \pim$} modes are corrected for the relative efficiencies of the neutral pion and photon modes as determined from simulation. As both of these modes are partially reconstructed with identical selection requirements, the relative efficiency is found to be close to unity and is varied within its uncertainty to determine the associated systematic uncertainty. In the measurement of $\mathcal{B}(\Dstar \to \D \piz)$, the assumption is made that \mbox{$\mathcal{B}(\Dstar \to \D \piz) + \mathcal{B}(\Dstar \to \D \gamma) = 1$}~\cite{PDG2020}. The branching fraction $\mathcal{B}(\Bm \to \Dstarz \pim)$ is determined from the total $\Bm \to \Dstar \pim$ and $\Bm \to \D \pim$ yields, the relative efficiencies determined from simulation, and the $\Bm \to \D \pim$ branching fraction~\cite{PDG2020}. Both the efficiencies and external input branching fraction are varied to determine the associated systematic uncertainty.

\begingroup
\renewcommand*{\arraystretch}{1.2}
\begin{table}[!h]
\small
\begin{center}
\caption{Yields for the 24 signal modes. \label{tab:Signal_Yields}}
\begin{tabular}{l l l}
\toprule
Decay & $D$ mode & Yield \\ \midrule
$\Bpm \to D\pipm$ & $\Km \pip$ & $1\,771\,385 \phantom{0}\pm\phantom{0} 2153$ \\
 & $\Kp\Km$ & $219\,584\phantom{\,0} \phantom{0}\pm\phantom{0} 569$ \\
 & $\pip\pim$ & $70\,594\phantom{\,00} \phantom{0}\pm\phantom{0} 273$ \\
 & $\Kp\pim$ & $6518\phantom{\,\,000} \phantom{0}\pm\phantom{0} 99$ \\
$\Bpm \to D\Kpm$ & $\Km \pip$ & $136\,734\phantom{\,0} \phantom{0}\pm\phantom{0} 457$ \\
 & $\Kp\Km$ & $16\,107\phantom{\,00} \phantom{0}\pm\phantom{0} 147$ \\
 & $\pip\pim$ & $5178\phantom{\,\,000} \phantom{0}\pm\phantom{0} 49$ \\
 & $\Kp\pim$ & $2372\phantom{\,\,000} \phantom{0}\pm\phantom{0} 65$ \\
$\Bpm \to (D \piz)_{D^*}\pipm$ & $\Km \pip$ & $1\,106\,081 \phantom{0}\pm\phantom{0} 10\,828$ \\
 & $\Kp\Km$ & $137\,111\phantom{\,0} \phantom{0}\pm\phantom{0} 1378$ \\
 & $\pip\pim$ & $44\,080\phantom{\,00} \phantom{0}\pm\phantom{0} 461$ \\
 & $\Kp\pim$ & $5292\phantom{\,\,000} \phantom{0}\pm\phantom{0} 543$ \\
$\Bpm \to (D \piz)_{D^*}\Kpm$ & $\Km \pip$ & $90\,031\phantom{\,00} \phantom{0}\pm\phantom{0} 1197$ \\
 & $\Kp\Km$ & $11\,660\phantom{\,00} \phantom{0}\pm\phantom{0} 277$ \\
 & $\pip\pim$ & $3748\phantom{\,\,000} \phantom{0}\pm\phantom{0} 90$ \\
 & $\Kp\pim$ & $1124\phantom{\,\,000} \phantom{0}\pm\phantom{0} 231$ \\
$\Bpm \to (D \g)_{D^*}\pipm$ & $\Km \pip$ & $536\,615\phantom{\,0} \phantom{0}\pm\phantom{0} 6065$ \\
 & $\Kp\Km$ & $66\,519\phantom{\,00} \phantom{0}\pm\phantom{0} 769$ \\
 & $\pip\pim$ & $21\,385\phantom{\,00} \phantom{0}\pm\phantom{0} 254$ \\
 & $\Kp\pim$ & $2273\phantom{\,\,000} \phantom{0}\pm\phantom{0} 403$ \\
$\Bpm \to (D \g)_{D^*}\Kpm$ & $\Km \pip$ & $44\,255 \phantom{\,00} \phantom{0}\pm\phantom{0} 899$ \\
 & $\Kp\Km$ & $5310\phantom{\,\,000} \phantom{0}\pm\phantom{0} 361$ \\
 & $\pip\pim$ & $1707\phantom{\,\,000} \phantom{0}\pm\phantom{0} 116$ \\
 & $\Kp\pim$ & $674\phantom{\,\,0000} \phantom{0}\pm\phantom{0} 931$ \\
 \bottomrule
\end{tabular}
\end{center}
\end{table}
\endgroup

\FloatBarrier

\begingroup
\renewcommand*{\arraystretch}{1.3}
\begin{table}
\centering
\begin{tabular}{l r r r r r r r}
\toprule
Observable & PID & PDF & Rates & Asym & Eff & Veto & Total \\ \midrule
$A_K^{CP}$ & 6 & 10 & 11 & 5 & 1 & 10 & 16 \\
$A_\pi^{CP}$ & 4 & 14 & 15 & 70 & 3 & 10 & 74 \\
$A_K^{K\pi}$ & 12 & 7 & 11 & 49 & 1 & 10 & 52 \\
$R^{CP}$ & 24 & 88 & 58 & 0 & 16 & 10 & 109 \\
$R_{K/\pi}^{K\pi}$ & 47 & 243 & 104 & 1 & 402 & 10 & 483 \\
$R_{K^-}^{\pi K}$ & 2 & 48 & 30 & 3 & 2 & 10 & 57 \\
$R_{\pi^-}^{\pi K}$ & 2 & 41 & 15 & 13 & 4 & 10 & 46 \\
$R_{K^+}^{\pi K}$ & 3 & 47 & 23 & 2 & 6 & 10 & 53 \\
$R_{\pi^+}^{\pi K}$ & 2 & 44 & 15 & 15 & 6 & 10 & 50 \\
$A_K^{CP,\gamma}$ & 9 & 34 & 40 & 18 & 9 & 10 & 57 \\
$A_K^{CP,\pi^0}$ & 9 & 28 & 31 & 16 & 10 & 10 & 47 \\
$A_K^{K\pi,\gamma}$ & 4 & 8 & 14 & 15 & 2 & 10 & 22 \\
$A_K^{K\pi,\pi^0}$ & 9 & 12 & 19 & 34 & 5 & 10 & 42 \\
$R^{CP,\gamma}$ & 2 & 87 & 55 & 2 & 22 & 10 & 105 \\
$R^{CP,\pi^0}$ & 30 & 87 & 76 & 0 & 33 & 10 & 124 \\
$R_{K/\pi}^{K\pi,\gamma/\pi^0}$ & 58 & 292 & 187 & 25 & 185 & 10 & 398 \\
$R_{K^-}^{\pi K,\gamma}$ & 13 & 117 & 82 & 14 & 21 & 10 & 146 \\
$R_{K^-}^{\pi K,\pi^0}$ & 4 & 39 & 48 & 4 & 22 & 10 & 66 \\
$R_{K^+}^{\pi K,\gamma}$ & 11 & 117 & 83 & 7 & 21 & 10 & 146 \\
$R_{K^+}^{\pi K,\pi^0}$ & 3 & 41 & 47 & 3 & 16 & 10 & 64 \\
$A_\pi^{CP,\gamma}$ & 2 & 18 & 39 & 11 & 3 & 10 & 45 \\
$A_\pi^{CP,\pi^0}$ & 2 & 16 & 16 & 31 & 3 & 10 & 39 \\
$A_\pi^{K\pi,\gamma}$ & 4 & 22 & 19 & 18 & 4 & 10 & 34 \\
$A_\pi^{K\pi,\piz}$ & 2 & 2 & 13 & 32 & 1 & 10 & 34 \\
$R_{\pi^-}^{\pi K,\gamma}$ & 13 & 114 & 57 & 11 & 6 & 10 & 128 \\
$R_{\pi^-}^{\pi K,\pi^0}$ & 1 & 86 & 60 & 16 & 15 & 10 & 107 \\
$R_{\pi^+}^{\pi K,\gamma}$ & 14 & 115 & 45 & 12 & 8 & 10 & 125 \\
$R_{\pi^+}^{\pi K,\pi^0}$ & 2 & 85 & 57 & 16 & 9 & 10 & 104 \\
$\mathcal{B}(D^* \to D \pi^0)$ & 27 & 281 & 76 & 8 & 177 & 10 & 342 \\
$\mathcal{B}(B^\pm \to D^{*0} \pipm)$ & 17 & 257 & 148 & 2 & 329 & 10 & 444 \\
\bottomrule
\end{tabular}
\captionof{table}{Systematic uncertainties for all observables, where values are quoted as a percentage of the statistical uncertainty for a given observable. The total uncertainty is given by the sum in quadrature of each contribution. PID refers to fixed PID efficiencies, PDF to fixed PDF parameters, Rates to fixed background contributions, Asym to the use of fixed detection asymmetries and background \CP asymmetries, Eff to the use of fixed efficiencies from simulation, and Veto to the procedure used to veto fully reconstructed $\Bm \to D^* h^-$ candidates. \label{tab:systematics_summary}}
\end{table}
\endgroup

\section{Systematic uncertainties}
\label{sec:Systematics}

The 30 observables of interest (28 \CP observables and two branching fractions) are subject to a set of systematic uncertainties resulting from the use of fixed parameters in the fit.
The systematic uncertainties associated with using these fixed parameters are assessed by repeating the fit 1000 times, varying the value of each external parameter within its uncertainty according to a Gaussian distribution. The resulting standard deviation of each observable under this variation is taken as the systematic uncertainty. The systematic uncertainties, grouped into six categories, are shown for each observable in Table~\ref{tab:systematics_summary}. Correlations between the categories are negligible, but correlations within categories are accounted for. The total systematic uncertainties are determined by the sum in quadrature of each category.


\section{Results}
\label{sec:results}

The \CP observable and branching fraction results are
\begin{align*}
A_K^{CP} &= \phantom{-}0.136\phantom{00} \pm 0.009\phantom{00} \pm 0.001, \\ 
A_\pi^{CP} &= -0.008\phantom{00} \pm 0.002\phantom{00} \pm 0.002, \\ 
A_K^{K\pi} &= -0.011\phantom{00} \pm 0.003\phantom{00} \pm 0.002, \\ 
R^{CP} &= \phantom{-}0.950\phantom{00} \pm 0.009\phantom{00} \pm 0.010, \\ 
R_{K/\pi}^{K\pi} &= \phantom{-}0.0796\phantom{0} \pm 0.0003\phantom{0} \pm 0.0013, \\ 
R_{K^-}^{\pi K} &= \phantom{-}0.0095\phantom{0} \pm 0.0005\phantom{0} \pm 0.0003,\\ 
R_{\pi^-}^{\pi K} &= \phantom{-}0.00415 \pm 0.00008 \pm 0.00004, \\ 
R_{K^+}^{\pi K} &= \phantom{-}0.0252\phantom{0} \pm 0.0008\phantom{0} \pm 0.0004, \\ 
R_{\pi^+}^{\pi K} &= \phantom{-}0.00320 \pm 0.00007 \pm 0.00004, \\ 
A_K^{CP,\gamma} &= \phantom{-}0.123\phantom{00} \pm 0.054\phantom{00} \pm 0.031, \\ 
A_K^{CP,\pi^0} &= -0.115\phantom{00} \pm 0.019\phantom{00} \pm 0.009, \\ 
A_K^{K\pi,\gamma} &= -0.004\phantom{00} \pm 0.014\phantom{00} \pm 0.003, \\ 
A_K^{K\pi,\pi^0} &= \phantom{-}0.020\phantom{00} \pm 0.007\phantom{00} \pm 0.003, \\ 
R^{CP,\gamma} &= \phantom{-}0.952\phantom{00} \pm 0.062\phantom{00} \pm 0.065, \\ 
R^{CP,\pi^0} &= \phantom{-}1.051\phantom{00} \pm 0.022\phantom{00} \pm 0.028, \\ 
R_{K/\pi}^{K\pi,\gamma/\pi^0} &= \phantom{-}0.0851\phantom{0} \pm 0.0012\phantom{0} \pm 0.0048, \\ 
R_{K^-}^{\pi K,\gamma} &= \phantom{-}0.0117\phantom{0} \pm 0.0215\phantom{0} \pm 0.0313, \\ 
R_{K^-}^{\pi K,\pi^0} &= \phantom{-}0.0202\phantom{0} \pm 0.0035\phantom{0} \pm 0.0023, \\ 
R_{K^+}^{\pi K,\gamma} &= \phantom{-}0.0292\phantom{0} \pm 0.0214\phantom{0} \pm 0.0312, \\ 
R_{K^+}^{\pi K,\pi^0} &= \phantom{-}0.0033\phantom{0} \pm 0.0035\phantom{0} \pm 0.0022, \\ 
A_\pi^{CP,\gamma} &= \phantom{-}0.000\phantom{00} \pm 0.014\phantom{00} \pm 0.006, \\ 
A_\pi^{CP,\pi^0} &= \phantom{-}0.013\phantom{00} \pm 0.007\phantom{00} \pm 0.003, \\ 
A_\pi^{K\pi,\gamma} &= -0.004\phantom{00} \pm 0.004\phantom{00} \pm 0.001, \\ 
A_\pi^{K\pi,\piz} &= \phantom{-}0.001\phantom{00} \pm 0.002\phantom{00} \pm 0.001, \\ 
R_{\pi^-}^{\pi K,\gamma} &= \phantom{-}0.00472 \pm 0.00092 \pm 0.00118, \\ 
R_{\pi^-}^{\pi K,\pi^0} &= \phantom{-}0.00405 \pm 0.00056 \pm 0.00059, \\ 
R_{\pi^+}^{\pi K,\gamma} &= \phantom{-}0.00403 \pm 0.00091 \pm 0.00114, \\ 
R_{\pi^+}^{\pi K,\pi^0} &= \phantom{-}0.00536 \pm 0.00056 \pm 0.00058, \\ 
\mathcal{B}(D^* \to D \pi^0) &= \phantom{-}0.655\phantom{00} \pm 0.003\phantom{00} \pm 0.012, \\ 
\mathcal{B}(B^\pm \to D^{*0} \pipm) &= \phantom{-}0.00535 \pm 0.00004 \pm 0.00016 \pm 0.00015,
\end{align*}
where the first uncertainties quoted are statistical and the second systematic; the third uncertainty on $\mathcal{B}(\Bpm \to D^{*0} \pipm)$ accounts for the use of the external branching fraction $\mathcal{B}(\Bpm \to D^0 \pipm) = (4.68 \pm 0.13) \times 10^{-3}$~\cite{PDG2020}. The statistical and systematic correlation matrices are given in App.~\ref{sec:corr_matrices}. The $R_{h^-}^{\pi K}$ and $R_{h^+}^{\pi K}$ ADS \CP observables can be expressed in terms of a charge-averaged rate $R_h^{\pi K}$ and an asymmetry $A_h^{\pi K}$
\begin{align*}
    R_h^{\pi K} &= (R_{h^-}^{\pi K} + R_{h^+}^{\pi K})/2\,, \\
    A_h^{\pi K} &= (R_{h^-}^{\pi K} - R_{h^+}^{\pi K})/(R_{h^-}^{\pi K} + R_{h^+}^{\pi K})\,.
\end{align*}
The values of these derived observables are
\begin{align*}
R_{K}^{\pi K} &= \phantom{-}0.0173\phantom{0} \pm 0.0006,\\ 
R_{K}^{\pi K,\gamma} &= \phantom{-}0.0163\phantom{0} \pm 0.0373,\\ 
R_{K}^{\pi K,\piz} &= \phantom{-}0.0118\phantom{0} \pm 0.0034,\\ 
R_{\pi}^{\pi K} &= \phantom{-}0.00368 \pm 0.00007,\\ 
R_{\pi}^{\pi K,\gamma} &= \phantom{-}0.00420 \pm 0.00138,\\ 
R_{\pi}^{\pi K,\piz} &= \phantom{-}0.00471 \pm 0.00077,\\ 
A_{K}^{\pi K} &= -0.451\phantom{00} \pm 0.026,\\ 
A_{K}^{\pi K,\gamma} &= -0.558\phantom{00} \pm 1.349,\\ 
A_{K}^{\pi K,\piz} &= \phantom{-}0.717\phantom{00} \pm 0.286,\\ 
A_{\pi}^{\pi K} &= \phantom{-}0.129\phantom{00} \pm 0.014,\\ 
A_{\pi}^{\pi K,\gamma} &= \phantom{-}0.079\phantom{00} \pm 0.128,\\ 
A_{\pi}^{\pi K,\piz} &= -0.140\phantom{00} \pm 0.059,
\end{align*}
where the statistical and systematic uncertainties are combined according to the correlations between the $R_{h^-}^{\pi K}$ and $R_{h^+}^{\pi K}$ observables.

World-best measurements of \CP observables in $\Bm \to D^{(*)} h^-$ decays are obtained with the \D meson reconstructed in the $\Kp\pim$, $\Kp\Km$, and $\pip\pim$ final states; these supersede earlier work on the GLW modes presented in Ref.~\cite{LHCb-PAPER-2017-021}. Updated world-best measurements of \CP observables in ADS $\Bm \to D h^-$ decays are also made, which supersede the results in Ref.~\cite{LHCb-PAPER-2016-003}. Measurements of \CP observables in ADS $\Bm \to D^* h^-$ decays are made for the first time at LHCb. The ADS $\Bm \to (D \piz)_{\Dstar} \Km$ signal is measured with a  significance of 3.5 standard deviations ($\sigma$, where both the statistical and systematic uncertainties are considered), with \CP violation measured to be non-zero at the $2.5\sigma$ level. The $\Bm \to (D \gamma)_{\Dstar} \Km$ signal is measured to be consistent with zero, which is due to the large uncertainties incurred as a result of correlations with several partially reconstructed background contributions. The value of $A_K^{\pi K,\piz}$ is consistent with the BaBar result~\cite{delAmoSanchez:2010dz}, while $R_K^{\pi K, \piz}$ is found to be smaller but consistent within measurement uncertainties. The values of $A_K^{\pi K,\gamma}$ and $R_K^{\pi K,\gamma}$ are consistent with the results from Ref.~\cite{delAmoSanchez:2010dz}. A first observation of the ADS $\Bm \to (D \piz)_{\Dstar} \pim$ decay is made with a significance of $6.1\sigma$, with \CP violation measured to be non-zero at the $2.4\sigma$ level. The ADS \mbox{$\Bm \to (D \gamma)_{\Dstar} \pim$} signal is measured with a significance of $3.0\sigma$, where the degree of \CP violation measured is consistent with zero. 

In general, good agreement is found with previous results from LHCb and the $B$-factories. However, the value of $R^{CP}$ has decreased from $R^{CP} = 0.989 \pm 0.013 \pm 0.010$ in Ref.~\cite{LHCb-PAPER-2017-021} due to the veto applied to remove background from $\Bm \to [h_1^- h^+]_D h_2^-$ decays where $h_1^-$ and $h_2^-$ are swapped. In Refs.~\cite{LHCb-PAPER-2017-021} and~\cite{LHCb-PAPER-2016-003} this veto was not applied, resulting in peaking background contamination from favoured $\Bm \to [\Km \pip]_D \pim$ decays in the $\Bm \to [\pip \pim]_D \Km$ sample which artificially increased the value of $R^{CP}$. The value of $R_K^{\pi K}$ has also reduced due to this veto and the modelling of additional background sources, such as $\Lb \to D p \pim$, which were not previously considered.

The values of $R_{K/\pi}^{K\pi,\gamma/\pi^0}$, $\mathcal{B}(\Dstarz \to \Dz \piz)$, and $\mathcal{B}(\Bpm \to \Dstarz \pipm)$ are found to agree well with the current world average values, ignoring previous LHCb inputs to the averages. These measurements demonstrate that the method of partial reconstruction accurately measures the $\Bm \to (\D \piz)_{\Dstar} h^-$ and $\Bm \to (\D \gamma)_{\Dstar} h^-$ signals, despite the presence of several partially reconstructed background sources which decrease the purity and introduce anti-correlations in the fit.

\section{Interpretation and conclusion}
\label{sec:conclusions}

Using the $\Bm \to D^{(*)} h^-$ \CP observable results as input, profile likelihood contours in the fundamental parameters $(\g,r_B^{DK},\delta_B^{DK},r_B^{D\pi},\delta_B^{D\pi},r_B^{D^*K},\delta_B^{D^*K},r_B^{D^*\pi},\delta_B^{D^*\pi})$ are constructed using Eqs.~\ref{masterADS} and~\ref{masterFAV} following Ref.~\cite{LHCb-PAPER-2016-032}. The parameters $r_D^{K\pi}$ and $\delta_D^{K\pi}$ are the amplitude ratio and strong phase difference for the $D \to K \pi$ decay, which are taken from Ref.~\cite{HFLAV18}. Similar expressions can be written for $\Bm \to D^* h^-$ decays, where the exact strong phase difference of $\pi$ between the $D^* \to D \piz$ and $D^* \to D \g$ decays is taken into account~\cite{PhysRevD.70.091503}. The effects of $\Dz-\Dzb$ mixing on the measured \CP observable values are accounted for in Eqs.~\ref{masterADS} and~\ref{masterFAV} within the terms proportional to the decay-time acceptance coefficient, $\alpha$, and the charm mixing parameters, $x$ and $y$~\cite{HFLAV18}. 
The experimental $D$ lifetime acceptance is studied using a fit to the $D$-candidate lifetime distribution in favoured $\Bm \to D \pim$ data, where $\alpha = 1.045 \pm 0.008$ is found.

\begin{figure}[!h]
  \begin{center}
   \includegraphics[width=0.35\linewidth]{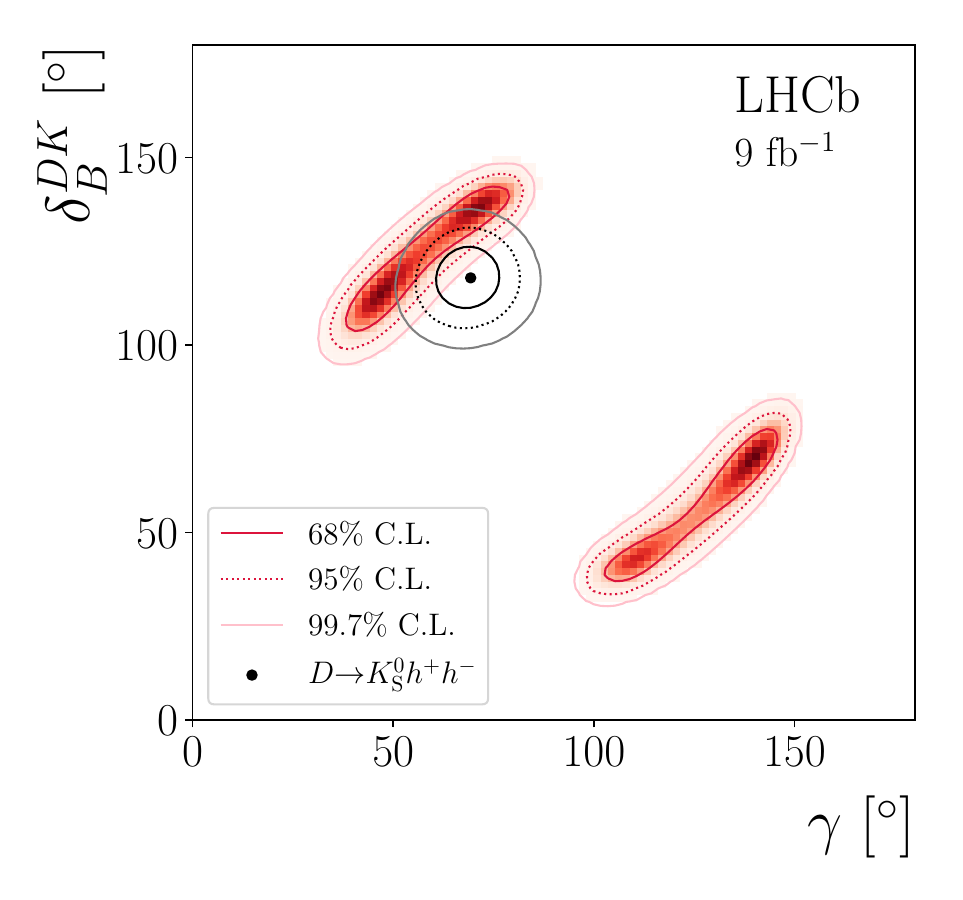}
   \includegraphics[width=0.35\linewidth]{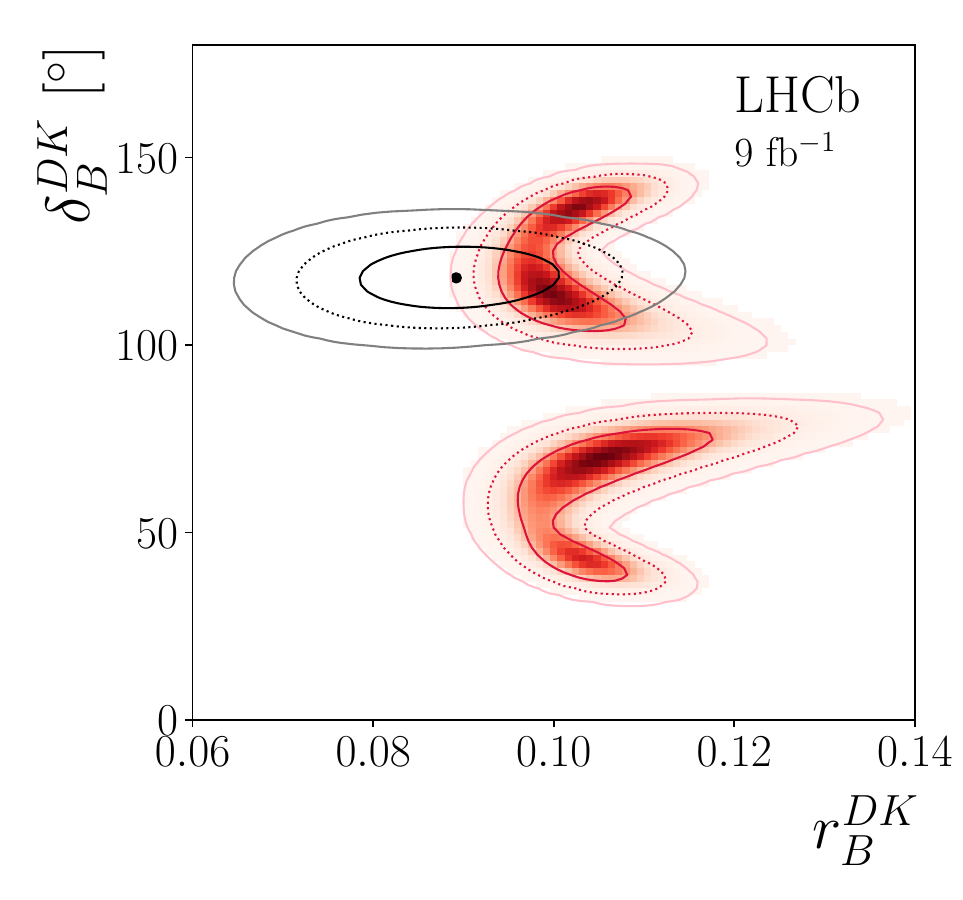}\\
   \includegraphics[width=0.35\linewidth]{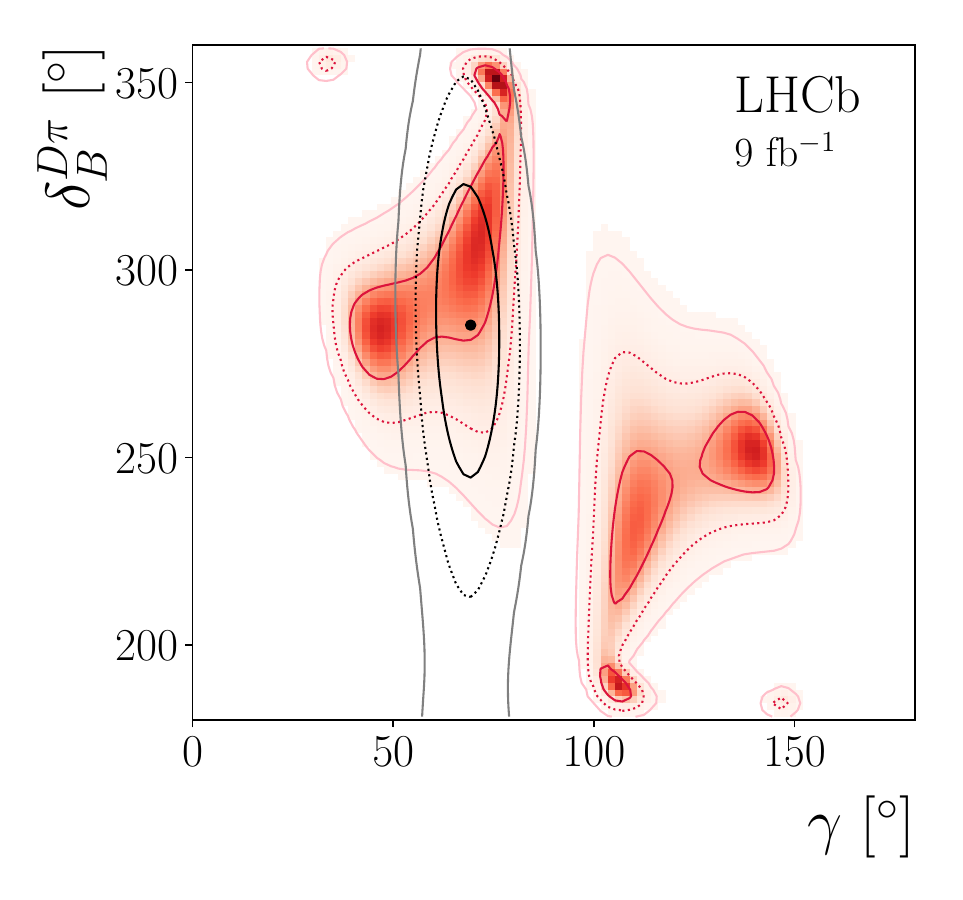}
   \includegraphics[width=0.35\linewidth]{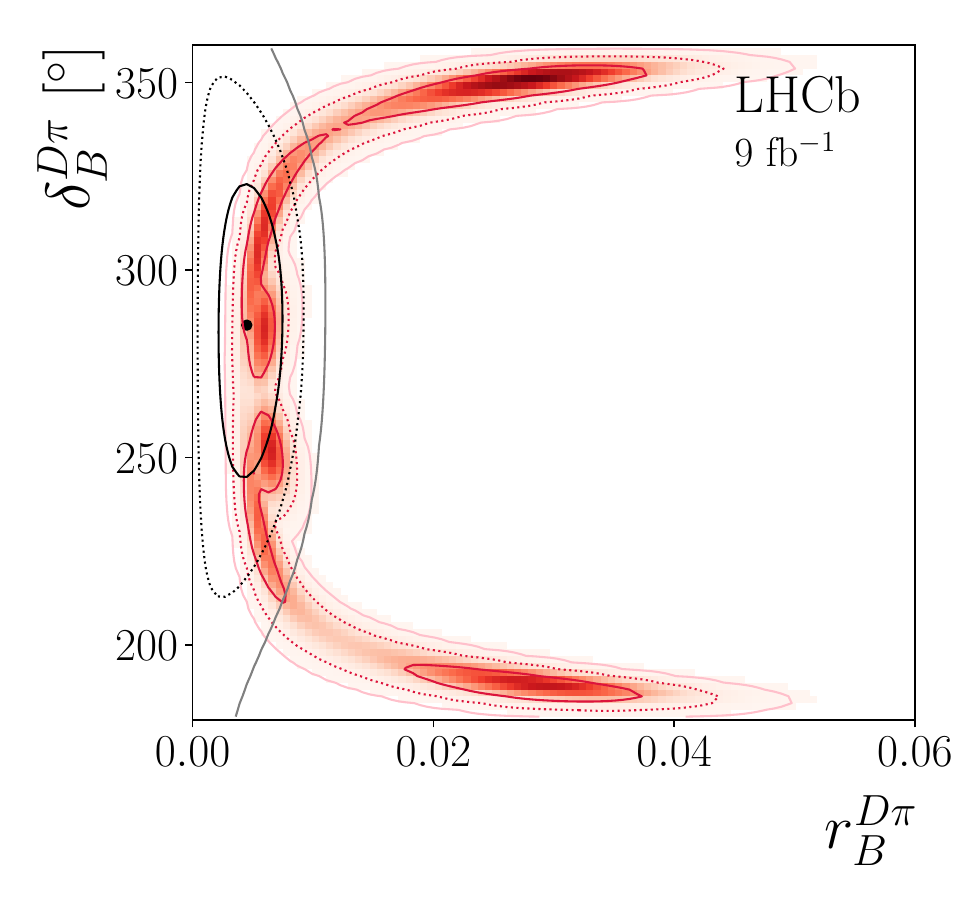}\\
   \includegraphics[width=0.35\linewidth]{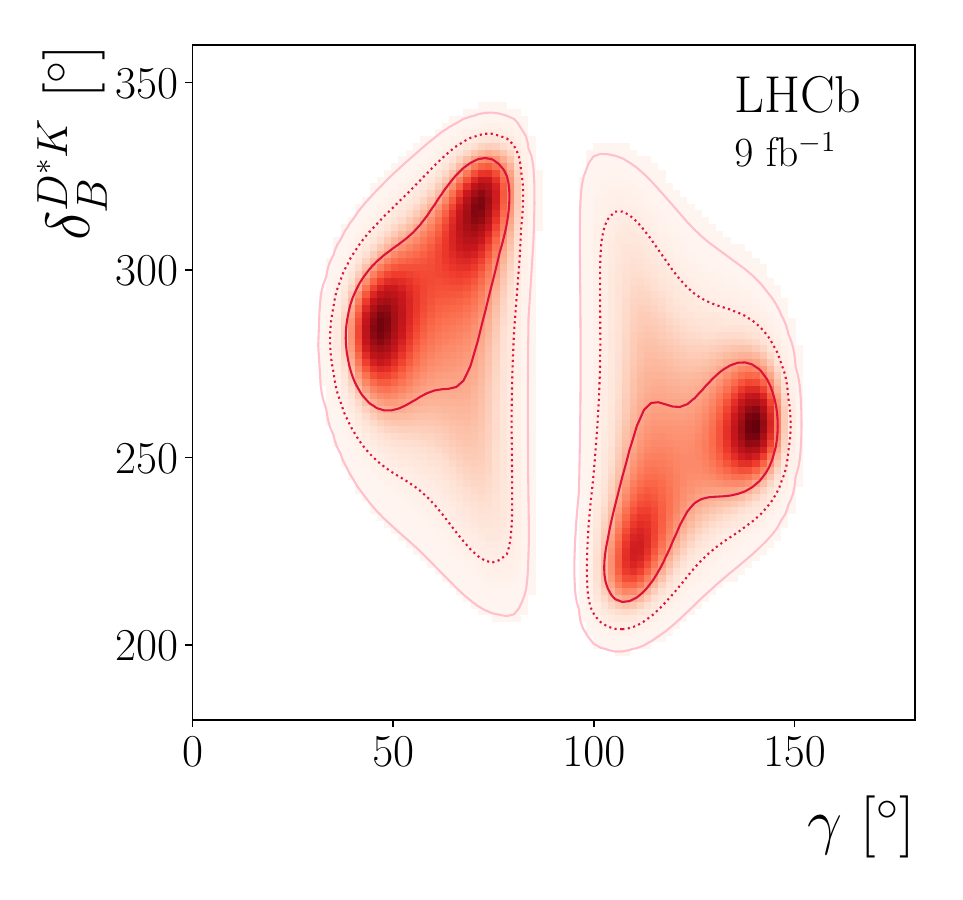}
   \includegraphics[width=0.35\linewidth]{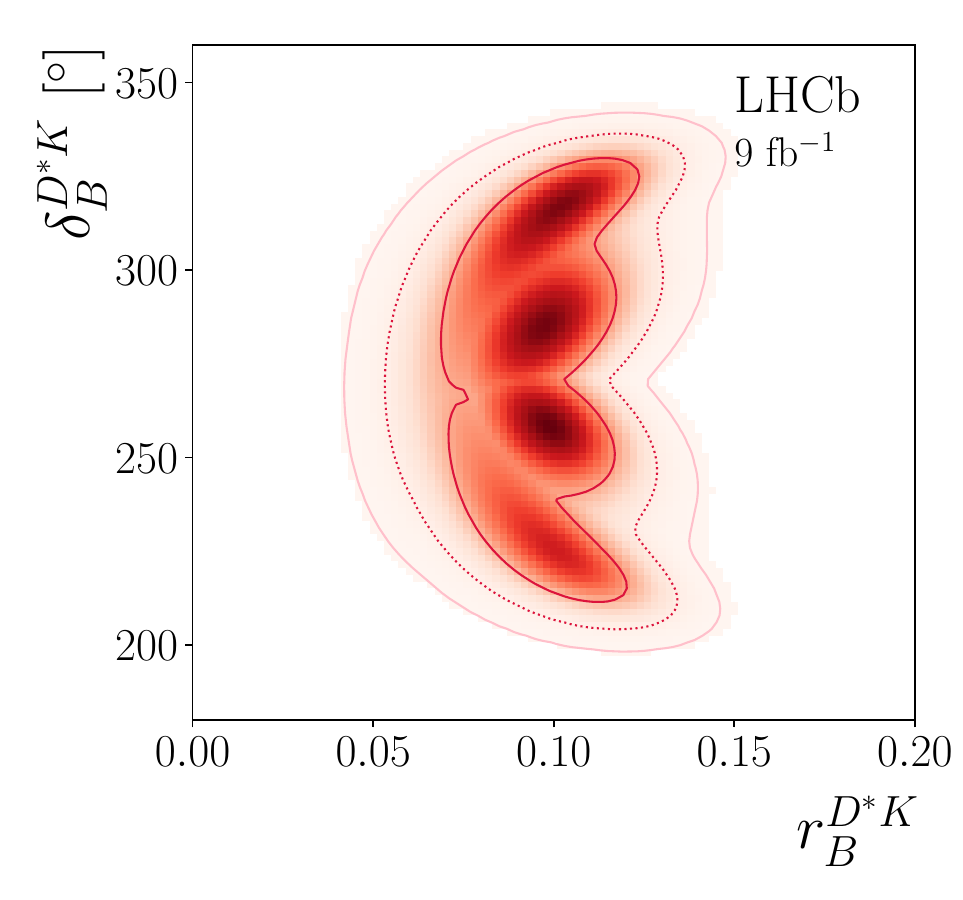}\\
   \includegraphics[width=0.35\linewidth]{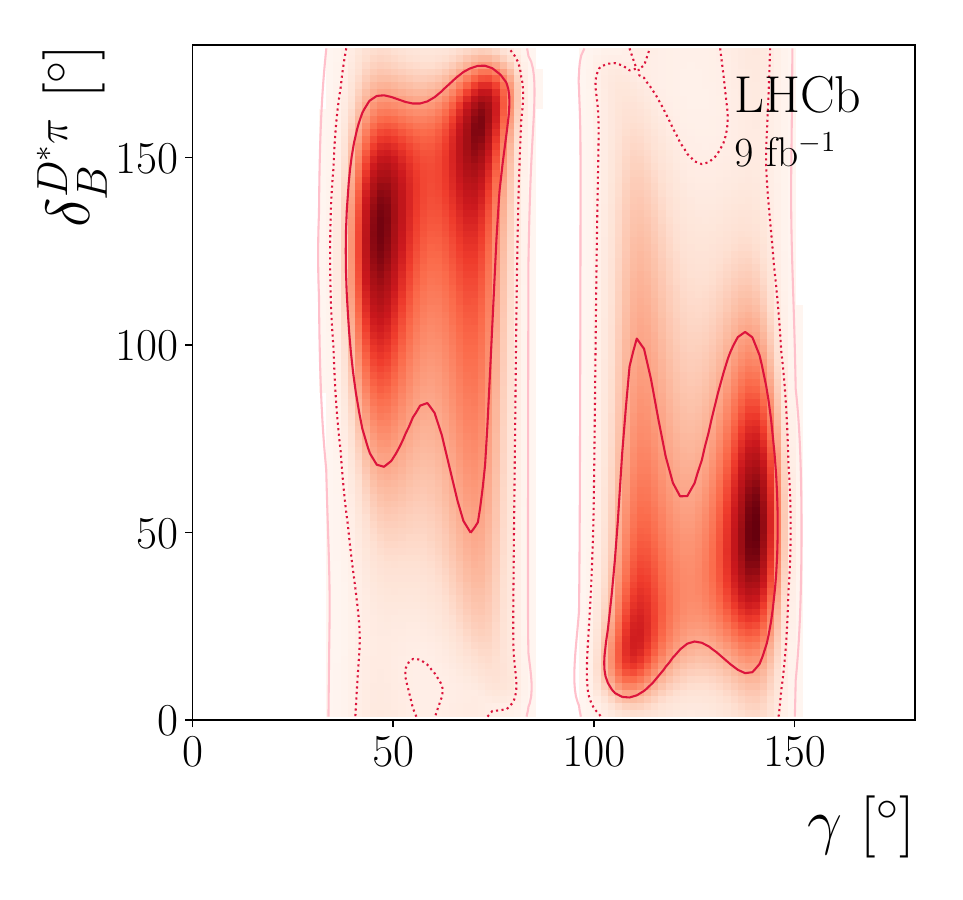}
   \includegraphics[width=0.35\linewidth]{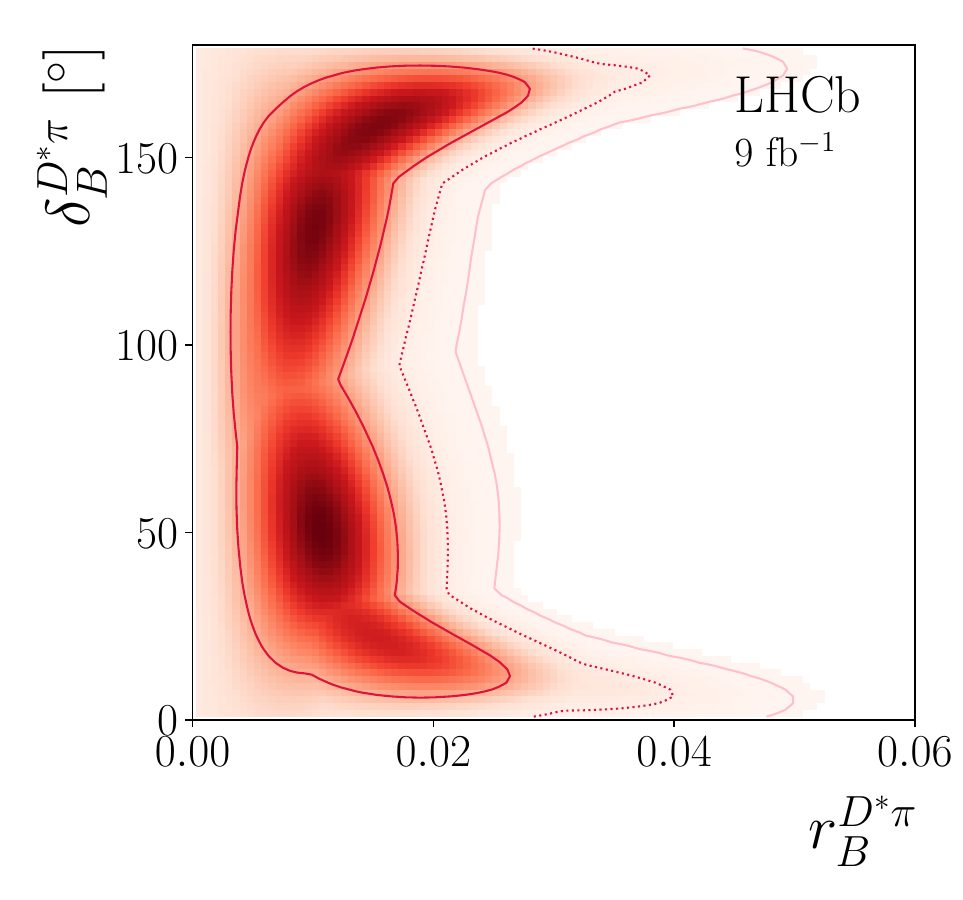}
   \end{center}
   \caption{Confidence regions for the (top row) $\Bm \to D \Km$ , (second row) $\Bm \to D \pim$ , (third row) $\Bm \to D^* \Km$ , and (fourth row) $\Bm \to D^* \pim$ fundamental parameters. The $\Bm \to D h^-$ with $D \to \KS h^+ h^-$ contours from Ref.~\cite{LHCb-PAPER-2020-019} are also overlaid.
  \label{fig:gammadini}}
\end{figure}

The profile likelihood contours for all fundamental parameters at 68\%, 95\%, and 99.7\% confidence level are shown in Fig.~\ref{fig:gammadini}. The contours found are dominated by the $\Bm \to D \Km$ measurements, although information from $\Bm \to D \pim$, $\Bm \to D^* \Km$, and $\Bm \to D^* \pim$ is also used in all cases. Compared to the ADS/GLW likelihood contours constructed using previous LHCb $\Bm \to D \Km$ results~\cite{LHCb-CONF-2018-002}, the favoured values of $r_B^{DK}$ are lower. This is due to the lower value of $R^{CP}$ measured in this analysis. As a result of this change in $R^{CP}$, the four distinct solutions visible in the $(\g,\delta_B^{DK})$ plane have merged into two distinct bands, which reduces the standalone sensitivity to $\g$ of the ADS/GLW $\Bm \to D \Km$ modes. 
The corresponding contours for $\Bm \to D h^-$ from Ref.~\cite{LHCb-PAPER-2020-019} are overlaid in Fig.~\ref{fig:gammadini}, and show good agreement with the results of this analysis both for $\Bm \to D \Km$ and $\Bm \to D \pim$.

The preferred value of $r_B^{D^*K}$ is around 0.1, which is consistent with the BaBar combination for $\Bm \to D^* \Km$~\cite{Lees:2013nha}. The favoured values of $\delta_B^{D^*K}$ are also consistent with those found in Ref.~\cite{Lees:2013nha}, with values around $300^\circ$ for $\gamma < 90^\circ$. Values of $r_B^{D^*\pi} \sim 0.01$ are favoured, with $\delta_B^{D^*\pi}$ around $150^\circ$ for $\gamma < 90^\circ$.  

When constructing these confidence regions, the charm parameters $x$, $y$, $r_D^{K\pi}$, and $\delta_D^{K\pi}$ are provided with their correlations as external constraints from Ref.~\cite{HFLAV18}. Alternatively, it is possible to make a measurement of $\delta_D^{K\pi}$ and $y$ by allowing them to vary freely in a combination of results. Following such a strategy using the fully-reconstructed $\Bm \to D h^-$ results in this analysis as well as recent studies of $\Bm\to[\KS h^+h^-]_Dh^-$ decays~\cite{LHCb-PAPER-2020-019}, \mbox{$\gamma=(61.8\pm4.0)^\circ$}, \mbox{$\delta_B^{DK}=(123.8\pm4.8)^\circ$}, and \mbox{$r_B^{DK}=0.0964\pm0.0028$} are found. This combination also finds  \mbox{$y=(0.76\pm0.24)\%$} and \mbox{$\delta_D^{K\pi} = (192.3 \pm 6.0)^\circ$} with a correlation of $+0.42$, where a $0.6^\circ$ systematic uncertainty on $\delta_D^{K\pi}$ is included from the necessary constraints on $x$ and $r_D^{K\pi}$. This compares favourably to the current world average, $\delta_D^{K\pi}=196.1^{+\ 7.9}_{-10.1}$~\cite{HFLAV18}. The fact that \mbox{$B \to D X$} measurements provide significant input to the understanding of charm parameters motivates a comprehensive combination of all $B\to DX$ results together with relevant charm results. 

In summary, measurements of \CP observables in $B^\pm \rightarrow D^{(*)} K^\pm$ and $B^\pm \rightarrow D^{(*)} \pi^\pm$ decays are made, where decays of the $\D$ meson are reconstructed in the $\Kpm \pimp$, $\Kp\Km$, and $\pip\pim$ final states. Decays of the $\Dstar$ meson to the $\D\piz$ and $\D\gamma$ final states are partially reconstructed without inclusion of the neutral pion or photon. The measurements of partially reconstructed $\Bpm \to \Dstar \Kpm$ and $\Bpm \to \Dstar \pipm$ with $D \to \Kmp \pipm$ decays are the first of their kind, and a first observation of the $\Bpm \to (D \pi^0)_{D^*} \pipm$ decay is made with a statistical significance of 6.1 standard deviations. All \CP observables are measured with world-best precision, and in combination with other LHCb results will provide strong constraints on the CKM angle $\gamma$.

\clearpage
\newpage

\appendix

\clearpage
\newpage

\section{Correlation matrices}
\label{sec:corr_matrices}

\begingroup
\renewcommand*{\arraystretch}{2.0}
\begin{table}[!h]
\centering
\caption{Statistical correlation matrix for $\Bpm \to D h^\pm$ observables.}
\scriptsize
\begin{tabular}{l | c c c c c c c c c }
\toprule
& $A_K^{CP}$ & $A_\pi^{CP}$ & $A_K^{K\pi}$ & $R^{CP}$ & $R_{K/\pi}^{K\pi}$ & $R_{K^-}^{\pi K}$ & $R_{\pi^-}^{\pi K}$ & $R_{K^+}^{\pi K}$ & $R_{\pi^+}^{\pi K}$ \\
\midrule
$A_K^{CP}$ & \phantom{$-$}$1.00$ & \phantom{$-$}$0.00$ & \phantom{$-$}$0.02$ & $-0.02$ & $\phantom{-}0.00$ & \phantom{$-$}$0.00$ & \phantom{$-$}$0.00$ & $\phantom{-}0.00$ & $\phantom{-}0.00$\\
$A_\pi^{CP}$ & \phantom{$-0.0$} & \phantom{$-$}$1.00$ & \phantom{$-$}$0.08$ & $\phantom{-}0.00$ & $\phantom{-}0.00$ & \phantom{$-$}$0.01$ & \phantom{$-$}$0.01$ & $\phantom{-}0.00$ & $-0.01$\\
$A_K^{K\pi}$ & \phantom{$-0.0$} & \phantom{$-0.0$} & \phantom{$-$}$1.00$ & \phantom{$-$}$0.00$ & $\phantom{-}0.00$ & \phantom{$-$}$0.00$ & \phantom{$-$}$0.01$ & $\phantom{-}0.00$ & $-0.01$\\
$R^{CP}$ & \phantom{$-0.0$} & \phantom{$-0.0$} & \phantom{$-0.0$} & \phantom{$-$}$1.00$ & $-0.33$ & \phantom{$-$}$0.03$ & $\phantom{-}0.00$ & \phantom{$-$}$0.04$ & $\phantom{-}0.00$\\
$R_{K/\pi}^{K\pi}$ & \phantom{$-0.0$} & \phantom{$-0.0$} & \phantom{$-0.0$} & \phantom{$-0.0$} & \phantom{$-$}$1.00$ & $-0.05$ & \phantom{$-$}$0.01$ & $-0.10$ & \phantom{$-$}$0.00$\\
$R_{K^-}^{\pi K}$ & \phantom{$-0.0$} & \phantom{$-0.0$} & \phantom{$-0.0$} & \phantom{$-0.0$} & \phantom{$-0.0$} & \phantom{$-$}$1.00$ & $-0.03$ & \phantom{$-$}$0.05$ & \phantom{$-$}$0.02$\\
$R_{\pi^-}^{\pi K}$ & \phantom{$-0.0$} & \phantom{$-0.0$} & \phantom{$-0.0$} & \phantom{$-0.0$} & \phantom{$-0.0$} & \phantom{$-0.0$} & \phantom{$-$}$1.00$ & \phantom{$-$}$0.02$ & \phantom{$-$}$0.08$\\
$R_{K^+}^{\pi K}$ & \phantom{$-0.0$} & \phantom{$-0.0$} & \phantom{$-0.0$} & \phantom{$-0.0$} & \phantom{$-0.0$} & \phantom{$-0.0$} & \phantom{$-0.0$} & \phantom{$-$}$1.00$ & $-0.04$\\
$R_{\pi^+}^{\pi K}$ & \phantom{$-0.0$} & \phantom{$-0.0$} & \phantom{$-0.0$} & \phantom{$-0.0$} & \phantom{$-0.0$} & \phantom{$-0.0$} & \phantom{$-0.0$} & \phantom{$-0.0$} & \phantom{$-$}$1.00$\\
\end{tabular}
\label{tab:stat_corr_full}
\end{table}
\endgroup
\begingroup
\renewcommand*{\arraystretch}{2.0}
\begin{table}[!h]
\centering
\caption{Systematic correlation matrix for $\Bpm \to D h^\pm$ observables.}
\scriptsize
\begin{tabular}{l | c c c c c c c c c }
\toprule
& $A_K^{CP}$ & $A_\pi^{CP}$ & $A_K^{K\pi}$ & $R^{CP}$ & $R_{K/\pi}^{K\pi}$ & $R_{K^-}^{\pi K}$ & $R_{\pi^-}^{\pi K}$ & $R_{K^+}^{\pi K}$ & $R_{\pi^+}^{\pi K}$ \\
\midrule
$A_K^{CP}$ & \phantom{$-$}$1.00$ & \phantom{$-$}$0.13$ & \phantom{$-$}$0.07$ & $-0.22$ & $-0.06$ & $-0.03$ & \phantom{$-$}$0.00$ & \phantom{$-$}$0.02$ & $-0.07$\\
$A_\pi^{CP}$ & \phantom{$-0.0$} & \phantom{$-$}$1.00$ & $-0.74$ & \phantom{$-$}$0.18$ & $-0.05$ & \phantom{$-$}$0.04$ & \phantom{$-$}$0.31$ & $-0.07$ & $-0.21$\\
$A_K^{K\pi}$ & \phantom{$-0.0$} & \phantom{$-0.0$} & \phantom{$-$}$1.00$ & $-0.02$ & \phantom{$-$}$0.00$ & \phantom{$-$}$0.05$ & $-0.24$ & \phantom{$-$}$0.11$ & \phantom{$-$}$0.22$\\
$R^{CP}$ & \phantom{$-0.0$} & \phantom{$-0.0$} & \phantom{$-0.0$} & \phantom{$-$}$1.00$ & $-0.20$ & \phantom{$-$}$0.09$ & \phantom{$-$}$0.11$ & \phantom{$-$}$0.17$ & \phantom{$-$}$0.14$\\
$R_{K/\pi}^{K\pi}$ & \phantom{$-0.0$} & \phantom{$-0.0$} & \phantom{$-0.0$} & \phantom{$-0.0$} & \phantom{$-$}$1.00$ & \phantom{$-$}$0.08$ & $-0.04$ & $-0.06$ & $-0.10$\\
$R_{K^-}^{\pi K}$ & \phantom{$-0.0$} & \phantom{$-0.0$} & \phantom{$-0.0$} & \phantom{$-0.0$} & \phantom{$-0.0$} & \phantom{$-$}$1.00$ & \phantom{$-$}$0.16$ & \phantom{$-$}$0.93$ & \phantom{$-$}$0.13$\\
$R_{\pi^-}^{\pi K}$ & \phantom{$-0.0$} & \phantom{$-0.0$} & \phantom{$-0.0$} & \phantom{$-0.0$} & \phantom{$-0.0$} & \phantom{$-0.0$} & \phantom{$-$}$1.00$ & \phantom{$-$}$0.21$ & \phantom{$-$}$0.84$\\
$R_{K^+}^{\pi K}$ & \phantom{$-0.0$} & \phantom{$-0.0$} & \phantom{$-0.0$} & \phantom{$-0.0$} & \phantom{$-0.0$} & \phantom{$-0.0$} & \phantom{$-0.0$} & \phantom{$-$}$1.00$ & \phantom{$-$}$0.26$\\
$R_{\pi^+}^{\pi K}$ & \phantom{$-0.0$} & \phantom{$-0.0$} & \phantom{$-0.0$} & \phantom{$-0.0$} & \phantom{$-0.0$} & \phantom{$-0.0$} & \phantom{$-0.0$} & \phantom{$-0.0$} & \phantom{$-$}$1.00$\\
\end{tabular}
\label{tab:syst_corr_full}
\end{table}
\endgroup

\begingroup
\renewcommand*{\arraystretch}{2.5}
\begin{sidewaystable}[!h]
\centering
\caption{Statistical correlation matrix for $\Bpm \to D^* h^\pm$ observables.}
\tiny
\begin{tabular}{l | p{0.75cm} p{0.75cm} p{0.75cm} p{0.75cm} p{0.75cm} p{0.75cm} p{0.75cm} p{0.75cm} p{0.75cm} p{0.75cm} p{0.75cm} p{0.75cm} p{0.75cm} p{0.75cm} p{0.75cm} p{0.75cm} p{0.75cm} p{0.75cm} p{0.75cm} }
\toprule
& $A_K^{CP,\gamma}$ & $A_K^{CP,\pi^0}$ & $A_K^{K\pi,\gamma}$ & $A_K^{K\pi,\pi^0}$ & $R^{CP,\gamma}$ & $R^{CP,\pi^0}$ & $R_{K/\pi}^{K\pi,\gamma/\pi^0}$ & $R_{K^-}^{\pi K,\gamma}$ & $R_{K^-}^{\pi K,\pi^0}$ & $R_{K^+}^{\pi K,\gamma}$ & $R_{K^+}^{\pi K,\pi^0}$ & $A_\pi^{CP,\gamma}$ & $A_\pi^{CP,\pi^0}$ & $A_\pi^{K\pi,\gamma}$ & $A_\pi^{K\pi,\pi^0}$ & $R_{\pi^-}^{\pi K,\gamma}$ & $R_{\pi^-}^{\pi K,\pi^0}$ & $R_{\pi^+}^{\pi K,\gamma}$ & $R_{\pi^+}^{\pi K,\pi^0}$ \\
\midrule
$A_K^{CP,\gamma}$ & \phantom{$-$}$1.00$ & $-0.61$ & \phantom{$-$}$0.00$ & \phantom{$-$}$0.00$ & $-0.15$ & \phantom{$-$}$0.07$ & \phantom{$-$}$0.03$ & $-0.03$ & \phantom{$-$}$0.01$ & $-0.03$ & \phantom{$-$}$0.01$ & $-0.01$ & $-0.02$ & \phantom{$-$}$0.01$ & \phantom{$-$}$0.01$ & $-0.01$ & \phantom{$-$}$0.00$ & $-0.02$ & $\phantom{-}0.00$\\
$A_K^{CP,\pi^0}$ & \phantom{$-0.0$} & \phantom{$-$}$1.00$ & \phantom{$-$}$0.00$ & \phantom{$-$}$0.01$ & $-0.05$ & \phantom{$-$}$0.08$ & \phantom{$-$}$0.05$ & $\phantom{-}0.00$ & \phantom{$-$}$0.00$ & $\phantom{-}0.00$ & \phantom{$-$}$0.00$ & \phantom{$-$}$0.03$ & \phantom{$-$}$0.05$ & \phantom{$-$}$0.02$ & \phantom{$-$}$0.03$ & \phantom{$-$}$0.00$ & \phantom{$-$}$0.00$ & $\phantom{-}0.00$ & $\phantom{-}0.00$\\
$A_K^{K\pi,\gamma}$ & \phantom{$-0.0$} & \phantom{$-0.0$} & \phantom{$-$}$1.00$ & $-0.59$ & $\phantom{-}0.00$ & $\phantom{-}0.00$ & $\phantom{-}0.00$ & \phantom{$-$}$0.00$ & \phantom{$-$}$0.00$ & $\phantom{-}0.00$ & $\phantom{-}0.00$ & \phantom{$-$}$0.00$ & \phantom{$-$}$0.01$ & \phantom{$-$}$0.00$ & \phantom{$-$}$0.01$ & \phantom{$-$}$0.00$ & \phantom{$-$}$0.00$ & $\phantom{-}0.00$ & $\phantom{-}0.00$\\
$A_K^{K\pi,\pi^0}$ & \phantom{$-0.0$} & \phantom{$-0.0$} & \phantom{$-0.0$} & \phantom{$-$}$1.00$ & $\phantom{-}0.00$ & \phantom{$-$}$0.00$ & $-0.02$ & \phantom{$-$}$0.00$ & \phantom{$-$}$0.00$ & $\phantom{-}0.00$ & $\phantom{-}0.00$ & \phantom{$-$}$0.01$ & \phantom{$-$}$0.02$ & \phantom{$-$}$0.05$ & \phantom{$-$}$0.08$ & \phantom{$-$}$0.00$ & \phantom{$-$}$0.01$ & $\phantom{-}0.00$ & $-0.01$\\
$R^{CP,\gamma}$ & \phantom{$-0.0$} & \phantom{$-0.0$} & \phantom{$-0.0$} & \phantom{$-0.0$} & \phantom{$-$}$1.00$ & $-0.44$ & $-0.02$ & \phantom{$-$}$0.23$ & $-0.08$ & \phantom{$-$}$0.23$ & $-0.08$ & \phantom{$-$}$0.00$ & $\phantom{-}0.00$ & $\phantom{-}0.00$ & \phantom{$-$}$0.00$ & \phantom{$-$}$0.10$ & $\phantom{-}0.00$ & \phantom{$-$}$0.10$ & $\phantom{-}0.00$\\
$R^{CP,\pi^0}$ & \phantom{$-0.0$} & \phantom{$-0.0$} & \phantom{$-0.0$} & \phantom{$-0.0$} & \phantom{$-0.0$} & \phantom{$-$}$1.00$ & $-0.07$ & $-0.04$ & \phantom{$-$}$0.03$ & $-0.04$ & \phantom{$-$}$0.02$ & $\phantom{-}0.00$ & \phantom{$-$}$0.00$ & \phantom{$-$}$0.00$ & $\phantom{-}0.00$ & $-0.02$ & $\phantom{-}0.00$ & $-0.02$ & \phantom{$-$}$0.00$\\
$R_{K/\pi}^{K\pi,\gamma/\pi^0}$ & \phantom{$-0.0$} & \phantom{$-0.0$} & \phantom{$-0.0$} & \phantom{$-0.0$} & \phantom{$-0.0$} & \phantom{$-0.0$} & \phantom{$-$}$1.00$ & \phantom{$-$}$0.03$ & $-0.03$ & \phantom{$-$}$0.04$ & \phantom{$-$}$0.01$ & $\phantom{-}0.00$ & \phantom{$-$}$0.00$ & \phantom{$-$}$0.01$ & $\phantom{-}0.00$ & \phantom{$-$}$0.02$ & $\phantom{-}0.00$ & \phantom{$-$}$0.02$ & \phantom{$-$}$0.01$\\
$R_{K^-}^{\pi K,\gamma}$ & \phantom{$-0.0$} & \phantom{$-0.0$} & \phantom{$-0.0$} & \phantom{$-0.0$} & \phantom{$-0.0$} & \phantom{$-0.0$} & \phantom{$-0.0$} & \phantom{$-$}$1.00$ & $-0.59$ & \phantom{$-$}$0.79$ & $-0.27$ & \phantom{$-$}$0.00$ & \phantom{$-$}$0.00$ & \phantom{$-$}$0.01$ & \phantom{$-$}$0.01$ & \phantom{$-$}$0.30$ & $-0.03$ & \phantom{$-$}$0.33$ & $-0.02$\\
$R_{K^-}^{\pi K,\pi^0}$ & \phantom{$-0.0$} & \phantom{$-0.0$} & \phantom{$-0.0$} & \phantom{$-0.0$} & \phantom{$-0.0$} & \phantom{$-0.0$} & \phantom{$-0.0$} & \phantom{$-0.0$} & \phantom{$-$}$1.00$ & $-0.27$ & \phantom{$-$}$0.10$ & \phantom{$-$}$0.00$ & \phantom{$-$}$0.00$ & $\phantom{-}0.00$ & \phantom{$-$}$0.00$ & $-0.07$ & \phantom{$-$}$0.05$ & $-0.10$ & \phantom{$-$}$0.05$\\
$R_{K^+}^{\pi K,\gamma}$ & \phantom{$-0.0$} & \phantom{$-0.0$} & \phantom{$-0.0$} & \phantom{$-0.0$} & \phantom{$-0.0$} & \phantom{$-0.0$} & \phantom{$-0.0$} & \phantom{$-0.0$} & \phantom{$-0.0$} & \phantom{$-$}$1.00$ & $-0.60$ & $\phantom{-}0.00$ & $\phantom{-}0.00$ & $-0.01$ & $-0.01$ & \phantom{$-$}$0.32$ & $-0.01$ & \phantom{$-$}$0.30$ & $-0.03$\\
$R_{K^+}^{\pi K,\pi^0}$ & \phantom{$-0.0$} & \phantom{$-0.0$} & \phantom{$-0.0$} & \phantom{$-0.0$} & \phantom{$-0.0$} & \phantom{$-0.0$} & \phantom{$-0.0$} & \phantom{$-0.0$} & \phantom{$-0.0$} & \phantom{$-0.0$} & \phantom{$-$}$1.00$ & \phantom{$-$}$0.00$ & $\phantom{-}0.00$ & $\phantom{-}0.00$ & $\phantom{-}0.00$ & $-0.09$ & \phantom{$-$}$0.04$ & $-0.07$ & \phantom{$-$}$0.05$\\
$A_\pi^{CP,\gamma}$ & \phantom{$-0.0$} & \phantom{$-0.0$} & \phantom{$-0.0$} & \phantom{$-0.0$} & \phantom{$-0.0$} & \phantom{$-0.0$} & \phantom{$-0.0$} & \phantom{$-0.0$} & \phantom{$-0.0$} & \phantom{$-0.0$} & \phantom{$-0.0$} & \phantom{$-$}$1.00$ & \phantom{$-$}$0.05$ & \phantom{$-$}$0.02$ & \phantom{$-$}$0.03$ & \phantom{$-$}$0.00$ & \phantom{$-$}$0.00$ & $\phantom{-}0.00$ & $\phantom{-}0.00$\\
$A_\pi^{CP,\pi^0}$ & \phantom{$-0.0$} & \phantom{$-0.0$} & \phantom{$-0.0$} & \phantom{$-0.0$} & \phantom{$-0.0$} & \phantom{$-0.0$} & \phantom{$-0.0$} & \phantom{$-0.0$} & \phantom{$-0.0$} & \phantom{$-0.0$} & \phantom{$-0.0$} & \phantom{$-0.0$} & \phantom{$-$}$1.00$ & \phantom{$-$}$0.05$ & \phantom{$-$}$0.08$ & \phantom{$-$}$0.00$ & \phantom{$-$}$0.00$ & $\phantom{-}0.00$ & $-0.01$\\
$A_\pi^{K\pi,\gamma}$ & \phantom{$-0.0$} & \phantom{$-0.0$} & \phantom{$-0.0$} & \phantom{$-0.0$} & \phantom{$-0.0$} & \phantom{$-0.0$} & \phantom{$-0.0$} & \phantom{$-0.0$} & \phantom{$-0.0$} & \phantom{$-0.0$} & \phantom{$-0.0$} & \phantom{$-0.0$} & \phantom{$-0.0$} & \phantom{$-$}$1.00$ & $-0.28$ & \phantom{$-$}$0.01$ & \phantom{$-$}$0.01$ & $-0.01$ & $-0.01$\\
$A_\pi^{K\pi,\pi^0}$ & \phantom{$-0.0$} & \phantom{$-0.0$} & \phantom{$-0.0$} & \phantom{$-0.0$} & \phantom{$-0.0$} & \phantom{$-0.0$} & \phantom{$-0.0$} & \phantom{$-0.0$} & \phantom{$-0.0$} & \phantom{$-0.0$} & \phantom{$-0.0$} & \phantom{$-0.0$} & \phantom{$-0.0$} & \phantom{$-0.0$} & \phantom{$-$}$1.00$ & \phantom{$-$}$0.01$ & \phantom{$-$}$0.02$ & $-0.01$ & $-0.02$\\
$R_{\pi^-}^{\pi K,\gamma}$ & \phantom{$-0.0$} & \phantom{$-0.0$} & \phantom{$-0.0$} & \phantom{$-0.0$} & \phantom{$-0.0$} & \phantom{$-0.0$} & \phantom{$-0.0$} & \phantom{$-0.0$} & \phantom{$-0.0$} & \phantom{$-0.0$} & \phantom{$-0.0$} & \phantom{$-0.0$} & \phantom{$-0.0$} & \phantom{$-0.0$} & \phantom{$-0.0$} & \phantom{$-$}$1.00$ & $-0.11$ & \phantom{$-$}$0.33$ & \phantom{$-$}$0.19$\\
$R_{\pi^-}^{\pi K,\pi^0}$ & \phantom{$-0.0$} & \phantom{$-0.0$} & \phantom{$-0.0$} & \phantom{$-0.0$} & \phantom{$-0.0$} & \phantom{$-0.0$} & \phantom{$-0.0$} & \phantom{$-0.0$} & \phantom{$-0.0$} & \phantom{$-0.0$} & \phantom{$-0.0$} & \phantom{$-0.0$} & \phantom{$-0.0$} & \phantom{$-0.0$} & \phantom{$-0.0$} & \phantom{$-0.0$} & \phantom{$-$}$1.00$ & \phantom{$-$}$0.19$ & \phantom{$-$}$0.58$\\
$R_{\pi^+}^{\pi K,\gamma}$ & \phantom{$-0.0$} & \phantom{$-0.0$} & \phantom{$-0.0$} & \phantom{$-0.0$} & \phantom{$-0.0$} & \phantom{$-0.0$} & \phantom{$-0.0$} & \phantom{$-0.0$} & \phantom{$-0.0$} & \phantom{$-0.0$} & \phantom{$-0.0$} & \phantom{$-0.0$} & \phantom{$-0.0$} & \phantom{$-0.0$} & \phantom{$-0.0$} & \phantom{$-0.0$} & \phantom{$-0.0$} & \phantom{$-$}$1.00$ & $-0.11$\\
$R_{\pi^+}^{\pi K,\pi^0}$ & \phantom{$-0.0$} & \phantom{$-0.0$} & \phantom{$-0.0$} & \phantom{$-0.0$} & \phantom{$-0.0$} & \phantom{$-0.0$} & \phantom{$-0.0$} & \phantom{$-0.0$} & \phantom{$-0.0$} & \phantom{$-0.0$} & \phantom{$-0.0$} & \phantom{$-0.0$} & \phantom{$-0.0$} & \phantom{$-0.0$} & \phantom{$-0.0$} & \phantom{$-0.0$} & \phantom{$-0.0$} & \phantom{$-0.0$} & \phantom{$-$}$1.00$\\
\end{tabular}
\label{tab:stat_corr_part}
\end{sidewaystable}
\endgroup
\begingroup
\renewcommand*{\arraystretch}{2.5}
\begin{sidewaystable}[!h]
\centering
\caption{Systematic correlation matrix for $\Bpm \to D^* h^\pm$ observables.}
\tiny
\begin{tabular}{l | p{0.75cm} p{0.75cm} p{0.75cm} p{0.75cm} p{0.75cm} p{0.75cm} p{0.75cm} p{0.75cm} p{0.75cm} p{0.75cm} p{0.75cm} p{0.75cm} p{0.75cm} p{0.75cm} p{0.75cm} p{0.75cm} p{0.75cm} p{0.75cm} p{0.75cm} }\toprule
& $A_K^{CP,\gamma}$ & $A_K^{CP,\pi^0}$ & $A_K^{K\pi,\gamma}$ & $A_K^{K\pi,\pi^0}$ & $R^{CP,\gamma}$ & $R^{CP,\pi^0}$ & $R_{K/\pi}^{K\pi,\gamma/\pi^0}$ & $R_{K^-}^{\pi K,\gamma}$ & $R_{K^-}^{\pi K,\pi^0}$ & $R_{K^+}^{\pi K,\gamma}$ & $R_{K^+}^{\pi K,\pi^0}$ & $A_\pi^{CP,\gamma}$ & $A_\pi^{CP,\pi^0}$ & $A_\pi^{K\pi,\gamma}$ & $A_\pi^{K\pi,\pi^0}$ & $R_{\pi^-}^{\pi K,\gamma}$ & $R_{\pi^-}^{\pi K,\pi^0}$ & $R_{\pi^+}^{\pi K,\gamma}$ & $R_{\pi^+}^{\pi K,\pi^0}$ \\
\midrule
$A_K^{CP,\gamma}$ & \phantom{$-$}$1.00$ & $-0.52$ & \phantom{$-$}$0.50$ & $-0.39$ & $-0.31$ & $-0.71$ & \phantom{$-$}$0.40$ & $-0.15$ & \phantom{$-$}$0.10$ & $-0.05$ & \phantom{$-$}$0.12$ & \phantom{$-$}$0.74$ & $-0.02$ & $-0.70$ & \phantom{$-$}$0.05$ & $-0.16$ & $-0.14$ & $-0.05$ & $-0.17$\\
$A_K^{CP,\pi^0}$ & \phantom{$-0.0$} & \phantom{$-$}$1.00$ & $-0.44$ & $-0.08$ & \phantom{$-$}$0.33$ & \phantom{$-$}$0.62$ & \phantom{$-$}$0.19$ & \phantom{$-$}$0.19$ & $-0.19$ & \phantom{$-$}$0.16$ & $-0.08$ & $-0.49$ & \phantom{$-$}$0.00$ & \phantom{$-$}$0.53$ & $-0.19$ & \phantom{$-$}$0.18$ & \phantom{$-$}$0.01$ & \phantom{$-$}$0.10$ & \phantom{$-$}$0.03$\\
$A_K^{K\pi,\gamma}$ & \phantom{$-0.0$} & \phantom{$-0.0$} & \phantom{$-$}$1.00$ & \phantom{$-$}$0.23$ & \phantom{$-$}$0.02$ & $-0.35$ & \phantom{$-$}$0.38$ & $-0.04$ & \phantom{$-$}$0.03$ & \phantom{$-$}$0.04$ & \phantom{$-$}$0.05$ & \phantom{$-$}$0.57$ & $-0.38$ & $-0.38$ & \phantom{$-$}$0.32$ & $-0.15$ & $-0.19$ & \phantom{$-$}$0.04$ & $-0.09$\\
$A_K^{K\pi,\pi^0}$ & \phantom{$-0.0$} & \phantom{$-0.0$} & \phantom{$-0.0$} & \phantom{$-$}$1.00$ & $-0.19$ & \phantom{$-$}$0.15$ & $-0.53$ & $-0.08$ & \phantom{$-$}$0.06$ & $-0.12$ & $-0.01$ & $-0.54$ & $-0.64$ & \phantom{$-$}$0.42$ & \phantom{$-$}$0.28$ & $-0.05$ & \phantom{$-$}$0.04$ & \phantom{$-$}$0.02$ & \phantom{$-$}$0.24$\\
$R^{CP,\gamma}$ & \phantom{$-0.0$} & \phantom{$-0.0$} & \phantom{$-0.0$} & \phantom{$-0.0$} & \phantom{$-$}$1.00$ & \phantom{$-$}$0.49$ & \phantom{$-$}$0.36$ & \phantom{$-$}$0.36$ & $-0.15$ & \phantom{$-$}$0.41$ & $-0.09$ & \phantom{$-$}$0.22$ & \phantom{$-$}$0.16$ & \phantom{$-$}$0.07$ & \phantom{$-$}$0.00$ & \phantom{$-$}$0.16$ & $-0.06$ & \phantom{$-$}$0.19$ & $-0.07$\\
$R^{CP,\pi^0}$ & \phantom{$-0.0$} & \phantom{$-0.0$} & \phantom{$-0.0$} & \phantom{$-0.0$} & \phantom{$-0.0$} & \phantom{$-$}$1.00$ & $-0.08$ & \phantom{$-$}$0.14$ & $-0.04$ & \phantom{$-$}$0.10$ & $-0.05$ & $-0.35$ & \phantom{$-$}$0.07$ & \phantom{$-$}$0.42$ & $-0.11$ & \phantom{$-$}$0.15$ & \phantom{$-$}$0.08$ & \phantom{$-$}$0.09$ & \phantom{$-$}$0.10$\\
$R_{K/\pi}^{K\pi,\gamma/\pi^0}$ & \phantom{$-0.0$} & \phantom{$-0.0$} & \phantom{$-0.0$} & \phantom{$-0.0$} & \phantom{$-0.0$} & \phantom{$-0.0$} & \phantom{$-$}$1.00$ & \phantom{$-$}$0.16$ & $-0.13$ & \phantom{$-$}$0.26$ & $-0.02$ & \phantom{$-$}$0.62$ & \phantom{$-$}$0.11$ & $-0.22$ & \phantom{$-$}$0.08$ & \phantom{$-$}$0.02$ & $-0.19$ & \phantom{$-$}$0.11$ & $-0.18$\\
$R_{K^-}^{\pi K,\gamma}$ & \phantom{$-0.0$} & \phantom{$-0.0$} & \phantom{$-0.0$} & \phantom{$-0.0$} & \phantom{$-0.0$} & \phantom{$-0.0$} & \phantom{$-0.0$} & \phantom{$-$}$1.00$ & $-0.55$ & \phantom{$-$}$0.98$ & $-0.51$ & \phantom{$-$}$0.00$ & \phantom{$-$}$0.07$ & \phantom{$-$}$0.10$ & \phantom{$-$}$0.00$ & \phantom{$-$}$0.37$ & $-0.04$ & \phantom{$-$}$0.40$ & $-0.04$\\
$R_{K^-}^{\pi K,\pi^0}$ & \phantom{$-0.0$} & \phantom{$-0.0$} & \phantom{$-0.0$} & \phantom{$-0.0$} & \phantom{$-0.0$} & \phantom{$-0.0$} & \phantom{$-0.0$} & \phantom{$-0.0$} & \phantom{$-$}$1.00$ & $-0.53$ & \phantom{$-$}$0.94$ & \phantom{$-$}$0.04$ & $-0.01$ & $-0.10$ & \phantom{$-$}$0.00$ & $-0.13$ & \phantom{$-$}$0.08$ & $-0.15$ & \phantom{$-$}$0.06$\\
$R_{K^+}^{\pi K,\gamma}$ & \phantom{$-0.0$} & \phantom{$-0.0$} & \phantom{$-0.0$} & \phantom{$-0.0$} & \phantom{$-0.0$} & \phantom{$-0.0$} & \phantom{$-0.0$} & \phantom{$-0.0$} & \phantom{$-0.0$} & \phantom{$-$}$1.00$ & $-0.48$ & \phantom{$-$}$0.12$ & \phantom{$-$}$0.04$ & \phantom{$-$}$0.03$ & \phantom{$-$}$0.02$ & \phantom{$-$}$0.37$ & $-0.06$ & \phantom{$-$}$0.41$ & $-0.06$\\
$R_{K^+}^{\pi K,\pi^0}$ & \phantom{$-0.0$} & \phantom{$-0.0$} & \phantom{$-0.0$} & \phantom{$-0.0$} & \phantom{$-0.0$} & \phantom{$-0.0$} & \phantom{$-0.0$} & \phantom{$-0.0$} & \phantom{$-0.0$} & \phantom{$-0.0$} & \phantom{$-$}$1.00$ & \phantom{$-$}$0.06$ & $-0.03$ & $-0.10$ & \phantom{$-$}$0.00$ & $-0.14$ & \phantom{$-$}$0.02$ & $-0.13$ & \phantom{$-$}$0.04$\\
$A_\pi^{CP,\gamma}$ & \phantom{$-0.0$} & \phantom{$-0.0$} & \phantom{$-0.0$} & \phantom{$-0.0$} & \phantom{$-0.0$} & \phantom{$-0.0$} & \phantom{$-0.0$} & \phantom{$-0.0$} & \phantom{$-0.0$} & \phantom{$-0.0$} & \phantom{$-0.0$} & \phantom{$-$}$1.00$ & \phantom{$-$}$0.26$ & $-0.67$ & \phantom{$-$}$0.17$ & $-0.10$ & $-0.18$ & \phantom{$-$}$0.02$ & $-0.25$\\
$A_\pi^{CP,\pi^0}$ & \phantom{$-0.0$} & \phantom{$-0.0$} & \phantom{$-0.0$} & \phantom{$-0.0$} & \phantom{$-0.0$} & \phantom{$-0.0$} & \phantom{$-0.0$} & \phantom{$-0.0$} & \phantom{$-0.0$} & \phantom{$-0.0$} & \phantom{$-0.0$} & \phantom{$-0.0$} & \phantom{$-$}$1.00$ & $-0.02$ & \phantom{$-$}$0.06$ & \phantom{$-$}$0.09$ & \phantom{$-$}$0.08$ & $-0.07$ & $-0.14$\\
$A_\pi^{K\pi,\gamma}$ & \phantom{$-0.0$} & \phantom{$-0.0$} & \phantom{$-0.0$} & \phantom{$-0.0$} & \phantom{$-0.0$} & \phantom{$-0.0$} & \phantom{$-0.0$} & \phantom{$-0.0$} & \phantom{$-0.0$} & \phantom{$-0.0$} & \phantom{$-0.0$} & \phantom{$-0.0$} & \phantom{$-0.0$} & \phantom{$-$}$1.00$ & \phantom{$-$}$0.43$ & \phantom{$-$}$0.13$ & \phantom{$-$}$0.09$ & \phantom{$-$}$0.06$ & \phantom{$-$}$0.17$\\
$A_\pi^{K\pi,\pi^0}$ & \phantom{$-0.0$} & \phantom{$-0.0$} & \phantom{$-0.0$} & \phantom{$-0.0$} & \phantom{$-0.0$} & \phantom{$-0.0$} & \phantom{$-0.0$} & \phantom{$-0.0$} & \phantom{$-0.0$} & \phantom{$-0.0$} & \phantom{$-0.0$} & \phantom{$-0.0$} & \phantom{$-0.0$} & \phantom{$-0.0$} & \phantom{$-$}$1.00$ & $-0.04$ & $-0.05$ & \phantom{$-$}$0.02$ & \phantom{$-$}$0.00$\\
$R_{\pi^-}^{\pi K,\gamma}$ & \phantom{$-0.0$} & \phantom{$-0.0$} & \phantom{$-0.0$} & \phantom{$-0.0$} & \phantom{$-0.0$} & \phantom{$-0.0$} & \phantom{$-0.0$} & \phantom{$-0.0$} & \phantom{$-0.0$} & \phantom{$-0.0$} & \phantom{$-0.0$} & \phantom{$-0.0$} & \phantom{$-0.0$} & \phantom{$-0.0$} & \phantom{$-0.0$} & \phantom{$-$}$1.00$ & \phantom{$-$}$0.53$ & \phantom{$-$}$0.79$ & \phantom{$-$}$0.16$\\
$R_{\pi^-}^{\pi K,\pi^0}$ & \phantom{$-0.0$} & \phantom{$-0.0$} & \phantom{$-0.0$} & \phantom{$-0.0$} & \phantom{$-0.0$} & \phantom{$-0.0$} & \phantom{$-0.0$} & \phantom{$-0.0$} & \phantom{$-0.0$} & \phantom{$-0.0$} & \phantom{$-0.0$} & \phantom{$-0.0$} & \phantom{$-0.0$} & \phantom{$-0.0$} & \phantom{$-0.0$} & \phantom{$-0.0$} & \phantom{$-$}$1.00$ & \phantom{$-$}$0.22$ & \phantom{$-$}$0.60$\\
$R_{\pi^+}^{\pi K,\gamma}$ & \phantom{$-0.0$} & \phantom{$-0.0$} & \phantom{$-0.0$} & \phantom{$-0.0$} & \phantom{$-0.0$} & \phantom{$-0.0$} & \phantom{$-0.0$} & \phantom{$-0.0$} & \phantom{$-0.0$} & \phantom{$-0.0$} & \phantom{$-0.0$} & \phantom{$-0.0$} & \phantom{$-0.0$} & \phantom{$-0.0$} & \phantom{$-0.0$} & \phantom{$-0.0$} & \phantom{$-0.0$} & \phantom{$-$}$1.00$ & \phantom{$-$}$0.41$\\
$R_{\pi^+}^{\pi K,\pi^0}$ & \phantom{$-0.0$} & \phantom{$-0.0$} & \phantom{$-0.0$} & \phantom{$-0.0$} & \phantom{$-0.0$} & \phantom{$-0.0$} & \phantom{$-0.0$} & \phantom{$-0.0$} & \phantom{$-0.0$} & \phantom{$-0.0$} & \phantom{$-0.0$} & \phantom{$-0.0$} & \phantom{$-0.0$} & \phantom{$-0.0$} & \phantom{$-0.0$} & \phantom{$-0.0$} & \phantom{$-0.0$} & \phantom{$-0.0$} & \phantom{$-$}$1.00$\\
\end{tabular}
\label{tab:syst_corr_part}
\end{sidewaystable}
\endgroup

\FloatBarrier

\section*{Acknowledgements}
%
%
\noindent We express our gratitude to our colleagues in the CERN
accelerator departments for the excellent performance of the LHC. We
thank the technical and administrative staff at the LHCb
institutes.
We acknowledge support from CERN and from the national agencies:
CAPES, CNPq, FAPERJ and FINEP (Brazil); 
MOST and NSFC (China); 
CNRS/IN2P3 (France); 
BMBF, DFG and MPG (Germany); 
INFN (Italy); 
NWO (Netherlands); 
MNiSW and NCN (Poland); 
MEN/IFA (Romania); 
MSHE (Russia); 
MICINN (Spain); 
SNSF and SER (Switzerland); 
NASU (Ukraine); 
STFC (United Kingdom); 
DOE NP and NSF (USA).
We acknowledge the computing resources that are provided by CERN, IN2P3
(France), KIT and DESY (Germany), INFN (Italy), SURF (Netherlands),
PIC (Spain), GridPP (United Kingdom), RRCKI and Yandex
LLC (Russia), CSCS (Switzerland), IFIN-HH (Romania), CBPF (Brazil),
PL-GRID (Poland) and OSC (USA).
We are indebted to the communities behind the multiple open-source
software packages on which we depend.
Individual groups or members have received support from
AvH Foundation (Germany);
EPLANET, Marie Sk\l{}odowska-Curie Actions and ERC (European Union);
A*MIDEX, ANR, Labex P2IO and OCEVU, and R\'{e}gion Auvergne-Rh\^{o}ne-Alpes (France);
Key Research Program of Frontier Sciences of CAS, CAS PIFI, CAS CCEPP, 
Fundamental Research Funds for Central Universities, 
and Sci. \& Tech. Program of Guangzhou (China);
RFBR, RSF and Yandex LLC (Russia);
GVA, XuntaGal and GENCAT (Spain);
the Royal Society
and the Leverhulme Trust (United Kingdom).


\addcontentsline{toc}{section}{References}
\bibliographystyle{LHCb}
\bibliography{main,standard,LHCb-PAPER,LHCb-CONF,LHCb-DP,LHCb-TDR}

\newpage
\centerline
{\large\bf LHCb collaboration}
\begin
{flushleft}
\small
R.~Aaij$^{32}$,
C.~Abell{\'a}n~Beteta$^{50}$,
T.~Ackernley$^{60}$,
B.~Adeva$^{46}$,
M.~Adinolfi$^{54}$,
H.~Afsharnia$^{9}$,
C.A.~Aidala$^{85}$,
S.~Aiola$^{26}$,
Z.~Ajaltouni$^{9}$,
S.~Akar$^{65}$
% ...

{\footnotesize \it

% ...
$ ^{9}$Universit{\'e} Clermont Auvergne, CNRS/IN2P3, LPC, Clermont-Ferrand, France\\
% ...
$ ^{26}$INFN Sezione di Milano, Milano, Italy\\
% ...
$ ^{32}$Nikhef National Institute for Subatomic Physics, Amsterdam, Netherlands\\
% ...
$ ^{46}$Instituto Galego de F{\'\i}sica de Altas Enerx{\'\i}as (IGFAE), Universidade de Santiago de Compostela, Santiago de Compostela, Spain\\
% ...
$ ^{50}$Physik-Institut, Universit{\"a}t Z{\"u}rich, Z{\"u}rich, Switzerland\\
% ...
$ ^{54}$H.H. Wills Physics Laboratory, University of Bristol, Bristol, United Kingdom\\
% ...
$ ^{60}$Oliver Lodge Laboratory, University of Liverpool, Liverpool, United Kingdom\\
% ...
$ ^{65}$University of Cincinnati, Cincinnati, OH, United States\\
% ...
$ ^{68}$Syracuse University, Syracuse, NY, United States\\
% ...
$ ^{85}$University of Michigan, Ann Arbor, United States, associated to $^{68}$\\
\bigskip
% ...
\medskip
}
\end{flushleft}

\end{document}